%% file: paper.tex
\documentclass[AMA,STIX1COL]{WileyNJD-v2}

\usepackage[figuresright]{rotating}

\usepackage{algorithm}
\usepackage{algpseudocode}
\usepackage{amsmath}
\usepackage{amssymb}
\usepackage{amsthm}
\usepackage{array}
\usepackage{caption}
\usepackage{csquotes}
\usepackage{enumitem}
\usepackage{graphicx}
\usepackage{hyperref}
\usepackage{latexsym}
\usepackage{mathtools}
\usepackage{multirow}
\usepackage{pdfpages}
\usepackage{textcomp}
\usepackage{xcolor}
\usepackage{xspace}

\usepackage{tikz}
\usetikzlibrary{calc}

\tikzset{
  SortingNetwork/.style={
    yscale=-1, thick,
    dot/.style={circle, fill, inner sep=5pt, transform shape},
    font=\footnotesize,
  },
}

\def\SNGrid#1#2{
  \foreach \x in {1,...,#1} {
    \draw (-0.75,\x) node[left] {\x} -- (#2 + 0.75,\x);
  }
  \def\X{0}
  \def\G{#1}
}
\def\SNConnect#1#2{
  \draw (\X,#1) node [dot] {} -- (\X,#2) node [dot] {};
  \pgfmathsetmacro{\X}{\X+1}
}
\def\SNLevel{
  \draw[semithick, densely dotted, line cap=round] (\X - 0.5,0.5) -- (\X - 0.5,\G + 0.5);
}
\def\SNBack{
  \pgfmathsetmacro{\X}{\X-1}
}

\usepackage{listings}

\makeatletter
\let\commentfullflexible\lst@column@fullflexible
\makeatother

\lstdefinestyle{InFigure}{
  aboveskip=0pt,
  belowskip=-1ex,
}

\newcommand{\ipso}{IPS${}^{\textrm{4}}$o\xspace}

\usepackage{hyperref}
\hypersetup{
  breaklinks=true,
  colorlinks=true,
  citecolor=blue,
  linkcolor=blue,
  urlcolor=blue,
  bookmarksnumbered,
  bookmarksopen,
}

\articletype{Article}%

\received{26 April 2016}
\revised{6 June 2016}
\accepted{6 June 2016}

\begin{document}

\title{Engineering Faster Sorters for Small Sets of Items}

\author[1]{Timo Bingmann*}

\author[1]{Jasper Marianczuk}

\author[1]{Peter Sanders}

\authormark{\textsc{Bingmann et al.}}

\address[1]{%
  \orgdiv{Institute of Theoretical Informatics}, %
  \orgname{Karlsruhe Institute of Technology}, %
  \orgaddress{\city{Karlsruhe}, \country{Germany}}}

\corres{%
  *Timo Bingmann, Karlsruhe Institute of Technology, %
  Am Fasanengarten 5, 76131 Karlsruhe, Germany. %
  \email{bingmann@kit.edu}}

\abstract[Summary]{
  Sorting a set of items is a task that can be useful by itself or as a building block for more complex operations.
  That is why a lot of effort has been put into finding sorting algorithms that sort large sets as efficiently as possible.
  But the more sophisticated and complex the algorithms become, the less efficient they are for small sets of items due to large constant factors.

  A relatively simple sorting algorithm that is often used as a base case sorter is insertion sort, because it has small code size and small constant factors influencing its execution time.

  We aim to determine if there is a faster way to sort small sets of items to provide an efficient base case sorter.
  We looked at sorting networks, at how they can improve the speed of sorting few elements, and how to implement them in an efficient manner using conditional moves.
  Since sorting networks need to be implemented explicitly for each set size, providing networks for larger sizes becomes less efficient due to increased code sizes.
  To also enable the sorting of slightly larger base cases, we adapted sample sort to Register Sample Sort, to break down those larger sets into sizes that can in turn be sorted by sorting networks.

  From our experiments we found that when sorting only small sets of integers, the sorting networks outperform insertion sort by a factor of at least 1.76 for any array size between six and sixteen, and by a factor of 2.72 on average across all machines and array sizes.
  When integrating sorting networks as a base case sorter into Quicksort, we achieved far less performance improvements over using insertion sort, which is probably due to the networks having a larger code size and cluttering the L1 instruction cache.
  The same effect occurs when including Register Sample Sort as a base case sorter for \ipso.
  But for x86 machines that have a larger L1 instruction cache of 64 KiB or more, we obtained speedups of 12.7\% when using sorting networks as a base case sorter in std::sort, and of 5--6\% when integrating Register Sample Sort as a base case sorter into \ipso, each in comparison to using insertion sort as the base case sorter.

  In conclusion, the desired improvement in speed could only be achieved under special circumstances, but the results clearly show the potential of using conditional moves in the field of sorting algorithms.}

\keywords{sorting, sorting algorithm, sorter, base case sorting, sorting networks, sample sort}


\maketitle

\section{Introduction}

\subsection{Motivation}
Sorting, that is rearranging elements of an input set into a specific order, is a fundamental algorithmic challenge.
At universities around the globe, basic sorting algorithms are taught in introductory computer science courses as examples for theoretical analysis of algorithms.
We learn that bubble sort and insertion sort have quadratic asymptotic running time, Quicksort expected $\mathcal{O}(n \log n)$ but worst-case quadratic time, and merge sort always runs in $\mathcal{O}(n \log n)$ time.
These algorithms are analyzed by the number of comparisons they require, both asymptotically, up to constant factors, and sometimes also exactly for small input sizes~$n$.
But later, practical experience shows that pure theoretical analysis cannot tell the whole story.
In real applications and on real hardware factors such as average cases, cache effects, modern CPU features like speculative execution, and of course constant factors actually matter, a lot.
Any well-founded choice on which sorting algorithm to use (for a particular use case) should be influenced by all factors.

The usual playing field for developing new sorting algorithms is to sort a \emph{large number} of items as quickly as possible.
We will call these \emph{complex} sorting algorithms and many follow the divide-and-conquer paradigm.
However, the algorithmic steps for large sets in these complex sorters do not perform well when sorting small sets of items, because they have good asymptotic properties but larger constant factors that become more important for the small sizes.
However also sorting small inputs can be relevant for performance.
This happens for the \emph{base case} of complex sorting algorithms
or when many sorting problems have to be solved. The latter case for example occurs when
the adjacency lists of a graph should be sorted by some criterion in order to support some greedy heuristics.
For example, the \emph{heavy edge matching} heuristics~\cite{karypis1998fast} repeatedly looks for the heaviest unmatched edge
of a vertex.

The most common choice for sorting small inputs is insertion sort, which has a worst-case running time of $\mathcal{O}(n^2)$, but small constant factors that make it suitable to use for small $n$.
When the sorter is executed many times, the total running time does add up to a substantial part of the computing time of an application.
In this paper we therefore attempt to optimize or replace this quadratic sorting algorithm at the heart of most complex sorters and other applications.

Put plainly: \emph{what is the fastest method to sort up to sixteen elements on actual modern hardware?}

We investigate two approaches: first to optimize sorting networks and the second to adapt sample sort to small numbers of items.
When optimizing sorting networks, the most important aspects are how to execute the conditional swap operations on modern CPUs and which sorting network instances to use: the best/shortest known ones, or recursive locality-aware constructions such as Bose-Nelson's~\cite{bose1962sorting}.
Optimizing sorting networks has previously been addressed by Codish, Cruz-Filipe, Nebel, and Schneider-Kamp~\cite{codish2017optimizing} in 2017, but we go much further into the hardware details of the conditional swap operations and use hand-coded assembly employing conditional move instructions.
We focused on using ``simple'' non-vectorized instructions, because they are available on a wider range of architectures.
For slightly larger sets of items (e.g. up to 256), we present Register Sample Sort (RSS), which is an adaptation of Super Scalar Sample Sort~\cite{sanders2004super} to use the registers in the CPU for splitters.
As items we focus on sorting pairs of an integer as key and an associated reference pointer or value.
This enables the sorting of items with complex payload values.
Integers keys fit into a register and can easily be compared using a simple instruction.
Our results can be applied directly to other keys like float and doubles, but not necessarily to larger objects.
However, some of our sorting network implementations are fully generic and can be translated to any data type.

For our experimental evaluation we used four machines: two with Intel CPUs, one with an AMD Ryzen, and a Rockchip RK3399 ARM system on a chip.
It turns out that sorting networks with hand-coded conditional move assembly instructions perform much better (a factor 2.4--5.3 faster) and have a much smaller variance in running time than insertion sort.
However, when integrating sorting networks into Quicksort we unexpectedly saw only a speedup of 7--13\% depending on the machine and base algorithm.
We attribute this to the larger code size of sorting networks and thus L1 instruction cache misses.
Register Sample Sort is also faster than insertion sort for 256 items: up to a factor 1.4 over \texttt{std::sort}.
We then integrated both sorting networks and Register Sample Sort into \ipso, a fast complex comparison-based sorter by Axtmann et al.~\cite{axtmann2017inplace}.
Our experiments validate the authors measurements that \ipso is some 40--60\% faster than \texttt{std::sort}, and were able to show that our better base case sorters improve this by another 1.3\% on Intel CPUs, 5\% on AMD CPUs, and 7\% on the ARM machine.
The ARM machine has higher variance in running time but less outliers than the Intel and AMD ones.
We conclude that the larger code size of sorting networks is a disadvantage, and that Intel and AMD's instruction scheduling and pipelining units are good at accelerating insertion sort.
For ARM machines, better algorithms however make a difference because the CPUs are simpler.

This paper is based on the bachelor's thesis of Jasper Marianczuk~\cite{marianczuk2019engineering}.

\subsection{Overview of the Paper}
Section~\ref{section:networks} is dedicated to sorting networks: Section \ref{section:preliminaries:networks} starts with the general basics of sorting networks and inline assembly code.
After that, we look at different ways of implementing sorting networks efficiently in C++ in Section~\ref{section:implementation-networks}.

In Section~\ref{section:samplesort} we regard Super Scalar Sample Sort and develop an efficient modified version for sets with 256 elements or less by holding the splitters in general purpose registers instead of an array.
The resulting algorithm is called Register Sample Sort.

Section~\ref{section:results} discusses the results and improvements of using sorting networks we achieved in our experiments, measuring the performance of the sorting networks and Register Sample Sort individually, and also including them as base cases into Quicksort and \ipso.
After that we conclude the results of this paper in Section~\ref{section:conclusion}.

\subsection{Related Work}

Sorting is a large and well-studied topic in computer science and the many results fill entire volumes~\cite{knuth1998sorting,mahmoud2011sorting} of related work.
We can focus only on the most relevant here.

Sorting networks are considered in more detail in the following section. The most well-known sorting networks are Bose-Nelson's recursive construction~\cite{bose1962sorting}, Batcher's bitonic and odd-even merge sorts~\cite{batcher1968sorting}, and the AKS (Ajtai, Koml\'os, and Szemer\'edi's) sorting network~\cite{ajtai1983nlogn}.
The publication closest to our work is the paper by Codish et al.\cite{codish2017optimizing}, who optimized sorting networks for locality and instruction-level parallelism.

Sorting networks were used by many authors to vectorize base-case sorting using single-instruction multiple-data (SIMD) instructions in wide registers, as part of larger vectorized sorters.
Inoue et al. proposed AA-sort~\cite{inoue2012high} first in 2007, which is both multi-core parallelized and SIMD vectorized but processes larger blocks of items which are then merged.
Furtak, Amaral, and Niewiadomski used code generation and developed three SIMD implementations~\cite{furtak2007using}, for x86-64's SSE2 and G5's AltiVec instruction sets.
Xiaochen, Rocki, and Suda proposed two algorithms for AVX-512 SIMD~\cite{xiaochen2013register}, one sorting sixteen items in one register, and one sorting 31 items in two registers, both based on bitonic sorting networks.
More recently, Bramas came up with an alternative way to used bitonic sorting networks to sort up to sixteen elements with AVX-512
SIMD instructions\cite{bramas2017novel}.
Hou, Wang, and Feng developed an entire framework~\cite{hou2018framework} which automatically translates abstract sorting networks into vectorized code for different instruction sets. For example, they support Bose-Nelson, Batcher's bitonic, and Hibbard~\cite{hibbard1963empirical} networks and can output AVX, AVX2, and IMCI code.
In 2019, Yin, et al. presented a both parallelized and vectorized sorting algorithm~\cite{yin2019efficient}, which contains a 16$\times$16 SIMD sorting kernel based on bitonic sorting for blocks of 256 items.
Beyond SIMD, sorting networks were also used in various FPGA hardware~\cite{mueller2012sorting,sklyarov2014high}.

Compared with previous work, we consider how to optimize sorting networks without special vectorized instructions or wide registers.
We also consider the variance or standard deviation of running time of our implementations, which is interesting for more accurately predicting code execution time.
None of the referenced papers consider this interesting aspect of sorting networks and branchless execution.

\section{Sorting Networks}\label{section:networks}

\subsection{Introduction} \label{section:preliminaries:networks}
Sorting algorithms can be classified into two groups: those of which the comparison behavior depends on the input and those of which the behavior is \emph{not} influenced by the particular configuration of the input.
Examples of the former are Quicksort~\cite{hoare1962quicksort}, merge sort, insertion sort, etc~\cite{sedgewick1983algorithms,knuth1998sorting}, while the latter are called \emph{data-oblivious}.

One example of data-oblivious sorting algorithms are \emph{sorting networks}.
A sorting network operates on a fixed number $n$ of \emph{channels} enumerated from $1$ to $n$, each representing one input variable, and connections between the channels, called \emph{comparators}.
When two channels are connected by a comparator then the values are compared and conditionally swapped:
if the channel with the lower number holds a value that is greater than the value of the channel with the higher number then the values in the variables are exchanged.

The comparators are given in a fixed order that determines the sequence of executing these \emph{conditional swaps}, such that in the end the channels contain \emph{a permutation} of the original input, and the values held by the channels are in \emph{non-decreasing order}.
Sorting networks are data-oblivious because all comparisons are always performed in the same order, no matter which permutation of an input is given.

The two most important metrics to quantify sorting networks are their \emph{size} and \emph{depth}.
A network's size refers to the \emph{total number of comparators} it contains, and a network's depth describes the \emph{minimal number of levels} a network can be divided into.

Each individual comparator is located on a singleton \emph{level}.
When two comparators do not share a channel and are consecutive, then they can be combined into a common \emph{level}.
Inductively, multiple consecutive comparators can be merged into a level, if their channels are not shared with any other comparator in the level.
Most importantly for performance, since all the comparators in a level are independent from one another, they can be executed in parallel.

\subsubsection{Networks in Practice}

There are many methods to come up with sorting networks which correctly order any input.
\begin{itemize}
\item \textbf{Best known networks:}
  Sorting networks with proven \emph{optimal sizes} and \emph{optimal depths} are known only for small numbers of input channels.
  To date, the optimal depth is known only for $n$ up to seventeen~\cite{knuth1998sorting, parberry1991computer, bundala2014optimal, codish2014twentyfive, ehlers2015new}, while the optimal size only for $n$ up to ten~\cite{codish2014twentyfive}.
  For example, a network for ten elements with optimal size twenty-nine has depth nine, and one with optimal depth seven has size thirty-one~\cite{knuth1998sorting, codish2014twentyfive}.
  For larger networks individual upper and lower bounds on size or depth are known.
  These optimal networks were initially optimized by hand and nowadays are searched for with the help of computers and evolutionary algorithms~\cite{dobbelaere2018sorterhunter}.

\item \textbf{Recursively generated networks:}
  Besides elaborate algorithms searching for optimal networks, there are also much simpler methods to generate correct (but non-optimal) networks.
  The most commonly used paradigm is recursive divide-and-conquer: split the input into two parts, sort each part recursively, and merge the two parts together in the end.
  Representatives for this kind of approach are the constructions of Nelson and Bose~\cite{bose1962sorting} and the algorithms by Batcher~\cite{batcher1968sorting}.

  Bose and Nelson split the input sequence into first and second half, while Batcher partitions it into elements with an even index and elements with an odd index.
  The advantage of these recursive networks over the specially optimized ones is that they can easily be created even for large network sizes.
  While the generated networks may have more comparators than the best known networks, the number of comparators in a network acquired from either Bose-Nelson or Batcher of size $n$ has an upper bound of $\mathcal{O}(n \,(\log n)^2)$.
  This was improved by Ajtai, Koml\'os, and Szemer\'edi~\cite{ajtai1983nlogn} to optimal $\mathcal{O}(\log n)$ depth with the much-cited AKS sorting network, which however have prohibitively high constants and are thus unusable for small $n$.
\end{itemize}

\begin{figure}
  \begin{minipage}{.4\linewidth}
    \centering
    \begin{tikzpicture}[SortingNetwork, scale=0.4]
      \SNGrid{6}{11}

      \SNConnect{2}{3}
      \SNConnect{1}{3}
      \SNConnect{1}{2}

      \SNConnect{5}{6}
      \SNConnect{4}{6}
      \SNConnect{4}{5}

      \SNConnect{1}{4}
      \SNConnect{2}{5}
      \SNConnect{3}{6}

      \SNConnect{3}{5}
      \SNConnect{2}{4}
      \SNConnect{3}{4}
    \end{tikzpicture}
    \caption{Recursively generated sorting network by Bose and Nelson for six
      elements.}
    \label{network:bosenelson:6}
  \end{minipage}%
  \begin{minipage}{.6\linewidth}
    \centering
    \begin{tikzpicture}[SortingNetwork, scale=0.3]
      \SNGrid{10}{28}

      \SNConnect{5}{10}
      \SNConnect{4}{9}
      \SNConnect{3}{8}
      \SNConnect{2}{7}
      \SNConnect{1}{6}
      \SNLevel
      \SNConnect{2}{5}
      \SNConnect{7}{10}
      \SNConnect{1}{4}
      \SNConnect{6}{9}
      \SNLevel
      \SNConnect{1}{3}
      \SNConnect{4}{7}
      \SNConnect{8}{10}
      \SNLevel
      \SNConnect{1}{2}
      \SNConnect{3}{5}
      \SNConnect{6}{8}
      \SNConnect{9}{10}
      \SNLevel
      \SNConnect{2}{3}
      \SNConnect{5}{7}
      \SNConnect{8}{9}
      \SNConnect{4}{6}
      \SNLevel
      \SNConnect{3}{6}
      \SNConnect{7}{9}
      \SNConnect{2}{4}
      \SNConnect{5}{8}
      \SNLevel
      \SNConnect{3}{4}
      \SNConnect{7}{8}
      \SNLevel
      \SNConnect{4}{5}
      \SNConnect{6}{7}
      \SNLevel
      \SNConnect{5}{6}
    \end{tikzpicture}
    \caption{Sorting network with optimal size for ten elements.}
    \label{network:optimal:10}
  \end{minipage}
\end{figure}

Sorting networks are customarily depicted by using horizontal lines for the channels, and undirected vertical connections between these lines for the comparators.
A network by Bose and Nelson for six elements is illustrated in this manner in Figure~\ref{network:bosenelson:6}, and Figure~\ref{network:optimal:10} shows a network with optimal size for ten elements.
The dotted vertical lines indicate the nine levels in the network.

\subsubsection{Improving the Speed of Sorting through Sorting Networks}
The central question to our investigation in this section is how sorting networks can improve the sorting speed on a set of elements (on average), if they can not take any shortcuts for \enquote{good} inputs, like an insertion sort that would leverage an already sorted input and do one comparison per element.
The answer to this question is avoiding penalties introduced by \emph{branching}.
Because the compiler and CPU know in advance which comparisons are going to be executed in which order, the control flow does not contain conditional branches, which in particular gets rid of expensive branch mispredictions and allows instruction level (superscalar) parallelism.
On uniformly distributed random inputs, the chances that any number is smaller than another is 50\% on average, making branches unpredictable.
In the case of insertion sort that means not knowing in advance with how many elements the next one has to be compared until it is inserted into the right place (on average it would be half of them).

Even though with sorting networks the compiler knows in advance when to execute which comparator, implementing the conditional-swap operation in a naive way (as seen in Section~\ref{section:preliminaries:conditional-swap}) the compiler might still generate branches. In that case, the sorting networks are no faster than insertion sort, or even slower. Hence, we investigated the use of assembly code in this paper.

Another interesting use of sorting networks may be in the field of cryptography and security-focused applications.
The time it takes to sort with non-data-oblivious algorithms (e.g. Quicksort) may introduce side channels allowing an attacker to infer the order of the input elements.

\subsubsection{Conditional Swap}\label{section:preliminaries:conditional-swap}
For sorting networks, the basic operation is to compare two values against each other and swap them if they are in the wrong order (the \enquote{smaller} element occurs after the \enquote{larger} one in the sequence).
This \emph{conditional-swap} operation can be implemented straight-forwardly in C++ with an \texttt{if} and a \texttt{swap}:
\begin{lstlisting}
void ConditionalSwap(Type& left, Type& right) {
    if (right < left) { std::swap(left, right); }
}
\end{lstlisting}
Here \texttt{Type} is a template and can be instantiated with any type that implements the \texttt{<} operator.
As suggested by Codish~et.~al.~\cite{codish2017optimizing}, the same piece of code can be rewritten like this:
\begin{lstlisting}
void ConditionalSwap2(Type& left, Type& right) {
    Type temp = left;
    if (right < temp) { left = right; }
    if (right < temp) { right = temp; }
}
\end{lstlisting}
At first glance it appears as if there are now two conditional branches.
But since the statement executed when the condition is true now only consists of a single assignment each, these can be expressed in x86-architecture with a \emph{conditional move} instruction.
In AT\&T syntax (see Section~\ref{section:preliminaries:asm}), a conditional move (\texttt{cmov a,b}) will write the value of register \texttt{a} into register \texttt{b}, if a condition is met.
If the condition is not met, no operation takes place (but still taking the same number of CPU cycles as the move operation would have).
Since the address of the next instruction no longer depends upon the previous condition, the control flow now does not contain branches.
This avoids the large cost of \emph{branch mispredictions} which require the execution pipeline of the CPU to be flushed and many speculative operations to be undone.
The only downside of conditional moves is that they may take longer to evaluate than a normal move instruction on certain architectures, and can only be executed when the comparison has performed and its result is available.
They can also only operate on register entries.

When the elements to be swapped are plain integers, some compilers do generate code with conditional moves for those operations while others default to jump branches.
In the experiments in Section~\ref{section:results} we consider pairs of an unsigned 64-bit integer key and an unsigned 64-bit reference value, which could be a pointer to data or an address in an array.
To force the usage of conditional moves, we investigated use of \emph{inline assembly}~\cite{GccInlineAssembly}, a feature of gcc and other compilers that allows the programmer to specify small amounts of assembly code to be inserted into the generated machine code. This technique and the notation is further explained in the following Section~\ref{section:preliminaries:asm}.

\subsubsection{Inline Assembly Code} \label{section:preliminaries:asm}
In this section we introduce the reader to a relatively obscure feature in modern C/C++ compilers: inline assembly code.
We will use it in the next section to hard-code conditional swap operations and thus bypass the compiler's optimizations for these operations.

The machine instructions executed by the CPU are also called \emph{assembly code}, which can be expressed as the actual op-codes or as human-readable text.
There are two competing conventions for the textual representation: the Intel syntax or MASM syntax and the AT\&T syntax.

The main differences are the parameter order and operand size.
In Intel syntax the destination parameter is written first, then the source of the value: \texttt{mov dest,src}, while the size of the operands need not be specified.
In AT\&T syntax on the other hand, the source parameter is written first, followed by the destination: \texttt{movq src,dest}.
The size of the operand must be appended to the instruction: \enquote{\texttt{b}} (byte = 8 bit), \enquote{\texttt{l}} (long = 32 bit), \enquote{\texttt{q}} (quad-word = 64 bit).
In this paper only the \emph{AT\&T syntax} will be used, because it is used internally by gcc.

The gcc and clang C++ compilers have a feature that allows the programmer to write assembly instructions in between regular C++ code, called \enquote{inline assembly} (\texttt{asm}) \cite{GccInlineAssembly}.
This inline code consists of a continuous piece of assembly code, together with a specification of how it should interact with the surrounding C++ code.
This specification communicates to the compiler what happens inside the \texttt{asm} block and consists of a definition for \emph{input} and \emph{output} variables and a list of \emph{clobbered registers}.
Gcc does not optimize the given assembly statements, they are added into the generated assembly code verbatim and translated to machine code by the GNU Assembler.

A variable listed as output means that the value will be modified, a clobbered register is one where gcc cannot assume that the value it held before the \texttt{asm} block will be the same as after the block.
In this paper, the clobbered registers will almost always be the conditional-codes registers (\enquote{\texttt{cc}}), which include the carry flag, zero flag and the signed flag, which are modified during a compare-instruction.
This way of specifying the input, output, and clobbered registers is also called \emph{extended asm}.

Taking the code from Section~\ref{section:preliminaries:conditional-swap}, and assuming \texttt{Type = uint64\_t}, the statement
\begin{lstlisting}
uint64_t temp = left;
if (right < temp) {
    left = right;
}
\end{lstlisting}
can now be written with extended inline x86 assembly as
\begin{lstlisting}
uint64_t temp = left;
__asm__(
    "cmpq %[temp],%[right]\n\t"                       // compare right and temp
    "cmovbq %[right],%[left]\n\t"                     // left = right, if right < temp
    : [left] "=&r"(left)                              // output variables
    : "0"(left), [right] "r"(right), [temp] "r"(temp) // input variables
    : "cc"                                            // clobbered registers
);
\end{lstlisting}
In extended \texttt{asm}, one can define C++ variables as input or output operands.
For inputs the compiler will assign a register if it has the \enquote{\texttt{r}} modifier and load the value into it.
For outputs, a register is allocated and the value is written back to the given variable after the \texttt{asm} block or used immediately in further steps.
The names in square brackets are symbolic names only valid in the context of the assembly instructions and independent from the names in the C++ code before.
The link between the C++ names and the symbolic names is defined in the input and output declarations, which may or may not be the same.

For conditional moves it is important to properly declare the input and output variables, because they perform a task that is a bit unusual: an output variable may or may not be overwritten.
In the case of the output register for \texttt{left} used above, two things must apply: if the condition is false, it must hold the value of \texttt{left}, and if the condition is true, it must hold the value of \texttt{right}.

For optimizations purposes, the compiler might reduce the number of registers used by placing the output of one operation into a register that previously held the input for some other operation.
To prevent this, the declaration for the output \texttt{[left] "=\&r"(left)} has the \enquote{\texttt{\&}} modifier added to it, meaning it is an \enquote{early clobber} register and that no other input can be placed in that register.
In combination with \texttt{"0"(left)} in the input definition, the same register is additionally tied to an input, such that the previous value of \texttt{left} is loaded beforehand, in case the conditional move is not executed.
Because \texttt{left} was already declared as output, instead of giving it a new symbolic name we tie it to the input by referencing its index \texttt{"0"} in the output list.

The \enquote{\texttt{=}} in the output declaration only means that this register will be written to.
Any output needs to have the \enquote{\texttt{=}} modifier.
Each assembly instruction is postfixed with \enquote{\texttt{\textbackslash{}n\textbackslash{}t}} because the strings are appended into a single long line by the C++ compiler and the line breaks separate instructions for the assembler.

The \texttt{cmov} instruction is postfixed with \enquote{\texttt{b}} in this example, which stands for \emph{below}, such that the move is executed if \texttt{right} is below \texttt{temp} (unsigned comparison \texttt{right < temp}).
Apart from \emph{below} we will also see \emph{not equal} (\enquote{\texttt{ne}}) and \emph{carry} (\enquote{\texttt{c}}) as a postfix in further examples.
Furthermore, both the \texttt{cmp} and the \texttt{cmovb} are postfixed with \enquote{\texttt{q}} (quad-word) to indicate that the operands are 64-bit values.

When a subtraction $(\texttt{minuend} - \texttt{subtrahend})$ is performed and \texttt{subtrahend} is larger than \texttt{minuend} (interpreted as unsigned numbers), the operation causes an underflow which results in the carry flag being set.
The carry flag can be used as a condition by itself (postfix \enquote{\texttt{c}}) and it also influences condition checks like \emph{below}.
This property of the comparison setting the carry flag will be used in Section~\ref{section:samplesort:impl}.

\subsection{Implementation of Sorting Networks}\label{section:implementation-networks}

We now consider how to actually implement sorting networks for performance.

\subsubsection{Providing the Network Frame}\label{section:network-frame}

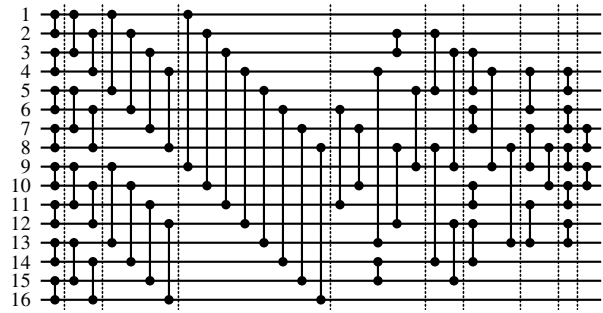
\begin{figure}
  \centering
  \begin{tikzpicture}[SortingNetwork, scale=0.25]
    \SNGrid{16}{28}

    \SNConnect{1}{2}\SNBack
    \SNConnect{3}{4}\SNBack
    \SNConnect{5}{6}\SNBack
    \SNConnect{7}{8}\SNBack
    \SNConnect{9}{10}\SNBack
    \SNConnect{11}{12}\SNBack
    \SNConnect{13}{14}\SNBack
    \SNConnect{15}{16}
    \SNLevel
    \SNConnect{1}{3}\SNBack
    \SNConnect{5}{7}\SNBack
    \SNConnect{9}{11}\SNBack
    \SNConnect{13}{15}
    \SNConnect{2}{4}\SNBack
    \SNConnect{6}{8}\SNBack
    \SNConnect{10}{12}\SNBack
    \SNConnect{14}{16}
    \SNLevel
    \SNConnect{1}{5}\SNBack
    \SNConnect{9}{13}
    \SNConnect{2}{6}\SNBack
    \SNConnect{10}{14}
    \SNConnect{3}{7}\SNBack
    \SNConnect{11}{15}
    \SNConnect{4}{8}\SNBack
    \SNConnect{12}{16}
    \SNLevel
    \SNConnect{1}{9}
    \SNConnect{2}{10}
    \SNConnect{3}{11}
    \SNConnect{4}{12}
    \SNConnect{5}{13}
    \SNConnect{6}{14}
    \SNConnect{7}{15}
    \SNConnect{8}{16}
    \SNLevel
    \SNConnect{6}{11}
    \SNConnect{7}{10}
    \SNConnect{4}{13}\SNBack
    \SNConnect{14}{15}
    \SNConnect{8}{12}\SNBack
    \SNConnect{2}{3}
    \SNConnect{5}{9}
    \SNLevel
    \SNConnect{2}{5}\SNBack
    \SNConnect{8}{14}
    \SNConnect{3}{9}\SNBack
    \SNConnect{12}{15}
    \SNLevel
    \SNConnect{3}{5}\SNBack
    \SNConnect{6}{7}\SNBack
    \SNConnect{10}{11}\SNBack
    \SNConnect{12}{14}
    \SNConnect{4}{9}
    \SNConnect{8}{13}
    \SNLevel
    \SNConnect{7}{9}\SNBack
    \SNConnect{11}{13}\SNBack
    \SNConnect{4}{6}
    \SNConnect{8}{10}
    \SNLevel
    \SNConnect{4}{5}\SNBack
    \SNConnect{6}{7}\SNBack
    \SNConnect{8}{9}\SNBack
    \SNConnect{10}{11}\SNBack
    \SNConnect{12}{13}
    \SNLevel
    \SNConnect{7}{8}\SNBack
    \SNConnect{9}{10}
  \end{tikzpicture}
  \caption{Sorting network with optimal size for sixteen elements.}
  \label{network:best:16}
\end{figure}

\begin{figure}
  \centering
  \begin{tikzpicture}[SortingNetwork, scale=0.25]
    \SNGrid{16}{54}

    \SNConnect{1}{2}\SNBack
    \SNConnect{3}{4}
    \SNConnect{1}{3}
    \SNConnect{2}{4}
    \SNConnect{2}{3}\SNBack
    \SNConnect{5}{6}\SNBack
    \SNConnect{7}{8}
    \SNConnect{5}{7}
    \SNConnect{6}{8}
    \SNConnect{6}{7}\SNBack
    \SNConnect{1}{5}
    \SNConnect{2}{6}
    \SNConnect{2}{5}
    \SNConnect{3}{7}
    \SNConnect{4}{8}
    \SNConnect{4}{7}
    \SNConnect{3}{5}
    \SNConnect{4}{6}
    \SNConnect{4}{5}\SNBack
    \SNConnect{9}{10}\SNBack
    \SNConnect{11}{12}
    \SNConnect{9}{11}
    \SNConnect{10}{12}
    \SNConnect{10}{11}\SNBack
    \SNConnect{13}{14}\SNBack
    \SNConnect{15}{16}
    \SNConnect{13}{15}
    \SNConnect{14}{16}
    \SNConnect{14}{15}\SNBack
    \SNConnect{9}{13}
    \SNConnect{10}{14}
    \SNConnect{10}{13}
    \SNConnect{11}{15}
    \SNConnect{12}{16}
    \SNConnect{12}{15}
    \SNConnect{11}{13}
    \SNConnect{12}{14}
    \SNConnect{12}{13}\SNBack
    \SNConnect{1}{9}
    \SNConnect{2}{10}
    \SNConnect{2}{9}
    \SNConnect{3}{11}
    \SNConnect{4}{12}
    \SNConnect{4}{11}
    \SNConnect{3}{9}
    \SNConnect{4}{10}
    \SNConnect{4}{9}
    \SNConnect{5}{13}
    \SNConnect{6}{14}
    \SNConnect{6}{13}
    \SNConnect{7}{15}
    \SNConnect{8}{16}
    \SNConnect{8}{15}
    \SNConnect{7}{13}
    \SNConnect{8}{14}
    \SNConnect{8}{13}
    \SNConnect{5}{9}
    \SNConnect{6}{10}
    \SNConnect{6}{9}
    \SNConnect{7}{11}
    \SNConnect{8}{12}
    \SNConnect{8}{11}
    \SNConnect{7}{9}
    \SNConnect{8}{10}
    \SNConnect{8}{9}

    \def\SNOutline#1#2#3#4#5#6{
      \draw [blue,rounded corners=2pt]
      ($(#1) + (#2) + (-\m,-\m)$) -- ($(#1) + (#3) + (\m,-\m)$) --
      ($(#1) + (#4) + (\m,0)$) -- ($(#1) + (#5) + (\m,\m)$) -- ($(#1) + (#6) + (-\m,\m)$) -- cycle;
    }
    \def\m{10pt}
    \SNOutline{0,0}{0,1}{2,1}{4,2.5}{2,4}{0,4}
    \SNOutline{3,4}{0,1}{2,1}{4,2.5}{2,4}{0,4}
    \SNOutline{14,8}{0,1}{2,1}{4,2.5}{2,4}{0,4}
    \SNOutline{17,12}{0,1}{2,1}{4,2.5}{2,4}{0,4}
    \def\m{16pt}
    \SNOutline{0,0}{0,1}{12,1}{15,4.5}{12,8}{0,8}
    \SNOutline{14,8}{0,1}{12,1}{15,4.5}{12,8}{0,8}
  \end{tikzpicture}
  \caption{Bose-Nelson network for sixteen elements optimizing locality
    (unrolled or recursive implementations). The subnetworks marked blue sort
    eight and four items, respectively.}
  \label{network:bosenelson:16}
\end{figure}
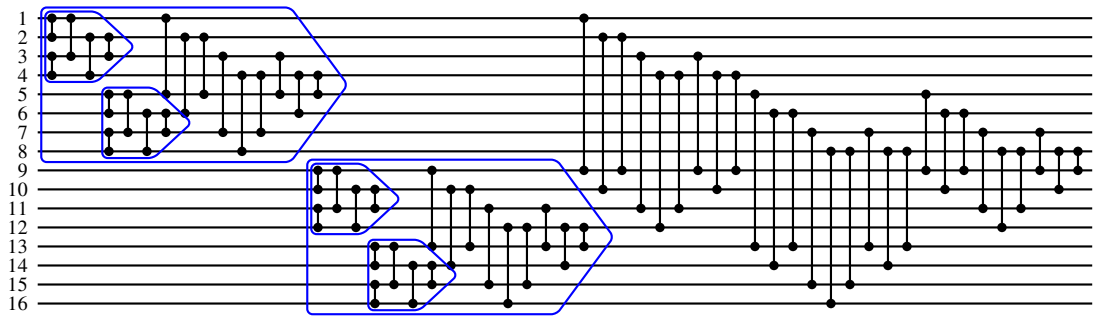

\begin{figure}
  \centering
  \begin{tikzpicture}[SortingNetwork, scale=0.25]
    \SNGrid{16}{47}

    \SNConnect{1}{2}\SNBack
    \SNConnect{3}{4}\SNBack
    \SNConnect{5}{6}\SNBack
    \SNConnect{7}{8}\SNBack
    \SNConnect{9}{10}\SNBack
    \SNConnect{11}{12}\SNBack
    \SNConnect{13}{14}\SNBack
    \SNConnect{15}{16}
    \SNConnect{1}{3}
    \SNConnect{2}{4}\SNBack
    \SNConnect{5}{7}
    \SNConnect{6}{8}\SNBack
    \SNConnect{9}{11}
    \SNConnect{10}{12}\SNBack
    \SNConnect{13}{15}
    \SNConnect{14}{16}
    \SNConnect{2}{3}\SNBack
    \SNConnect{6}{7}
    \SNConnect{1}{5}\SNBack
    \SNConnect{10}{11}\SNBack
    \SNConnect{14}{15}
    \SNConnect{9}{13}
    \SNConnect{2}{6}
    \SNConnect{3}{7}
    \SNConnect{4}{8}\SNBack
    \SNConnect{10}{14}
    \SNConnect{11}{15}
    \SNConnect{12}{16}
    \SNConnect{2}{5}
    \SNConnect{4}{7}\SNBack
    \SNConnect{10}{13}
    \SNConnect{12}{15}
    \SNConnect{3}{5}
    \SNConnect{4}{6}\SNBack
    \SNConnect{11}{13}
    \SNConnect{12}{14}
    \SNConnect{4}{5}\SNBack
    \SNConnect{12}{13}
    \SNConnect{1}{9}
    \SNConnect{2}{10}
    \SNConnect{3}{11}
    \SNConnect{4}{12}
    \SNConnect{5}{13}
    \SNConnect{6}{14}
    \SNConnect{7}{15}
    \SNConnect{8}{16}
    \SNConnect{2}{9}
    \SNConnect{4}{11}
    \SNConnect{6}{13}
    \SNConnect{8}{15}
    \SNConnect{3}{9}
    \SNConnect{4}{10}
    \SNConnect{7}{13}
    \SNConnect{8}{14}
    \SNConnect{4}{9}
    \SNConnect{8}{13}
    \SNConnect{5}{9}
    \SNConnect{6}{10}
    \SNConnect{7}{11}
    \SNConnect{8}{12}
    \SNConnect{6}{9}
    \SNConnect{8}{11}
    \SNConnect{7}{9}
    \SNConnect{8}{10}
    \SNConnect{8}{9}
  \end{tikzpicture}
  \caption{Bose-Nelson network for sixteen elements optimizing parallelism}
  \label{network:bosenelson:parl:16}
\end{figure}
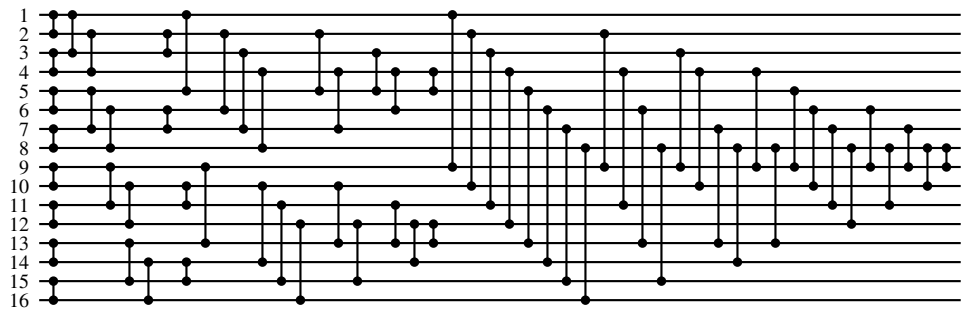

We collected the following sorting networks for small inputs:
For sizes of up to sixteen elements the best known networks were taken from John Gamble's Website~\cite{gamble2019networks} and are size-optimal up to ten elements.
The Bose-Nelson networks have been generated using the instructions from their paper~\cite{bose1962sorting}.
We did not use any Batcher odd-even network because Codish~et.~al.~\cite{codish2017optimizing} showed that there was no difference between Batcher and Bose-Nelson in practice.

For sizes of eight and below the best and generated networks have the same amount of comparators and levels.
For sizes larger than eight the generated networks are at a disadvantage because they have more comparators and/or levels.
As a trade-off their recursive structure makes it possible to leverage a different trait: \emph{locality}.
Instead of optimizing them to sort in as few levels as possible, we can first sort the first half of the set, then the second half, and then apply the merger.
Thus chances are higher that all $\frac{n}{2}$ elements of the first half may fit into the processor's general purpose registers.
To determine if there is an achievable speedup, the networks were generated optimizing for (a) \emph{locality} and (b) \emph{parallelism}.

Furthermore, we investigated two implementations of locality-optimizing recursively defined Bose-Nelson networks: one where the entire network is rolled out for each input size individually, and a second were the functions of each input size may call smaller sorters for parts of the input.
The first unrolled variant retains the name \emph{locality}, while we call the second one a \emph{recursive} implementation.

Examples of networks for sixteen elements can be seen in Figures~\ref{network:best:16}, \ref{network:bosenelson:16} and \ref{network:bosenelson:parl:16}.

All networks are implemented such that they have an entry method that takes a pointer to an array \texttt{A} and an array size \texttt{n} as input and delegates the call to the specific method for that number of elements.
To measure different implementations for the conditional swaps, the sorting networks are templated with both the swap and the item type.

Our approach differs from most previous work~\cite{codish2017optimizing} in the type of elements that were sorted.
While most experiments measured the sorting of plain \texttt{int}s, which are usually 32-bit sized integers, we made the decision to sort elements that consist of a 64-bit integer key and a 64-bit integer reference value.
This enables not only sorting of numbers but also of complex payloads, by giving a pointer or an array index as the reference value.
This was implemented by creating a \texttt{struct} that contain a key and reference value each, having the following structure:
\begin{lstlisting}
struct SortableRef {
    uint64_t key, ref;
}
\end{lstlisting}
We also defined the operators \texttt{>}, \texttt{>=}, \texttt{==}, \texttt{<}, \texttt{<=}, and \texttt{!=} for usability.
Other keys than integers are also possible: floats or doubles also fit into a single register and can be compared using a single CPU instruction.

\subsubsection{Implementing the Conditional Swap} \label{section:implementation-conditionalswap}

\texttt{ConditionalSwap} is implemented as a templated method like this:
\begin{lstlisting}
template <typename Type>
inline void ConditionalSwap(Type& left, Type& right) {
    // body
}
\end{lstlisting}

Our goal is thus to find the best instructions to implement this method, either by convincing the compiler to produce good code or by writing inline assembly instructions directly.
The following variants will represent the body of one specialization of the template function for a specific struct.
Each of them was given a three to four letter abbreviation to name them in the results.
We implemented the following approaches:

\def\P{\ }
\def\ISwp{\texttt{ISwp} (if and std::swap)}
\def\TCOp{\texttt{TCOp} (ternary conditional operators)}
\def\Tie{\texttt{Tie\P} (std::tie with std::tuple)}
\def\JXhg{\texttt{JXhg} (jmp and xchg)}
\def\FCm{\texttt{4Cm\P} (four cmovs and temp variables)}
\def\FCmS{\texttt{4CmS} (four cmovs split and temp variables)}
\def\SCm{\texttt{6Cm\P} (six cmovs and temp variables)}
\def\TCPm{\texttt{2CPm} (move pointers with two cmovs)}
\def\TCPp{\texttt{2CPp} (move pointers with two cmovs and predicate)}

\begin{description}
\item[\ISwp]:\quad
  This is the simplest way of writing the conditional-swap operation, without any inline assembly as a C/C++ \texttt{if} statement with a \texttt{std::swap} (Figure~\ref{fig:ISwp}). We already saw this code in Section~\ref{section:preliminaries:conditional-swap}.

\item[\TCOp]:\quad
  This is an alternative portable C/C++ implementation without inline assembly.
  It uses two \emph{ternary conditional operators} (``\texttt{?:}'') and a temporary variable (Figure~\ref{fig:TCOp}).
  The goal of this method is to try to convince the compiler to generate conditional moves, which seems to happen more often for the ternary operator.

\item[\Tie]:\quad
  This is another portable C/C++ implementation using an assignment of a pair of variables using \emph{structured bindings}.
  In the current C++ versions this can be expressed with \texttt{std::tuple} and \texttt{std::tie} (Figure~\ref{fig:Tie}).
  Again this is portable, and the compiler is tasked to generate efficient code from it.

\item[\JXhg]:\quad
  This is a straight-forward inline assembly version implemented using a comparison, a conditional jump, and two exchange (\texttt{xchg}) instructions.
  If \texttt{right\_key} is above \texttt{left\_key} or equal to it, the \texttt{xchg} instructions are skipped (\textbf{j}ump \textbf{a}bove or \textbf{e}qual, \texttt{jae}). \texttt{xchg} swaps the contents of two registers.
  The \enquote{\texttt{\%=}} directive generates a unique label for each \texttt{asm} instance (Figure~\ref{fig:JXhg}).

\item[\FCm]:\quad
  This is the shortest assembly implementation with a comparison and four conditional move (\texttt{cmov}) operations.
  Since we need to swap a key-reference pair of 64-bit integers, we need four \texttt{cmov}s and two temporary variables (Figure~\ref{fig:4Cm}).
  The left key and value are stored in the temporary variables, unconditionally.
  Then the two keys are compared in assembly and the result is stored in the flags register.
  If the items need to swapped, then the four \texttt{cmov}s first move the right item to the left, and then the temporary item into the right.
  The issue with this version is that there is an unconditional copy of the left pair into the temporary variables.

\item[\SCm]:\quad
  In \texttt{4Cm} the temporary variables are unconditionally assigned outside the assembly block.
  This is unnecessary if the condition is false, such that we proposed this variant with six \texttt{cmov}s (Figure~\ref{fig:6Cm}).
  As before, the keys are compared in assembly, but this time, six \texttt{cmov}s are necessary to swap the items: first copy the left pair into temporary variables, then replace the left with the right pair, and then move the temporaries into the right.
  If the items need not be swapped, then the temporary variable is not changed.

\item[\FCmS]:\quad
  Since the C++ compiler cannot reorder operations inside inline assembly blocks, we attempted to split these such that the compiler can \emph{interleave} load/store operations from multiple consecutive conditional-swaps to avoid memory stalls.
  There is obviously a limit to this reordering, which requires the \texttt{asm} blocks to be declared \texttt{volatile} such that these stay in order among themselves.
  The \texttt{cmov}s are the same as in the \texttt{4Cm} version, except they are split into different blocks (Figure~\ref{fig:4CmS}).

\item[\TCPm]:\quad
  The following two variants are based on applying \texttt{cmov} to \emph{pointers} to the \texttt{SortableRef} instead of moving a pair of key and value.
  The idea is based on assembly code generated by the clang compiler for the \texttt{ConditionalSwap2} method in Section~\ref{section:preliminaries:conditional-swap}.
  First we declare two pointers to the items, and copy the left item into a temporary variable.
  After the comparison, the right pointer is copied into the left pointer using a \texttt{cmov}, and then the temporary pointer into the right pointer.
  After the assembly block, the pointers are used to actually swap the key and value pairs using assignments in C/C++, unconditionally.
  We split the assembly into two blocks and let the compiler optimize and interleave load and store operations (Figure~\ref{fig:2CPm}).

\item[\TCPp]:\quad
  Instead of performing the comparison inside the \texttt{asm} block, which requires knowledge of the datatype of the key, it can also be done using an indicator predicate.
  This predicate is simply an integer variable that is passed into the assembly code.
  Combining predicates with the pointer swapping technique from \texttt{2CPm} delivers this variant which can operate on keys with custom comparators and any data payload (Figure~\ref{fig:2CPp}).
  The goal is simply to see how much slower the external predicate is.

\end{description}

\def\P{}

\begin{figure}
  \begin{minipage}{.5\linewidth}
    \begin{lstlisting}[style=InFigure]
if (right < left)
    std::swap(left, right);
    \end{lstlisting}
    \caption{\ISwp}\label{fig:ISwp}
    \bigskip

    \begin{lstlisting}[style=InFigure]
bool r = (right < left);
auto temp = left;
left = r ? right : left;
right = r ? temp : right;
    \end{lstlisting}
    \caption{\TCOp}\label{fig:TCOp}
    \bigskip

    \begin{lstlisting}[style=InFigure]
std::tie(left, right) = (right < left)
    ? std::make_tuple(right, left)
    : std::make_tuple(left, right);
    \end{lstlisting}
    \caption{\Tie}\label{fig:Tie}
  \end{minipage}%
  \begin{minipage}{.5\linewidth}
    \begin{lstlisting}[style=InFigure]
__asm__(
    "cmpq %[left_key],%[right_key]\n\t"
    "jae %=f\n\t"
    "xchg %[left_key],%[right_key]\n\t"
    "xchg %[left_ref],%[right_ref]\n\t"
    "%=:\n\t"
    : [left_key] "=&r"(left.key),
      [right_key] "=&r"(right.key),
      [left_ref] "=&r"(left.ref),
      [right_ref] "=&r"(right.ref)
    : "0"(left.key), "1"(right.key),
      "2"(left.ref), "3"(right.ref)
    : "cc"
);
    \end{lstlisting}
    \caption{\JXhg}\label{fig:JXhg}
  \end{minipage}
\end{figure}

\begin{figure}
  \begin{minipage}[t]{.5\linewidth}
    \medskip

    \begin{lstlisting}[style=InFigure]
uint64_t tmp = left.key;
uint64_t tmp_ref = left.ref;
__asm__(
    "cmpq %[left_key],%[right_key]\n\t"
    "cmovbq %[right_key],%[left_key]\n\t"
    "cmovbq %[right_ref],%[left_ref]\n\t"
    "cmovbq %[tmp],%[right_key]\n\t"
    "cmovbq %[tmp_ref],%[right_ref]\n\t"
    : [left_key] "=&r"(left.key),
      [right_key] "=&r"(right.key),
      [left_ref] "=&r"(left.ref),
      [right_ref] "=&r"(right.ref)
    : "0"(left.key), "1"(right.key),
      "2"(left.ref), "3"(right.ref),
      [tmp] "r"(tmp), [tmp_ref] "r"(tmp_ref)
    : "cc"
);
    \end{lstlisting}
    \caption{\FCm}\label{fig:4Cm}
    \bigskip

    \begin{lstlisting}[style=InFigure]
uint64_t tmp;
uint64_t tmp_ref;
__asm__ (
    "cmpq %[left_key],%[right_key]\n\t"
    "cmovbq %[left_key],%[tmp]\n\t"
    "cmovbq %[left_ref],%[tmp_ref]\n\t"
    "cmovbq %[right_key],%[left_key]\n\t"
    "cmovbq %[right_ref],%[left_ref]\n\t"
    "cmovbq %[tmp],%[right_key]\n\t"
    "cmovbq %[tmp_ref],%[right_ref]\n\t"
    : [left_key] "=&r"(left.key),
      [right_key] "=&r"(right.key),
      [left_ref] "=&r"(left.ref),
      [right_ref] "=&r"(right.ref),
      [tmp] "=&r"(tmp), [tmp_ref] "=&r"(tmp_ref)
    : "0"(left.key), "1"(right.key), "2"(left.ref),
      "3"(right.ref), "4"(tmp), "5"(tmp_ref)
    : "cc"
);
    \end{lstlisting}
    \caption{\SCm}\label{fig:6Cm}
  \end{minipage}%
  \begin{minipage}[t]{.5\linewidth}
    \begin{lstlisting}[style=InFigure]
uint64_t tmp = left.key;
uint64_t tmp_ref = left.ref;
__asm__ volatile (
    "cmpq %[left_key],%[right_key]\n\t"
    :
    : [left_key] "r"(left.key),
      [right_key] "r"(right.key)
    : "cc"
);
__asm__ volatile (
    "cmovbq %[right_key],%[left_key]\n\t"
    : [left_key] "=&r"(left.key)
    : "0"(left.key), [right_key] "r"(right.key)
    :
);
__asm__ volatile (
    "cmovbq %[right_ref],%[left_ref]\n\t"
    : [left_ref] "=&r"(left.ref)
    : "0"(left.ref), [right_ref] "r"(right.ref)
    :
);
__asm__ volatile (
    "cmovbq %[tmp],%[right_key]\n\t"
    : [right_key] "=&r"(right.key)
    : "0"(right.key), [tmp] "r"(tmp)
    :
);
__asm__ volatile (
    "cmovbq %[tmp_ref],%[right_ref]\n\t"
    : [right_ref] "=&r"(right.ref)
    : "0"(right.ref), [tmp_ref] "r"(tmp_ref)
    :
);
    \end{lstlisting}
    \caption{\FCmS}\label{fig:4CmS}
  \end{minipage}
\end{figure}

\begin{figure}
  \begin{minipage}[t]{.5\linewidth}
    \begin{lstlisting}[style=InFigure]
Type* left_pointer = &left;
Type* right_pointer = &right;
Type temp = left;
__asm__ volatile(
    "cmpq %[tmp_key],%[right_key]\n\t"
    "cmovbq %[right_pointer],%[left_pointer]\n\t"
    : [left_pointer] "=&r"(left_pointer)
    : "0"(left_pointer),
      [right_pointer] "r"(right_pointer),
      [tmp_key] "r"(temp.key),
      [right_key] "r"(right.key)
    : "cc"
);
left = *left_pointer;
left_pointer = &temp;
__asm__ volatile(
    "cmovbq %[left_pointer],%[right_pointer]\n\t"
    : [right_pointer] "=&r"(right_pointer)
    : "0"(right_pointer),
      [left_pointer] "r"(left_pointer)
    :
);
right = *rightPointer;
    \end{lstlisting}
    \caption{\TCPm}\label{fig:2CPm}
  \end{minipage}%
  \begin{minipage}[t]{.5\linewidth}
    \begin{lstlisting}[style=InFigure]
Type* left_pointer = &left;
Type* right_pointer = &right;
Type temp = left;
int cmp_result = (int)(right < temp);
__asm__ volatile(
    "cmp $0,%[cmp_result]\n\t"
    "cmovneq %[right_pointer],%[left_pointer]\n\t"
    : [left_pointer] "=&r"(left_pointer)
    : "0"(left_pointer),
      [right_pointer] "r"(right_pointer),
      [cmp_result] "r"(cmp_result)
    : "cc"
);
left = *left_pointer;
left_pointer = &temp;
__asm__ volatile(
    "cmovneq %[left_pointer],%[right_pointer]\n\t"
    : [right_pointer] "=&r"(right_pointer)
    : "0"(right_pointer),
      [left_pointer] "r"(left_pointer)
    :
);
right = *rightPointer;
  \end{lstlisting}
    \caption{\TCPp}\label{fig:2CPp}
  \end{minipage}
\end{figure}

In Section~\ref{section:experiments:normal} and \ref{section:experiments:inrow} we report on our experiments with these conditional swap operations.

We also tried a variant with XOR (see also Codish~et.al.~\cite{codish2017optimizing}), where the left item is first stored in a temporary variable and then conditionally overwritten with the right in case the values have to be swapped.
The right item is then calculated using an XOR of the right, the original left, and current left (which was possibly replaced with the right).
In case no swap occurs, the original left and current left are the same and cancel out.
In case a swap occurs, the current left and right are the same and cancel out, leaving the original left as the new value of right.
We experimented with this variant, but it turned out to be much slower (by a factor of 2--3) in preliminary results.
Another variant with XOR would be to use $x = x \oplus y$, $y = x \oplus y$, $x = x \oplus y$.
This however requires a conditional branch and is thus only a more complicated replacement for a swap.

\section{Register Sample Sort}\label{section:samplesort}

After initial positive experimental results of our investigation of sorting networks for small inputs, we decided to turn to \emph{sample sort} for slightly larger inputs.
Our preliminary experiments showed that sorting networks were indeed faster, but they also required a lot of instructions, leading to large instruction decoding times and L1 cache misses.
Another motivation was that the base cases issued by \textit{In-Place Parallel Super Scalar Samplesort} (\ipso)~\cite{axtmann2017inplace} were considerably larger than sixteen items.
Hence, instead of extending sorting networks beyond sixteen elements, we close this gap by providing a completely different basic algorithm for small to medium size inputs: \emph{Register Sample Sort} (RSS).
Our new algorithm is based on Super Scalar Sample Sort (S$^4$)~\cite{sanders2004super} and can reduce large base case sizes down to blocks of sixteen or less in an efficient manner.
The central idea is to place the splitters not into an array, as described in the original S$^4$, but to hold them in general purpose registers for the whole duration of the element classification.

\paragraph{Basic Sequential Sample Sort}\label{section:samplesort:preliminaries}
Sample sort~\cite{frazer1970samplesort} is a sorting algorithm that follows the divide-and-conquer principle.
The input is split into $k$ disjoint intervals of the total ordering defined by $k+1$ splitters $s_0,\ldots,s_k$.
These are chosen by first selecting a sample subset $S$ of $a \cdot k$ items with oversampling factor $a$ and sorting the sample $S$.
Afterwards the splitters $\{s_0, s_1, \ldots, s_{k-1}, s_k\} = \{-\infty, S_a, S_{2a}, \ldots, S_{(k-1)a}, \infty\}$ are taken equidistant from $S$.
Oversampling is used to get better splitters to achieve more evenly-sized partitions, trading balance for the additional time to select and sort the larger sample.

Given the splitters, all elements $e_i$ are then \emph{classified} by placing them into buckets $b_j$, where $j \in \{1, \ldots, k\}$ and $s_{j-1} < e_i \leq s_j$.
If $k$ is a power of 2, this placement can be achieved by viewing the splitters as a perfect binary tree, with $s_{k/2}$ being the root, all $s_l$ with $l < k/2$ representing the left subtree and those with $l > k/2$ the right one.
To classify an element, one must only traverse this binary tree in logarithmic time, resulting in a binary search on $k+1$ elements instead of a linear one \cite{sanders2004super}.

Quicksort~\cite{hoare1962quicksort} can be seen as a specialization of sample sort with fixed parameter $k = 2$.
Sample sort is very popular for sorting large amounts of items on distributed systems~\cite{blelloch1991comparison,gerbessiotis1994direct,axtmann2017robust}, on GPUs~\cite{leischner2010gpu}, and also for strings~\cite{bingmann2013parallel,bingmann2018scalable}.

\subsection{Implementing Register Sample Sort for Medium-Sized Sets}\label{section:samplesort:impl}

In this section we explain how to implement sample sort using registers in a CPU.
The main issue is that, other than memory, registers \emph{cannot be accessed using an index} on most popular architectures.

\paragraph{Traversing A Tree in Registers}
In implementations of S$^4$ splitters are organized in memory as a perfect binary search tree $t := [\, s_{k/2},$ $s_{k/4},$ $s_{3k/4},$ $s_{k/8},$ $s_{3k/8},$ $s_{5k/8},$ $s_{7k/8}, \ldots \,]$.
Traversal of the splitter tree is then performed using an index $j$: the children of splitter $j$ are at positions $2j$ and $2j + 1$.
If an element is smaller than $t_j$, it must be compared to $t_{2j}$, otherwise to $t_{2j+1}$, in the next step. This allows an easy branchless traversal of the tree by multiplying $j$ with two and conditionally incrementing it.
But this way of accessing the splitters does not work when they are placed in registers, because we cannot access registers by index.

Our solution is to create \emph{an unconditional complete copy} of the left subtree, and to use \texttt{cmov} operations to \emph{conditionally overwrite it} with the complete right subtree should the element be greater than the root node.
The next comparison is then performed against the root of the copied tree that now contains the correct splitters.
This way we traverse the tree until a leaf node is reached.
The copying of subtrees is obviously expensive and only viable for small trees.
For 3 splitters this requires 1 conditional move, and for 7 splitters it requires 3 conditional moves after the first comparison and 1 additional after the second comparison, per element.

\paragraph{Calculating the Final Bucket Index}
However, after finding the correct splitters to compare to, we are left with yet another problem: how to determine the index of the bucket the element is to be placed into.
In S$^4$~\cite{sanders2004super} this bucket index is calculated directly from the index of the last splitter.
For Register Sample Sort, we choose an approach similar to creating this index using the correlation between binary numbers and the tree-like structure of the splitters.
We view the splitters not as a binary tree but just as a list where the middle of the list represents the root node of the tree, its children being the middle element of the left and the middle element of the right list.

If an element $e_i$ is larger than the first splitter $s_{k/2}$ (with $k-1$ being the number of non-sentinel splitters, $s_0$ and $s_k$ are handled implicitly), it must be placed in a bucket $b_j$ with $j \geq \frac{k}{2}$ (assuming 0-based indexing for $b$).
This also means that the index of that bucket, represented as a binary number, must have its bit at position $l := \log \frac{k}{2}$ set to 1.
Hence the result of the comparison ($e_i > s_{k/2}$) can be interpreted as an integer (1 for \textsl{true}, 0 for \textsl{false}) and added to $j$.
If it was not the final comparison, $j$ is then multiplied by 2 (meaning its bits are shifted left by one position).
This means the bit from the first comparison makes its way \enquote{left} in the binary representation while the comparison traverses down the tree, and so forth with the other comparisons.
After traversing the splitter tree to the end, $e_i$ will have been compared to the correct splitters and $j$ will hold the index of the bucket that $e_i$ belongs into.
A similar method is used in S$^4$ when calculating the index during tree traversal.
These operations can be implemented without branches by making use of the way the comparisons are performed:

At the end of Section~\ref{section:preliminaries:asm} we explained that when comparing (unsigned) numbers (which is nothing but a subtraction), and the \texttt{subtrahend} being greater than the \texttt{minuend}, the operation causes an underflow and the carry flag is set.
We also notice that when converting the result of the predicate ($e_i > s_{k/2}$) to an integer value, the integer will be 1 for \texttt{true} and 0 for \texttt{false}.
So in assembly code, we can compare the result from evaluating the predicate to the value 0: \texttt{cmp \%[result],\%[zero]} where \texttt{zero} is just a register that holds the value 0.
This trick is needed because the \texttt{cmp} instruction needs the second operand to be a register.
This will execute $0 - \mathtt{result}$, which underflows for the predicate returning true.
This way we can postfix the \texttt{cmov} needed for moving the next splitters with \enquote{\texttt{c}} checking for a set carry flag. The second instruction we make use of is the \emph{rotate carry left} (\texttt{rcl}) instruction, which performs a \emph{rotate left} instruction on $j$, but includes the carry flag as an additional bit after the least significant bit of the integer.
This exactly takes the predicate result and puts it at the bottom of $j$, with the previous content being shifted one to the left beforehand. That means it performs two necessary operations at once.

\begin{table}
  \caption{Number of registers required by Register Sample Sort with three or seven splitters.}\label{table:samplesort:registerusage}
  \centering
  \begin{tabular}{ r | c c c c c | c c c c c}
                          & \multicolumn{5}{c |}{3 splitters} & \multicolumn{5}{c}{7 splitters}            \\ \hline
                          & \multicolumn{5}{c |}{block size}  & \multicolumn{5}{c}{block size}             \\
                          & 1                                 & 2  & 3  & 4  & 5  & 1  & 2  & 3  & 4  & 5  \\ \hline
    splitters             & 3                                 & 3  & 3  & 3  & 3  & 7  & 7  & 7  & 7  & 7  \\
    buckets pointer       & 1                                 & 1  & 1  & 1  & 1  & 1  & 1  & 1  & 1  & 1  \\
    current element index & 1                                 & 1  & 1  & 1  & 1  & 1  & 1  & 1  & 1  & 1  \\
    element count         & 1                                 & 1  & 1  & 1  & 1  & 1  & 1  & 1  & 1  & 1  \\ \hline
    index                 & 1                                 & 2  & 3  & 4  & 5  & 1  & 2  & 3  & 4  & 5  \\
    predicate result      & 1                                 & 2  & 3  & 4  & 5  & 1  & 2  & 3  & 4  & 5  \\
    \texttt{splitterX}    & 1                                 & 2  & 3  & 4  & 5  & 3  & 6  & 9  & 12 & 15 \\ \hline
    sum                   & 9                                 & 12 & 15 & 18 & 21 & 15 & 20 & 25 & 30 & 35
  \end{tabular}
\end{table}

\paragraph{Putting Things Together: Determining Parameters and Pseudocode}
With the previous two challenges of classification solved, we can now design the parameters for our Register Sample Sort.

As with S$^4$, when processing items from the input we can \emph{interleave classification} of multiple elements, allowing for all the registers in the machine to be used.
This additional parameter is called \emph{block size}.

The main constraint on the parameters of Register Sample Sort is the \emph{number of registers} in the CPU.
The keys of the splitters (since we only need a splitter's key for classifying an element) must be small enough to fit into a general purpose register.
Needing more than one register per key would mean running out of registers more quickly and also spending extra time to conditionally move the splitter keys around.
For three splitters the number of registers needed for block sizes 1 to 5 are shown in Table~\ref{table:samplesort:registerusage}.
We can see that the trade-off for classifying multiple elements at the same time is the amount of registers needed.

If we were to use 7 splitters instead of 3, the number of registers required for classifying just 1 element at a time would go up to 15.
Furthermore, if we get recursive subproblems with just slightly over sixteen items, classifying into 8 buckets would be greatly inefficient, resulting in many empty buckets.
This is why we decided to only use \emph{exactly three splitters} for this particular sorter.

Instead of inline assembly from our implementation, we present pseudocode of the classification for \texttt{block size = 1} in Algorithm~\ref{algo:samplesort:cversion}.
The index $j$ is here called \texttt{index}, and the temporary subtree consists in this case of one splitter, which we gave the name \texttt{splitterX}.
For our branchless implementation we used \texttt{cmovne} together with a \texttt{test} instruction to move the right hand splitter into \texttt{splitterX} and just assigned the integer result of the comparison to index in the first step (lines \ref{algo:samplesort:cmov} and \ref{algo:samplesort:indexassign}). At the second and last level of classification no more movement of splitters is required, so instead of performing a comparison against the predicate's result and using \texttt{rcl}, we can just shift \texttt{index} left by one position and add the predicate's result to it (line \ref{algo:samplesort:lastindex}).
Alternatively we could use a bit-wise \texttt{OR} or \texttt{XOR} after the shift, which would have the same result.
But we decided that adding the predicate result was more readable.

For sorting the splitter sample, the same sorting method can be used as for the base case, which in our case is to use a sorting network from Section~\ref{section:networks}.

\begin{algorithm}
  \caption{Register Sample Sort Classification Pseudocode for Three Splitters}\label{algo:samplesort:cversion}
  \begin{lstlisting}[mathescape]
void RegisterSampleSortClassification(Type* array, unsigned array_size) {
    Type [splitter0, splitter1, splitter2] = determineSplitters();
    Type* [b$_{\texttt{0}}$,b$_{\texttt{1}}$,b$_{\texttt{2}}$,b$_{\texttt{3}}$] = allocateBuckets(array_size);
    for (int i = 0; i < array_size; ++i) {
      int index = 0;
      int cmp_result = (int)(splitter1 < array[i]);
      splitterx = splitter0;                  // copy left tree into $\texttt{splitterx}$
      if (cmp_result) {                       // first and only decision in a tree with three splitters
        splitterx = splitter2;                // overwrite using a $\texttt{cmovne}$$\label{algo:samplesort:cmov}$
      }
      index = cmp_result;$\label{algo:samplesort:indexassign}$
      cmp_result = (int)(splitterx < array[i]);
      index = (index << 1) + cmp_result;$\label{algo:samplesort:lastindex}$
      b$_{\texttt{index}}$.push_back(array[i]);
    }
}
  \end{lstlisting}
\end{algorithm}

\section{Experimental Results} \label{section:results}
In this section we report on four sets of experiments with sorting networks and Register Sample Sort.
The first are pure performance measurements of sorting networks, either sorting the same array repeatedly or sorting a larger array containing many independent small problems.
The second experiments are on the performance of Quicksort with sorting networks as base case,
the third on finding good parameters for Register Sample Sort,
and the last on integrating Register Sample Sort and sorting networks into \ipso.

\subsection{Parameters, Machines, Inputs, Methodology}\label{section:parameters}

\paragraph{Machines and Compiler}
We used four different machines to perform the measurements, including Intel, AMD, and ARM CPUs.
Their labels and hardware properties can be seen in Table~\ref{table:machines}.
In the table \enquote{I} and \enquote{D} refer to dedicated L1 instruction and data caches.
While the AMD Ryzen's L3 cache has a total size of 16\,MiB, it is divided into two 8\,MiB caches that are exclusive to 4 cores each.
Since all measurements were done on a single core, the L3 cache size in brackets is the one available to the program.

For compiling our experiment code we used the gcc C++ compiler in version 7.3.0 or 8.3.0 with \texttt{-O3} and \texttt{-march=native} flags.
We did not experiment with LLVM or other compilers because this increases the parameter space by another dimension.

The measurements were done with only essential processes running on the machine apart from the measurement.
To prevent the process from being swapped to another core during execution it was run with the command \enquote{\texttt{taskset 0x1}} as prefix, which pins it to the first core.

\begin{table}
  \caption{Hardware properties of the machines used in our experiments.}\label{table:machines}
  \centering\scriptsize\def\arraystretch{1.3}
  \setlength\tabcolsep{5.2pt}
  \begin{tabular}{r | r | r | r | r}
    Machine Name            & \multicolumn{1}{c|}{Intel-2650} & \multicolumn{1}{c|}{Intel-2670} & \multicolumn{1}{c|}{Ryzen-1800X} & \multicolumn{1}{c}{RK3399}              \\ \hline
    \multirow{3}{*}{CPU}    & 2 x Intel XeonE5-2650 v2        & 2 x Intel XeonE5-2670 v3        & AMD Ryzen 1800X                 & Rockchip RK3399                         \\
                            & 8-core, 2.6 GHz                 & 12-core, 2.3 GHz                & 8-core, 3.6 GHz                 & ARM Cortex-A53 | -A72                   \\
                            &                                 &                                 &                                 & 4 $\times$ 1.5 GHz | 2 $\times$ 2.0 GHz \\
    RAM                     & 128 GiB DDR3-1600               & 128 GiB DDR4-2133               & 32 GiB DDR4-2133                & 4 GiB LPDDR4                            \\
    L1 cache (KiB per core) & 32 I + 32 D (8-way)             & 32 I + 32 D (8-way)             & 64 I + 32 D  (4/8-way)          & 32 I + 32 D | 48 I + 32 D               \\
    L2 cache (KiB per core) & 256 (8-way)                     & 256  (8-way)                    & 512  (8-way)                    & 512 | 1024                              \\
    L3 cache (MiB total)    & 20                              & 30  (20-way)                    & 16 [8] (8-way)                  & -                                       \\
    Linux Distribution      & Ubuntu 18.04                    & Ubuntu 18.04                    & Ubuntu 18.04                    & Armbian (Debian) Buster                 \\
    Compiler                & gcc 7.3.0                       & gcc 7.3.0                       & gcc 7.3.0                       & gcc 8.3.0                               \\
  \end{tabular}
\end{table}

\paragraph{Inputs: Random Numbers}
In order to measure the time needed to sort some data, we first have to generate data.
For all experiments the data type consisted of a pair of one 64-bit unsigned integer key and one 64-bit unsigned integer reference value.
Items were generated as uniformly distributed random numbers by a lightweight implementation of the \texttt{std::minstd\_rand} generator from the C++ \texttt{<random>} library that works as follows:
First a \texttt{seed} is set, taken e.g. from the current time.
When a new random number is requested, the generator calculates $\mathtt{seed} = (\mathtt{seed} \cdot 48271) \bmod 2147483647$ and returns the current \texttt{seed}.
The numbers generated by this generator does not use all 64 bits available, which however has no effect on the experiment results.

For each measurement $i$, a new $\mathtt{seed}_i$ is taken from the current time.
The same $\mathtt{seed}_i$ is then set before the execution of each sorter, to provide all sorters with the same random inputs.

We only experimented with random inputs. In future work, ``easy'' inputs such as sorted and reverse sorted inputs, and others should also be included.

\paragraph{Measurement of Cycles with perf\_event}
The actual measurement was done via linux's \texttt{perf\_event} interface that allows to do fine-grained measurements using hardware counters.
We measured the number of CPU \emph{cycles} spent on sorting.
This also means that our results do not depend on clock speeds (e.g. when overclocking), but only on the CPU's architecture.
On the ARM machine RK3399 we resorted to simply measuring time, because the CPU cycles event interface was not available.

\paragraph{Checking Results: Permutation Check}
For compilation, the optimization flag \texttt{-O3} was used to achieve high optimization and speed.
This also meant that, without using the sorted data in some way, the compiler would deem the result unimportant and skip the sorting altogether.
That is why after each sort, to generate a side-effect, the set is checked for two properties: That it is sorted, and that it is a permutation of the input set.
The first can easily be done by checking for each value that it is not greater than the value before it.

The permutation check is done probabilistically: At design time, a (preferably large) prime number $p$ is chosen.
Before sorting, $v = \prod_{i = 1}^{n} (z - a_i) \bmod p$ is calculated for an arbitrary number $z$ and values $a = \{a_1, \ldots, a_n\}$.
To check the permutation after sorting and obtaining $a' = \{a'_1, \ldots, a'_n\}$, $w = \prod_{i = 1}^{n} (z - a'_i) \bmod p$ is calculated.
If $v \neq w$, $a'$ cannot be a permutation of $a$. If $v = w$, we claim that $a'$ is a permutation of $a$.

To minimize the chances of $a'$ not being a permutation of $a$, but $v$ being equal to $w$, $v = 0$ was disallowed in the first step.
If $v$ is zero, $z$ is incremented by one and the product calculated again, until $v \neq 0$.

We do this kind of check because in each iteration the input array is overwritten with new random numbers, and using an easier algorithm to sort the array afterwards and then checking for equality would include copying the unsorted array, sorting and checking, which adds extra time and variation to a measurement consisting of many of these iterations. The permutation check on the other hand is branch-free, and the sorted check is effectively branch-free for any sorted array, because the compiler is told to expect the sorted condition to be true, and will speculatively execute the branch in which it is, not leading to any flushing of the pipeline.
This way we can run the sorted and permutation check in a second measurement with the same number of iterations, and subtract that time from the original measurement, giving only the time needed for sorting.

\paragraph{Multiple Source File Compilation}
Initially, the experiments were a single source file (\texttt{.cpp}) with an increasing amount of headers that were all included in that single file.
This was mostly due to the fact that templated methods cannot be placed in source files because they need to be visible to all including files at compile time.
The increasing amount of code and the many different templates, however, brought the compiler to a point where it took over a minute to compile the program.
The problem we encountered was that the compiler apparently allots less time to optimization the longer the compilation runs on a source file.
Hence, once our experiment program became reasonably large, the optimizations became poor.
We saw measurements being slower for no apparent reason.
To solve that problem, we used code generation to create source files that contain a smaller amount of methods that initiate part of a measurement in a wrapper method.
From the main source file we thus only need to call the correct wrapper methods to perform the measurements, and this way we were able to achieve results that were more stable and reproducible.
In our case that means one file for the OneArrayRepeat experiment (\ref{section:experiments:normal}), one for ArrayInRow (\ref{section:experiments:inrow}), one for QuickSort (\ref{section:experiments:quicksort}), one for SampleSort (\ref{section:experiments:samplesort}), and an individual file for each configuration of the \ipso ~measurement (\ref{section:experiments:ipso})

\paragraph{Measurement Loops and Warm-Up}
As the sorting algorithm on a small number of items is very fast, we measured many iterations and divided by the number of repetitions.
Since in this scenario we have to generate new random inputs for each iteration, we decided to first measure the three steps (a) data generation, (b) sorting, and (c) checking, and then measure only (a) data generation and (c) checking on the same inputs.
By subtracting the two running times we receive the pure sorting time.
We call this measurement loop \emph{OneArrayRepeat}.

More details on the method are now discussed:
At the beginning a random seed is selected and the generator initialized.
To reduce the chance of cache misses at the beginning of the measurement, one warm-up run of random generation, sorting, and checking is performed before starting the clock.
After the warm-up round, the array is then filled, sorted, and checked \texttt{numberOfIterations} times.
The random generator is not reseeded each round.
After the main measurement, a second phase is run with the same data.
But this time only the generation of the random numbers and the checking is measured to later subtract the time from the previously measured one, resulting in the time needed for the sorting alone.
To generate and check the same input again, the random generator is reseeded with the previously selected seed.
Obviously, the checking of the unsorted data (usually) fails but it has to performed to measure the time.
Hence, we devised a simulated checking method, which does the exact same comparisons and permutation calculations, but ignores the result.

Nevertheless there are random non-deterministic fluctuations in running time even on the same code.
And since both measurement parts are subject to their own deviation, it can occasionally happen that the second measurement takes longer than the first, leading to negative running times for sorting.
We received negative values more often for the sorters with small array sizes, where the sorting itself takes relatively little time compared to the random generation and sorted checking.
The negative times show up as outliers in the results.

The measurement loop itself is repeated \texttt{numberOfMeasures} times for each \texttt{arraySize} that is sorted.

For the measurements shown in Section~\ref{section:experiments:inrow} the method was slightly modified.
The goal was to better highlight cache- and memory-effects by creating one longer array that does not fit into the CPU's L3-cache and then sorting disjoint short subsequences of size \texttt{arraySize} in order.
We call this modified measurement loop \emph{ArrayInRow}.

Because we can create the whole array at the beginning, we can generate the numbers before and check for correct sorting after measuring, hence there is no need to do a second measurement like in the OneArrayRepeat benchmark.
As in the prior benchmark, one warm-up round is performed prior to running the measured loop.

For ArrayInRow, instead of giving a \texttt{numberOfIterations} parameter to indicate how often the sorting is to be repeated, we provide a \texttt{numberOfArrays} value that prescribes how many arrays of size \texttt{arraySize} are to be created contiguously.
This parameter is chosen for each \texttt{arraySize} in a way that $(\mathtt{numberOfArrays} \cdot \mathtt{arraySize})$ does not fit into the L3 cache of the machine the measurement is performed on.

\paragraph{Generating Plots}
Due to the high number of dimensions in the measurements (machine the measurement is run on, type of sorting network, conditional swap implementation, array size) the results could not always be plotted two-dimensionally.
We used box-plots where applicable to show more than just an average value for a measurement.
The box encloses all values between the first quartile $q_1$ and third quartile $q_3$.
The line in the middle shows the median.
Further the inter-quartile-range $\bar{q}$ is calculated as the distance between first and third quartile.
The lines (called whiskers) left and right of the boxes extend to the smallest value greater than $q_1 - 1.5 \bar{q}$ and the greatest value smaller than $q_3 + 1.5 \bar{q}$ respectively.
Values below these ranges are called \emph{outlier} and shown as individual dots.

\subsection{Sorting Sets of 2--16 Items}\label{section:experiments:normal}
In this and the following subsection we report on experiments comparing sorting algorithms and conditional-swap implementations.
For the details about the different sorters and swaps refer to Section~\ref{section:implementation-networks}.

The algorithm variants in the tables and figures are labeled in an abbreviatory way such as \enquote{\texttt{SN BN-R 4CmS}}.
The abbreviations are composed from the following three parts:
\begin{enumerate}
\item First, \texttt{IS} or \texttt{SN} indicate if the algorithm is insertion sort or a sorting network. \label{enumeration:experimentnaming:algtype}

  For \texttt{IS} we evaluated four variants: \texttt{Def} is a textbook implementation with array indices, \texttt{POp} uses pointers as iterators, \texttt{STL} is copied from gcc's STL implementation, and \texttt{AIF} is \texttt{Def} with an additional check if the next item is smaller than the first.

\item In case of sorting networks, the algorithms are labeled as \texttt{Best} networks or Bose-Nelson networks (\texttt{BN}).
  The Bose-Nelson sorters are further available optimized for \emph{locality} (\texttt{BN-L}), \emph{parallelism} (\texttt{BN-P}), or written as \emph{recursively} called sorting functions (\texttt{BN-R}) (see Section~\ref{section:network-frame}).

\item And as last component, the name of the conditional swap implementation is appended for sorting networks (see Section~\ref{section:implementation-conditionalswap} for their abbreviations).
\end{enumerate}

As an example consider \enquote{\texttt{SN BN-R 4CmS}}. This implementation is a Bose-Nelson sorting network generated using recursively constructed functions which perform the conditional-swap using the \texttt{4CmS} variant.

Table~\ref{table:sort:codesize} shows the binary x86 code size of each insertion sort and sorting network variant (size 2--16).
We determined their binary x86 code size by compiling the code (with \texttt{-O3} optimization) and then disassembling the object file and identifying which functions therein belong to the algorithm.
In Table~\ref{table:sort:codesize} we also included the code size of \texttt{std::sort}, Register Sample Sort (RSS), and \ipso, each excluding any subsorters.

To determine the fastest algorithm variant, we ran the OneArrayRepeat experiment with the following parameters on all machines and algorithms:
$\mathtt{numberOfIterations} = 100$, $\mathtt{numberOfMeasures} = 500$, and $\mathtt{arraySize} \in \{2, \ldots, 16\}$.

Table~\ref{table:normalsort:avg:all} shows our results as a ranking of the algorithms by geometric mean, individually and across all machines, of the slowdown relative to the best on the particular machine.
Table~\ref{table:normalsort:speedup:all} summarizes the fastest sorting network over the fastest insertion sort implementation.
Further results are shown in the appendix Tables~\ref{table:normalsort:avg:A}, \ref{table:normalsort:avg:B}, \ref{table:normalsort:avg:C}, and  \ref{table:normalsort:avg:D}; there we show the name of the sorter and the average number of cycles per iteration, over the total of all measurements, for all machines.
The algorithm that performed best in a column is marked in bold font, and for each column the slowdown relative to the best in that column was calculated.
For each row the geometric mean \enquote{GeoM} is shown over these relative slowdown values and the mean is used to rank the algorithms.

\begin{table}
  \caption{Size in bytes of binary x86 assembly code of sorters generated by gcc (with optimization).}\label{table:sort:codesize}
  \scriptsize\centering
  \def\arraystretch{1.3}
  \setlength\tabcolsep{5.1pt}
  \begin{tabular}[t]{|l|r|} \hline
    Algorithm           & Size    \\ \hline
    \verb+IS   Def+     & 91      \\
    \verb+IS   POp+     & 122     \\
    \verb+IS   STL+     & 188     \\
    \verb+IS   AIF+     & 200     \\ \hline
    \verb+std::sort+    & 1283    \\
    RSS \verb+332+      & 1280    \\
    \ipso               & 42\,689 \\ \hline
  \end{tabular}
  \begin{tabular}[t]{|l|r|} \hline
    Algorithm           & Size    \\ \hline
    \verb+SN Best ISwp+ & 17\,936 \\
    \verb+SN Best Tie + & 33\,328 \\
    \verb+SN Best JXhg+ & 12\,080 \\
    \verb+SN Best 4Cm + & 19\,008 \\
    \verb+SN Best 4CmS+ & 19\,472 \\
    \verb+SN Best 2CPm+ & 24\,064 \\
    \verb+SN Best 2CPp+ & 26\,912 \\ \hline
  \end{tabular}
  \begin{tabular}[t]{|l|r|} \hline
    Algorithm           & Size    \\ \hline
    \verb+SN BN-L ISwp+ & 19\,264 \\
    \verb+SN BN-L Tie + & 34\,400 \\
    \verb+SN BN-L JXhg+ & 10\,160 \\
    \verb+SN BN-L 4Cm + & 17\,456 \\
    \verb+SN BN-L 4CmS+ & 17\,312 \\
    \verb+SN BN-L 2CPm+ & 25\,360 \\
    \verb+SN BN-L 2CPp+ & 28\,416 \\ \hline
  \end{tabular}
  \begin{tabular}[t]{|l|r|} \hline
    Algorithm           & Size    \\ \hline
    \verb+SN BN-P ISwp+ & 19\,216 \\
    \verb+SN BN-P Tie + & 35\,920 \\
    \verb+SN BN-P JXhg+ & 12\,880 \\
    \verb+SN BN-P 4Cm + & 20\,640 \\
    \verb+SN BN-P 4CmS+ & 20\,704 \\
    \verb+SN BN-P 2CPm+ & 25\,696 \\
    \verb+SN BN-P 2CPp+ & 28\,736 \\ \hline
  \end{tabular}
  \begin{tabular}[t]{|l|r|} \hline
    Algorithm           & Size    \\ \hline
    \verb+SN BN-R ISwp+ & 12\,880 \\
    \verb+SN BN-R Tie + & 19\,072 \\
    \verb+SN BN-R JXhg+ & 6\,144  \\
    \verb+SN BN-R 4Cm + & 9\,584  \\
    \verb+SN BN-R 4CmS+ & 10\,720 \\
    \verb+SN BN-R 2CPm+ & 14\,096 \\
    \verb+SN BN-R 2CPp+ & 14\,640 \\ \hline
  \end{tabular}
\end{table}

\begin{table}
  \caption{Ranking of sorting algorithms by geometric mean of relative slowdowns for the OneArrayRepeat experiment across all machines.}\label{table:normalsort:avg:all}
  \scriptsize\centering
  \def\arraystretch{1.1}
  \setlength\tabcolsep{10pt}
  \input{speedupTableNormalAllGeoms.tex}
\end{table}

\begin{table}
  \caption{Speedup factor of the fastest sorting network over the fastest insertion sort implementation for the OneArrayRepeat experiment and array sizes 2--16.}
  \label{table:normalsort:speedup:all}
  \scriptsize\centering
  \def\arraystretch{1.1}
  \setlength\tabcolsep{5.7pt}
  \input{speedupTableNormalAll.tex}
\end{table}

The tables show that the implementations without conditional branches (\texttt{4Cm}, \texttt{4CmS}, \texttt{2CPm}, \texttt{2CPp}) and those with (\texttt{ISwp}, \texttt{Tie}, \texttt{JXhg}) are clearly separated by rank in the overall result, the former occupy the lower share of the ranks, while the latter all the higher ranks.
To improve readability, the variants \texttt{TCOp} and \texttt{6Cm} are omitted.
Variant \texttt{6Cm} was similar to the other variants without conditional branches, and \texttt{TCOp} was close to the results of \texttt{JXhg}.
We omitted them because they are probably executed very similarly by the processors.

Comparing the machines in Table~\ref{table:normalsort:avg:all}, we see that our hypothesis that the \texttt{4CmS} conditional swap is better than the \texttt{4Cm} was shown to be true for machines Intel-2650 and Intel-2670, but not for machines Ryzen-1800X and RK3399.
We believe this is due to the first machines having better support for interleaving load and store operations.
We also see in Table~\ref{table:normalsort:avg:all} that the first five ranks have very similar geometric means, which implies the Bose-Nelson networks (\texttt{BN-L} and \texttt{BN-R}) can compete (due to their locality) with the optimized networks (\texttt{Best}) that have fewer comparators.

\begin{figure}
  \includegraphics{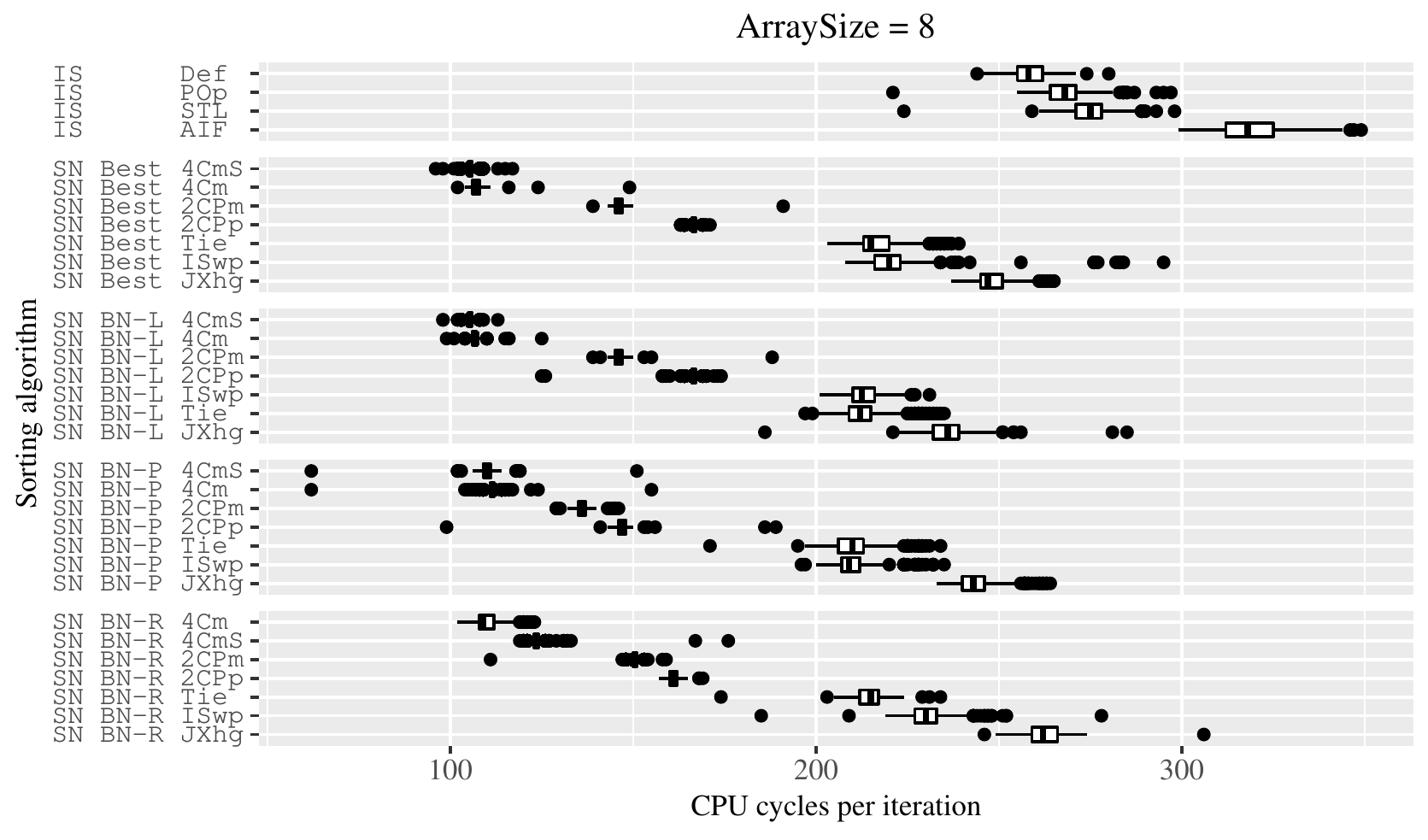}
  \caption{OneArrayRepeat experiment with array size = 8 on machine Intel-2650} \label{plot:normal:8:A}
  \bigskip
  \includegraphics{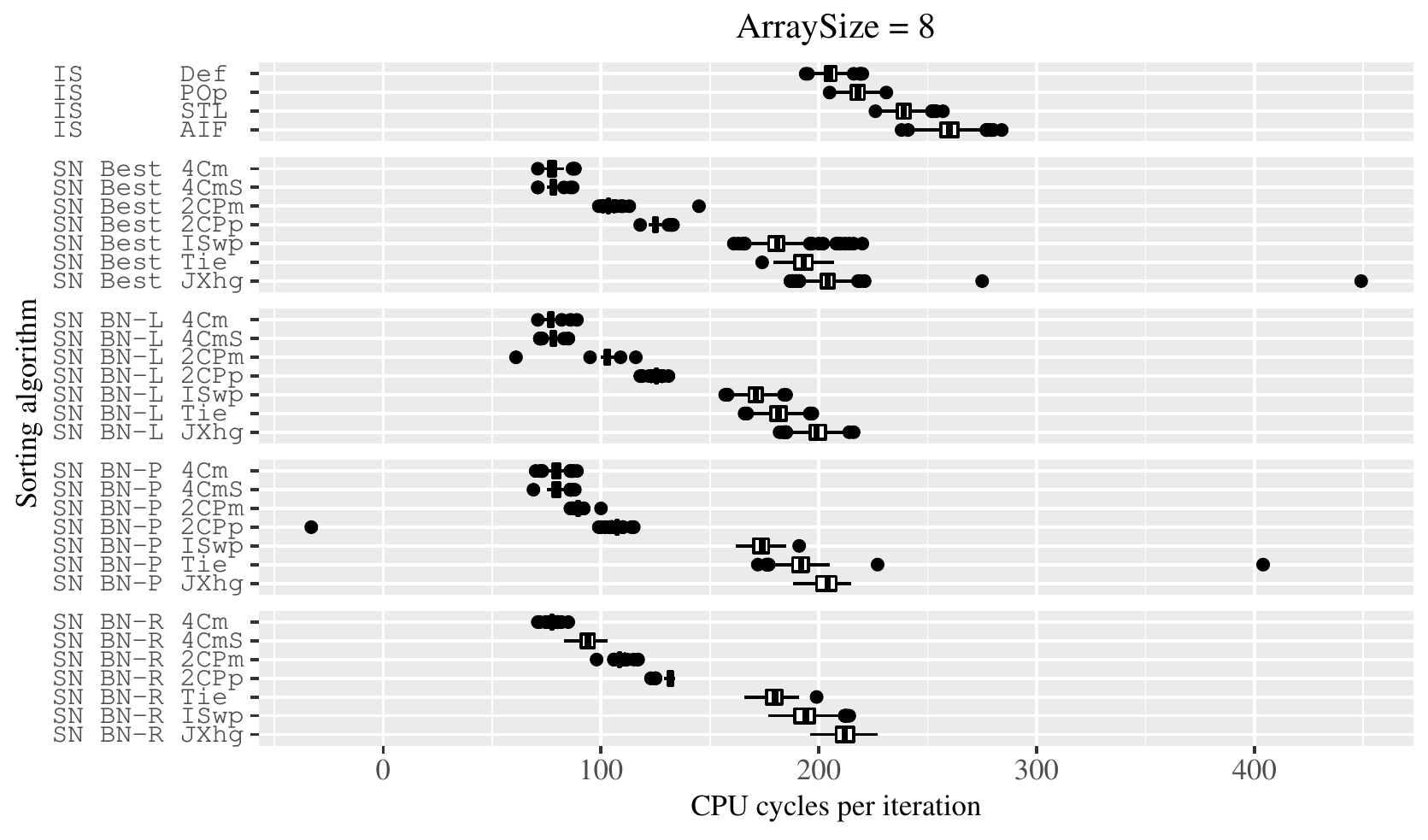}
  \caption{OneArrayRepeat experiment with array size = 8 on machine Intel-2670} \label{plot:normal:8:B}
\end{figure}

\begin{figure}
  \includegraphics{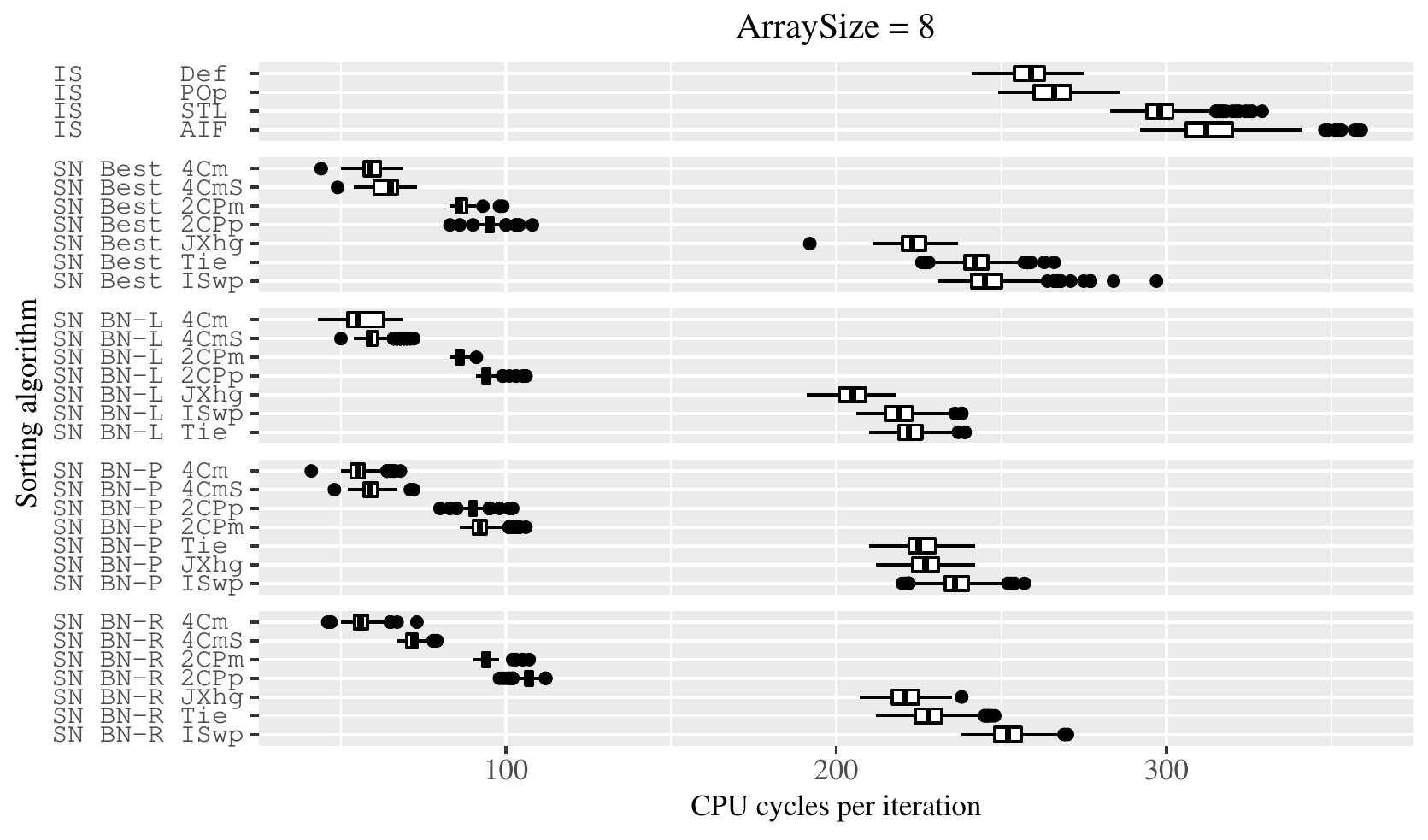}
  \caption{OneArrayRepeat experiment with array size = 8 on machine Ryzen-1800X} \label{plot:normal:8:C}
  \bigskip
  \includegraphics{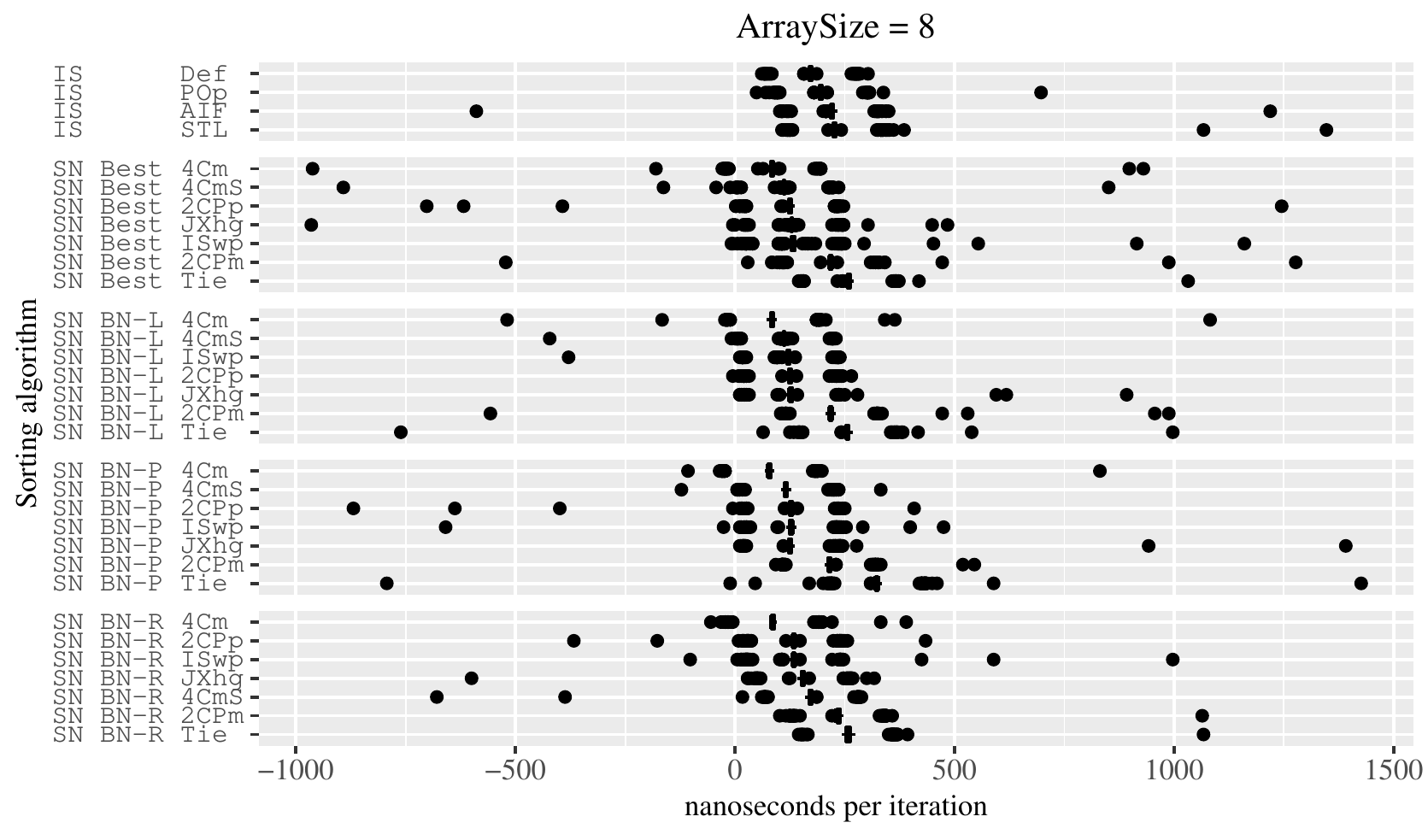}
  \caption{OneArrayRepeat experiment with array size = 8 on machine RK3399} \label{plot:normal:8:D}
\end{figure}

Figures~\ref{plot:normal:8:A}, \ref{plot:normal:8:B}, \ref{plot:normal:8:C}, and \ref{plot:normal:8:D} show box plots of the CPU cycle measurement for array size 8 on each machine.
These plots highlight that the best sorting network implementations are not only faster on average, but that their distribution is almost entirely faster than any of the insertion sort implementations, together with a lower variance.
As in the tables, the variants \texttt{TCOp} and \texttt{6Cm} are omitted to improve readability.
Furthermore, one outlier was removed from dataset of machine Intel-2670 for the \texttt{SN BN-P 4CmS} sorter with value $-228.55$ such that the plot has a scale similar to those of the other two machines, to improve comparability.
It is remarkable that on the RK3399 machine the variance of the sorting networks is much higher than on the other platforms.
However, this may be an effect due to measuring nanoseconds wallclock time on the machine (due to lack of performance counters or support for them) versus CPU cycles on the others.

\begin{figure}
  \centering

  \begin{minipage}[b]{0.5\linewidth}
    \includegraphics{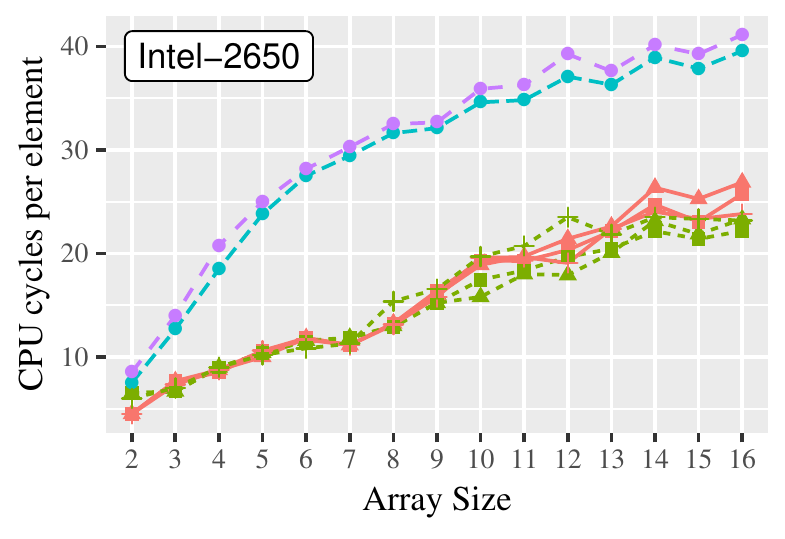}
    \vspace{-4ex}
  \end{minipage}%
  \begin{minipage}[b]{0.5\linewidth}
    \includegraphics{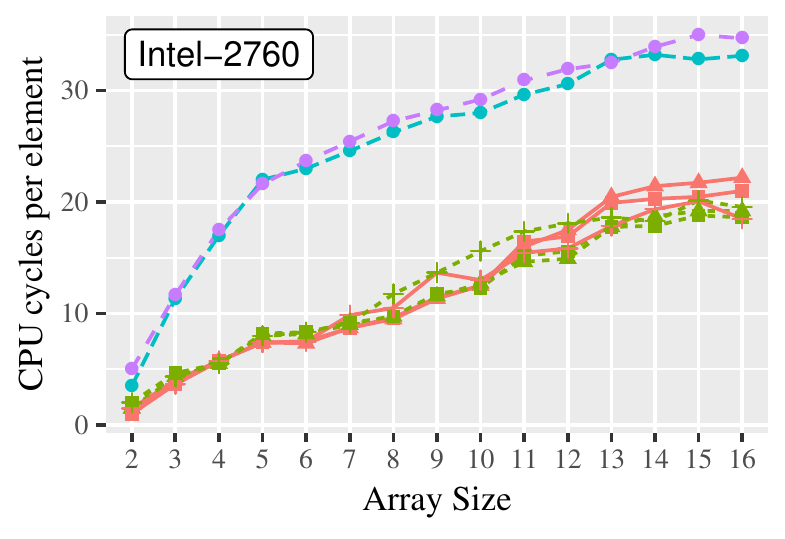}
    \vspace{-4ex}
  \end{minipage}

  \begin{minipage}[b]{0.5\linewidth}
    \includegraphics{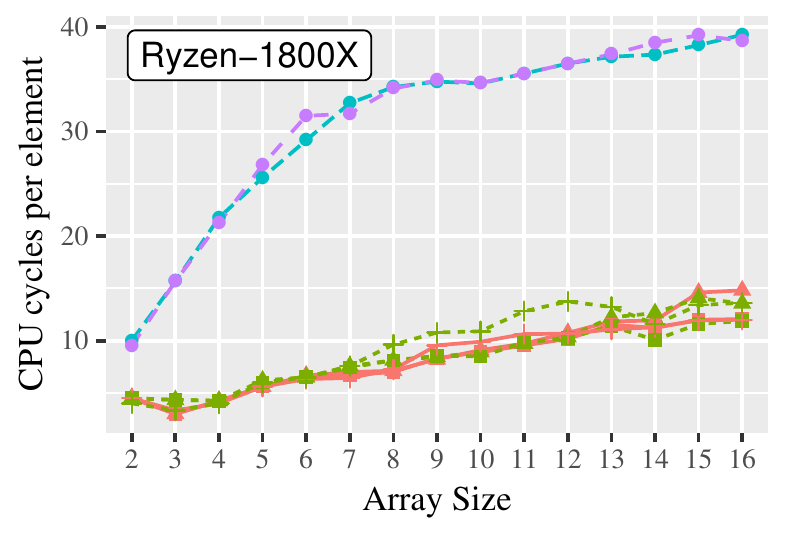}
    \vspace{-4ex}
  \end{minipage}%
  \begin{minipage}[b]{0.5\linewidth}
    \includegraphics{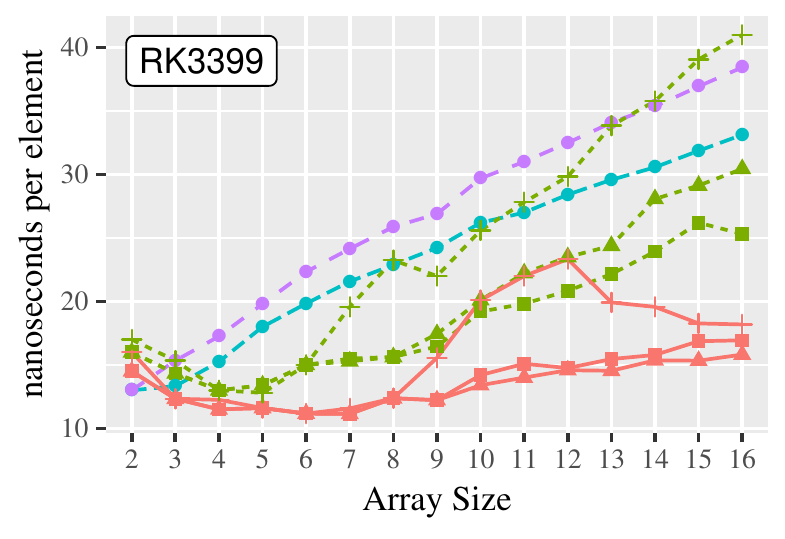}
    \vspace{-4ex}
  \end{minipage}

  \includegraphics[]{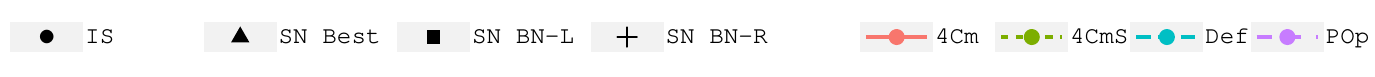}
  \caption{OneArrayRepeat experiment with array size 2--16 on all machines} \label{plot:normal:lineplot}

  \bigskip

  \begin{minipage}[b]{0.5\linewidth}
    \includegraphics{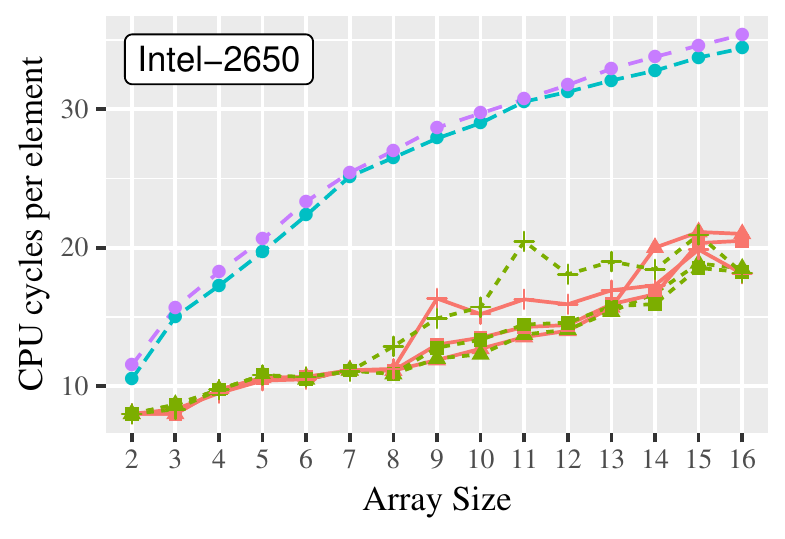}
    \vspace{-4ex}
  \end{minipage}%
  \hfill%
  \begin{minipage}[b]{0.5\linewidth}
    \includegraphics{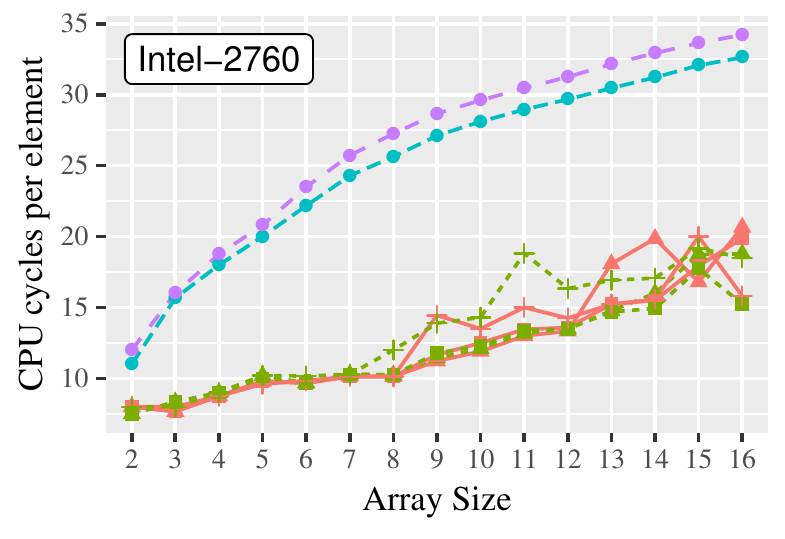}
    \vspace{-4ex}
  \end{minipage}

  \begin{minipage}[b]{0.5\linewidth}
    \includegraphics{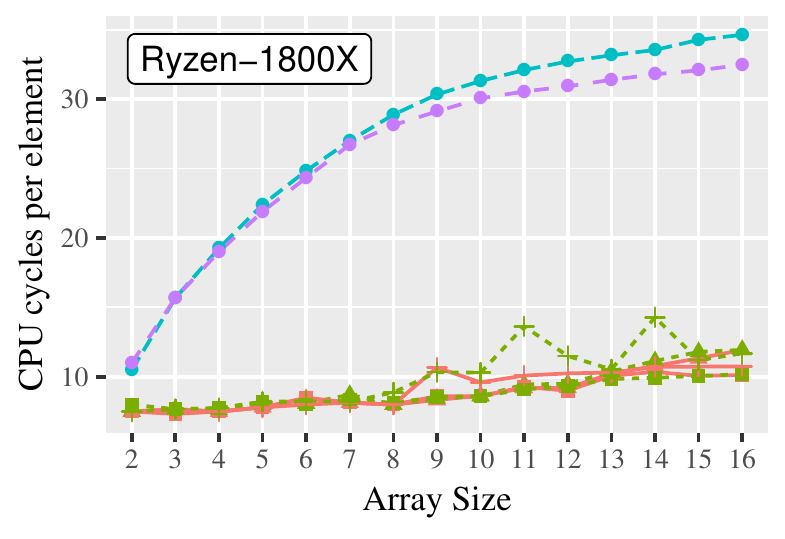}
    \vspace{-4ex}
  \end{minipage}%
  \hfill%
  \begin{minipage}[b]{0.5\linewidth}
    \includegraphics{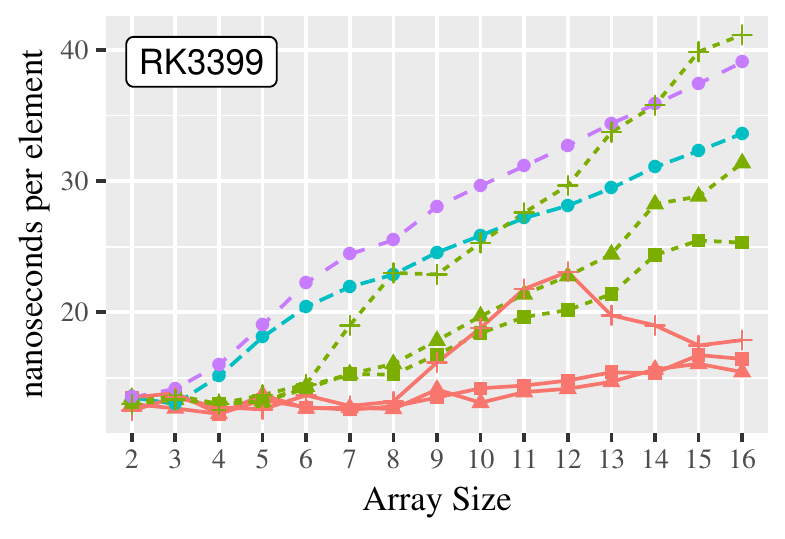}
    \vspace{-4ex}
  \end{minipage}

  \includegraphics[]{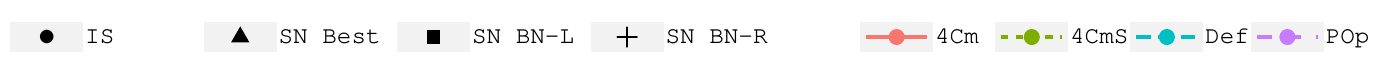}
  \caption{ArrayInRow experiment with array sizes 2--16 across all machines}\label{plot:inrow:lineplot}
\end{figure}

To see a trend in increasing array size, we chose a few conditional-swap implementations that do best for more than one network and array size on all machines.
These series increase as expected from the $\mathcal{O}(n \log^2 n)$ asymptotic growth rate of the sorting networks.
Their average sorting times with OneArrayRepeat can be seen in Figure~\ref{plot:normal:lineplot}.
For better readability, we omitted the Bose-Nelson parallel (\texttt{BN-P}) networks in these plots.
The figures underline the results already shown in the tables: the \texttt{4Cm} and \texttt{4CmS} implementations have good performance and are almost always faster on average than insertion sort (apart from \texttt{arraySize = 2} on machine RK3399).

These results indicate that there is potential in using sorting networks, showing that the best insertion sort is slower by a factor of 1.79 on average on machine RK3399, up to a factor of 4.47 on average on machine Ryzen-1800X (see Table~\ref{table:normalsort:speedup:all} for details).
The issue with the OneArrayRepeat experiment is that the same memory area is sorted over and over again, which is rarely a use case when sorting a base case.
Because of this, the results probably reflect unrealistic conditions regarding cache accesses and cache misses.
To get closer to realistic base case sorting, the next section regards the ArrayInRow pattern.


\begin{table}
  \caption{Ranking of sorting algorithms by geometric mean of relative slowdowns for the ArrayInRow experiment across all machines.}\label{table:inrowsort:avg:all}
  \scriptsize \centering
  \def\arraystretch{1.1}
  \setlength\tabcolsep{10pt}
  \input{inrowSortAvgTableAll.tex}
\end{table}

\subsection{Sorting Many Continuous Sets of 2--16 Items} \label{section:experiments:inrow}

In this section we consider the ArrayInRow benchmark: instead of sorting a single array multiple times, multiple arrays are created adjacent to each other and sorted in series.
The number of arrays used is chosen in a way that their concatenation does not fit into the CPU's L3 cache.
Before the actual measurement, the entire array is copied into a reference array and each subarray is sorted.
Since all items are touched in this reference sweep, the original array should not be present in the cache at the start of the actual sorting run.
After the timed run, the results are compared against the reference array for correctness.

Overall the results of ArrayInRow shown in Figure~\ref{plot:inrow:lineplot} are similar to the previous ones in Figure~\ref{plot:normal:lineplot}.
Table~\ref{table:inrowsort:avg:all} shows the algorithms ranked by geometric mean of the relative slow across all machines, and we can see that the order has only changed very slightly.
Due to the input not being in cache there is a performance penalty for all sorters in ArrayInRow, but the relative performances do not change much.
Table~\ref{table:inrowsort:speedup:all} also shows that insertion sort is slower by a factor of 2.26 on average across all machines

\begin{table}
  \caption{Speedup factor of the fastest sorting network over the fastest insertion sort implementation for the ArrayInRow experiment and array sizes 2--16.}
  \label{table:inrowsort:speedup:all}
  \scriptsize\centering
  \def\arraystretch{1.1}
  \setlength\tabcolsep{5.7pt}
  \input{speedupTableInrowAll.tex}
\end{table}

\subsection{Sorting a Large Set of Items with Quicksort} \label{section:experiments:quicksort}

After the encouraging performance of the sorting networks, we were interested in how they perform as base case sorters inside a sorting algorithm for larger sets.
To study this we modified Introsort~\cite{musser1997introspective}, the Quicksort implementation used in the current gcc's STL library.
In this particular implementation Introsort calls insertion sort only once at the end on the entire array.
Since this is not possible with our sorting networks, we modified the implementation to called the networks directly when the partitioning resulted in a set of 16 elements or less.
Also we determined the pivot using a 3-element Bose-Nelson network instead of using \texttt{if-else} and \texttt{std::swap}.

In the results we labeled the base case sorter variants with the same schema as described in Section~\ref{section:experiments:normal}, but note again that this time the label describes the \emph{base case sorter} integrated into Introsort.
To validate the results, we also include two unmodified Introsort implementations as references:
\texttt{QSort} is a direct source code copy of gcc's Introsort doing a final insertion sort at the end, and \texttt{StdSort} is a simple call to \texttt{std::sort}.
Theoretically these should perform identically but in practice there are differences due to the way gcc optimizes library and non-library code.

\begin{figure}
  \includegraphics{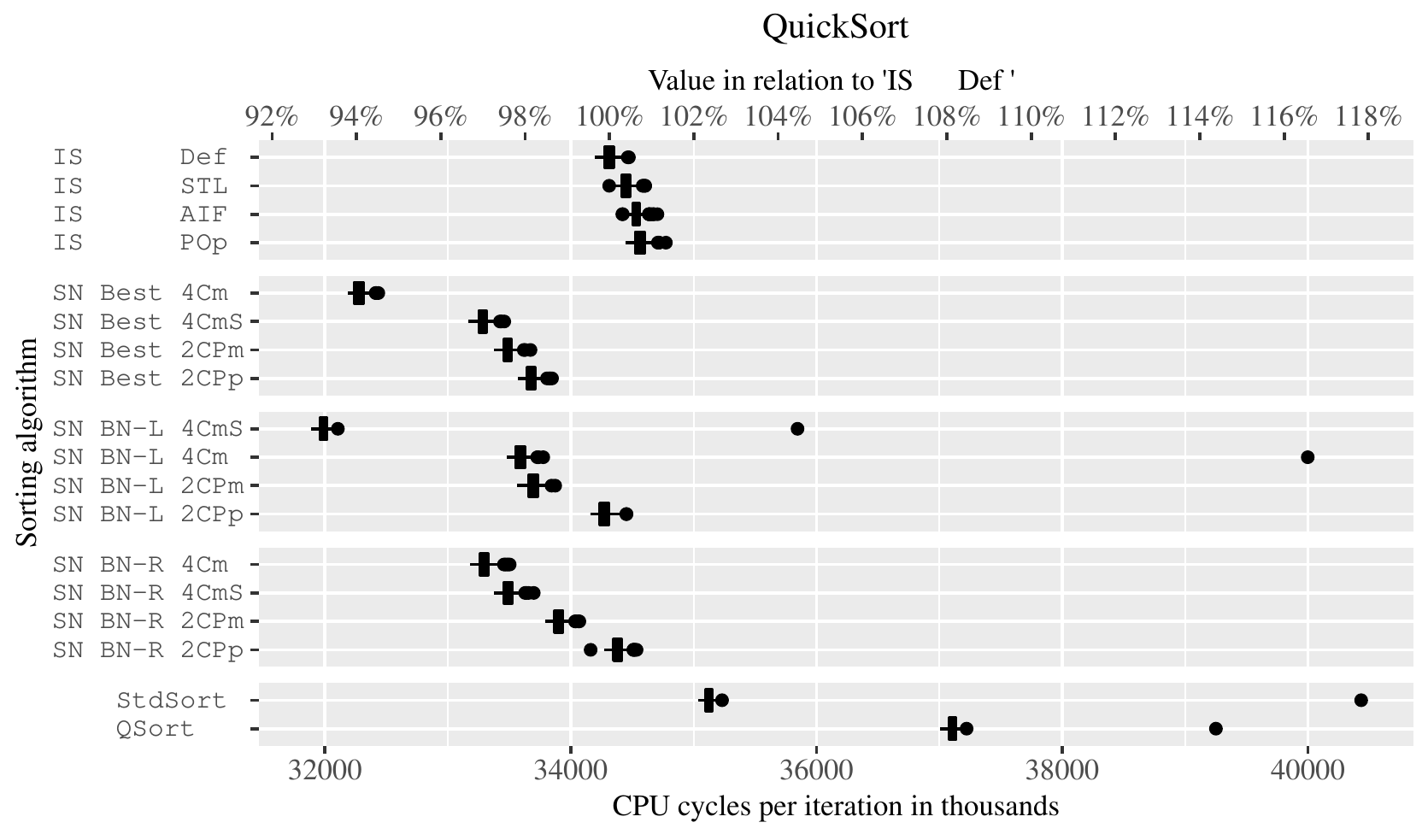}
  \caption{Running time of Quicksort with different base cases on machine Intel-2650} \label{plot:quicksort:A}
  \bigskip
  \includegraphics{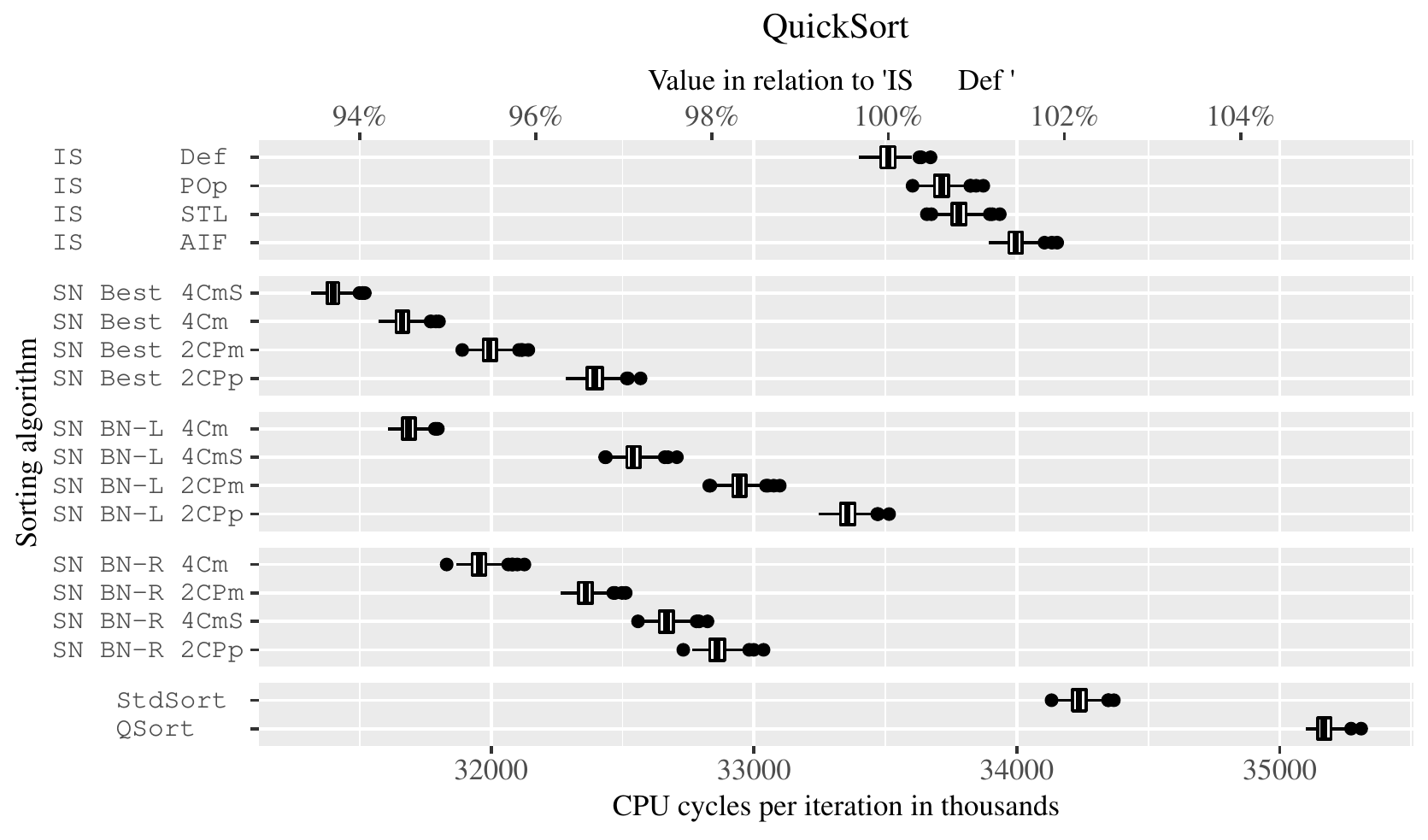}
  \caption{Running time of Quicksort with different base cases on machine Intel-2670} \label{plot:quicksort:B}
\end{figure}

\begin{figure}
  \includegraphics{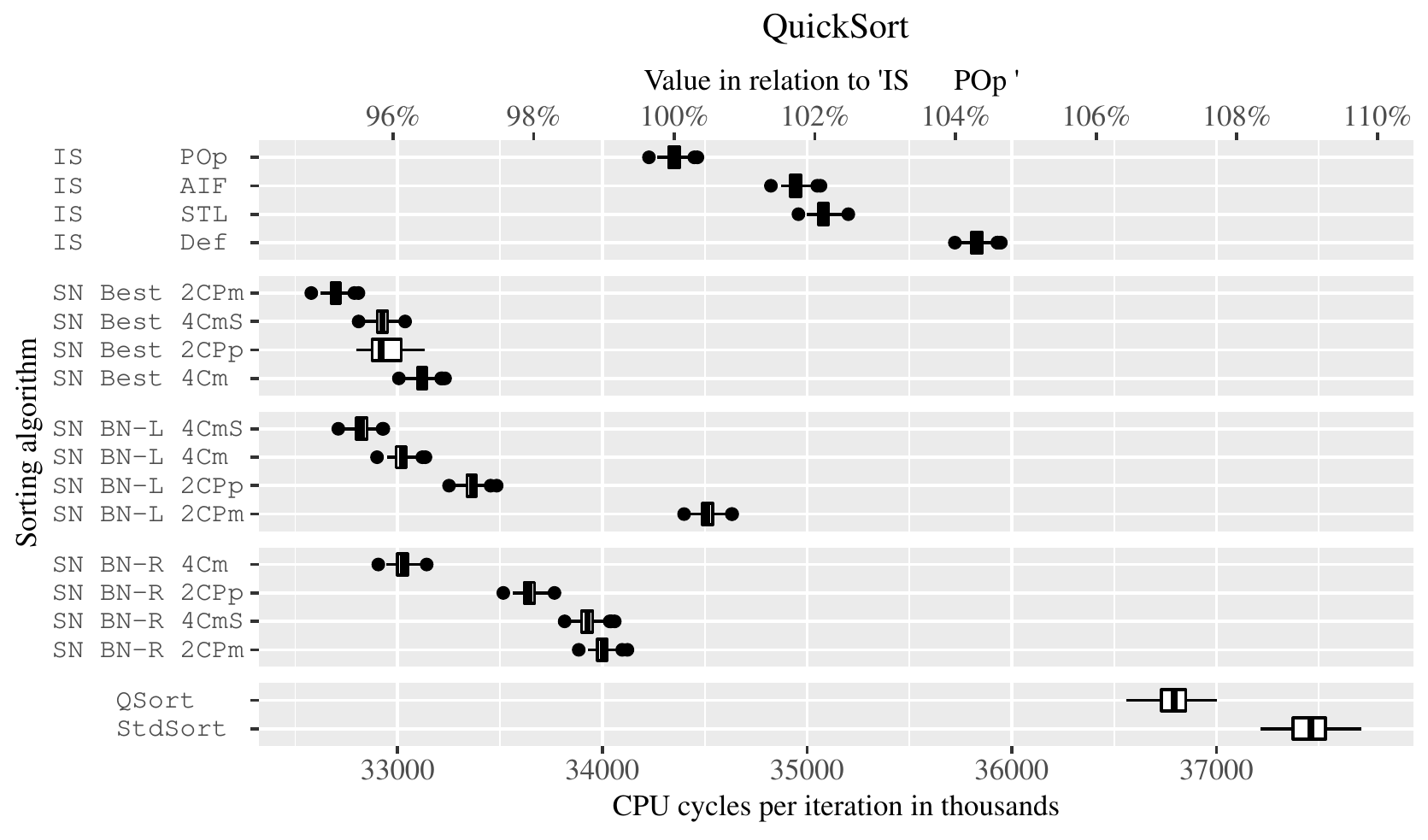}
  \caption{Running time of Quicksort with different base cases on machine Ryzen-1800X} \label{plot:quicksort:C}
  \bigskip
  \includegraphics{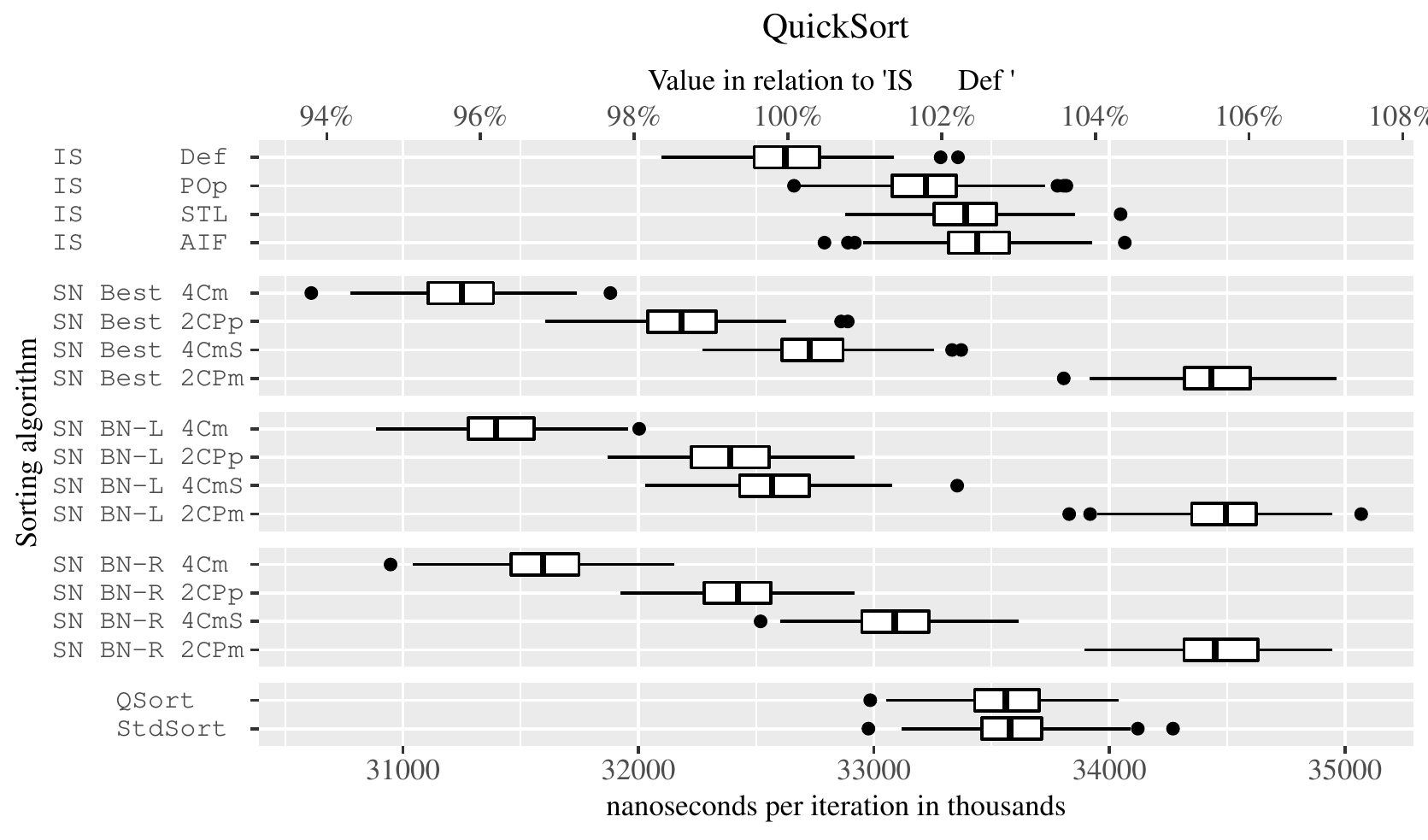}
  \caption{Running time of Quicksort with different base cases on machine RK3399} \label{plot:quicksort:D}
\end{figure}

The sorters were measured using the OneArrayRepeat experiment loop with parameters $\mathtt{numberOfIterations} = 50$, $\mathtt{numberOfMeasures} = 200$, $\mathtt{arraySize} = 2^{14} = 16384$.
The running times are shown as box plots in Figures~\ref{plot:quicksort:A}, \ref{plot:quicksort:B}, \ref{plot:quicksort:C}, and \ref{plot:quicksort:D}.
The speedups of the best modified Introsort variants with sorting networks are compared against the \texttt{std::sort} reference implementation and variants with insertion sort as base case in Table~\ref{table:completesort:speedups}.


\begin{table}
  \caption{Average speedups of the fastest sorting network over the fastest insertion sort as base case in Quicksort and unmodified \texttt{std::sort}} \label{table:completesort:speedups}
  \centering
  \begin{tabular}{ c | c | c | c | c }
                       & Intel-2650:           & Intel-2670:           & Ryzen-1800X:          & RK3399:              \\
                       & \texttt{SN BN-L 4CmS} & \texttt{SN Best 4CmS} & \texttt{N Best 2CPm} & \texttt{SN Best 4Cm} \\ \hline
    \texttt{IS Def}    & 6.7\%                 & 6.3\%                 & 8.7\%                & 4.26\%               \\
    \texttt{IS POp}    & 7.4\%                 & 6.8\%                 & 4.8\%                & 5.9\%                \\
    \texttt{std::sort} & 8.96\%                & 8.3\%                 & 12.7\%               & 6.9\%                \\
  \end{tabular}
\end{table}

Our first notable observation is that the variants with immediate insertion sort at the base case are faster than the one with the delayed final insertion sort, which probably comes from the fact that the elements are still present in the first- or second-level caches.
This also explains why the \texttt{2CPm} conditional swap performs the best with Quicksort, while we saw in the last section that this is not necessarily the case when we have a cache miss for loading the items.

Recalling the results from the previous sections, we expected large improvements by reducing the time needed for sorting sets of 2--16 items due to the branchless sorting networks.
However, the results with Quicksort highlight the networks' main weakness: the larger code size (see again Table~\ref{table:sort:codesize}).

By integrating the sorting networks into Quicksort for sorting the base cases, every time a partition has 16 elements or less, the executions switches from the code for Quicksort to the code for the sorting network.
Considering the code sizes shown in Table~\ref{table:sort:codesize}, we can conclude that Quicksort with insertion sort is around 1500 bytes, while Quicksort with sorting networks ranges from 12--37\,KiB.
We believe the code for Quicksort is partly removed from the L1 instruction cache and replaced with the code for the sorting network.
Because the network's code is a branchless sequence of conditional swaps, each line of code is accessed exactly once per base case sort.
This causes a lot of Quicksort's code to be removed from the instruction cache, counteracting the speedup of the sorting network which is then in the cache and will be partially removed again when Quicksort is handed back the flow of control.

This effect is more pronounced for machines Intel-2650, Intel-2670, and RK3399 which have 32 or 48~KiB of L1 instruction cache, where the speedup is 4.3--6.7\% for the best network base case over the best insertion sort base case, and 6.9--9\% over \texttt{std::sort}.
On machine Ryzen-1800X a larger improvement was achieved for \texttt{std::sort}, probably due to the 64~KiB L1 instruction cache.
Here we achieved a speedup of 12.7\% over \texttt{std::sort} when making use of the best networks.

Furthermore, it comes as no surprise that we do not see improvements as large as those in Section~\ref{section:experiments:normal} or \ref{section:experiments:inrow} because the partitioning steps of Quicksort take the same amount of time regardless of the base case sorter.
Some simple measurements showed that only 13--20\% of the time of Quicksort is spent in the base cases.
Hence, these are a considerable fixed part of the variants that is not optimized using sorting networks.

\subsection{Sorting a Medium-Sized Set of Items with Sample Sort} \label{section:experiments:samplesort}

In this section we focus on evaluating Register Sample Sort.
The measurements were done with two different goals in mind:
First, to determine which parameters work best for the machines and the array size set.
And second to see if the results from the preceding three Sections~\ref{section:experiments:normal}, \ref{section:experiments:inrow}, and \ref{section:experiments:quicksort} would relate to the results from sample sort with the sorting networks as base cases.

Register Sample Sort was measured using the OneArrayRepeat experiment loop with parameters: $\mathtt{numberOfIterations} = 50$, $\mathtt{numberOfMeasures} = 200$, and $\mathtt{arraySize} = 256$.
In the experimental results the parameters of Register Sample Sort are labeled with \enquote{\texttt{$x$$y$$z$}} where $x = \texttt{numberOfSplitters}$, $y =
\texttt{oversamplingFactor}$, and $z = \texttt{blockSize}$.

\begin{figure}
  \includegraphics{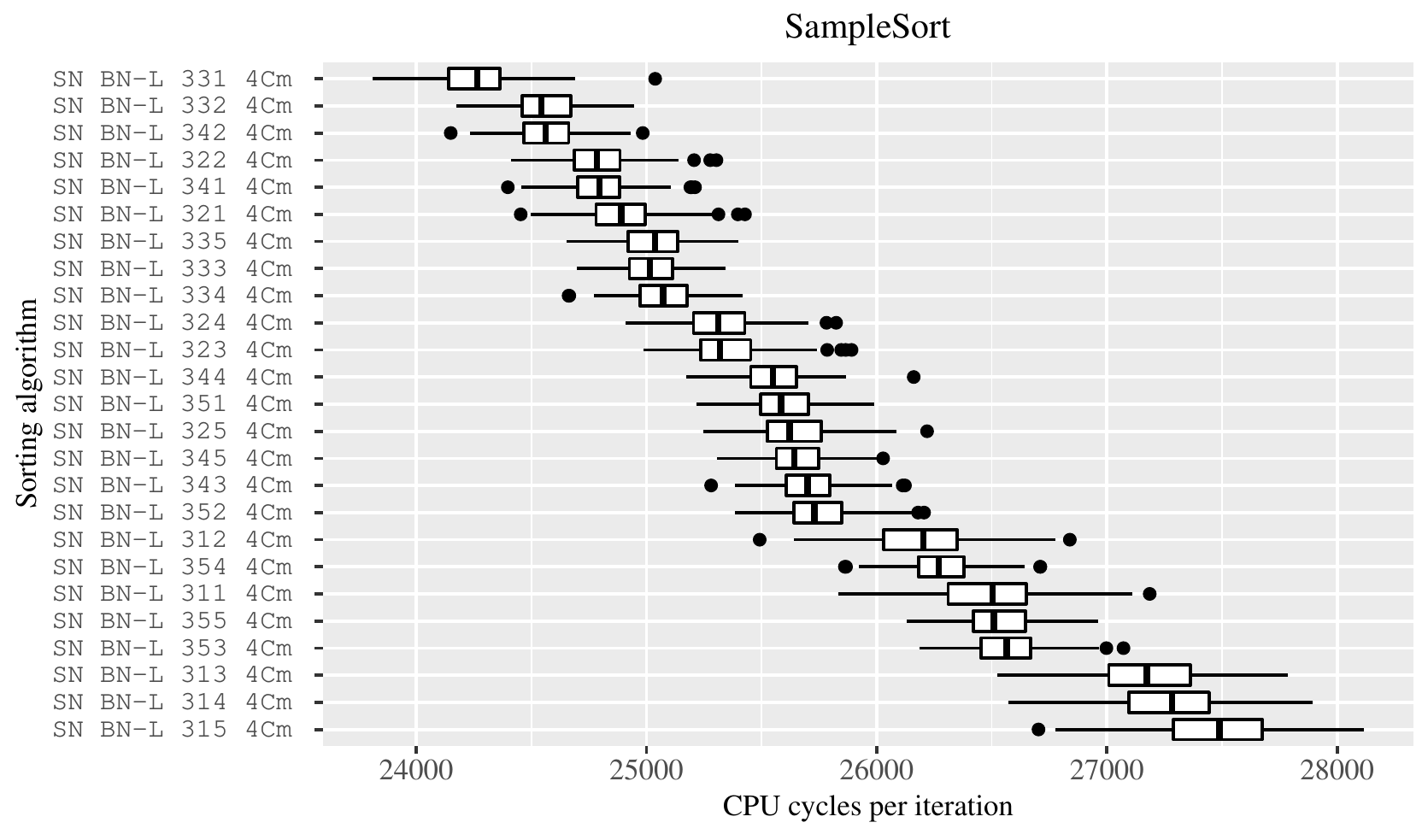}
  \caption{Register Sample Sort on machine Intel-2650 with 256 items and different configurations}\label{plot:samplesort:bonel:A}
  \bigskip
  \includegraphics{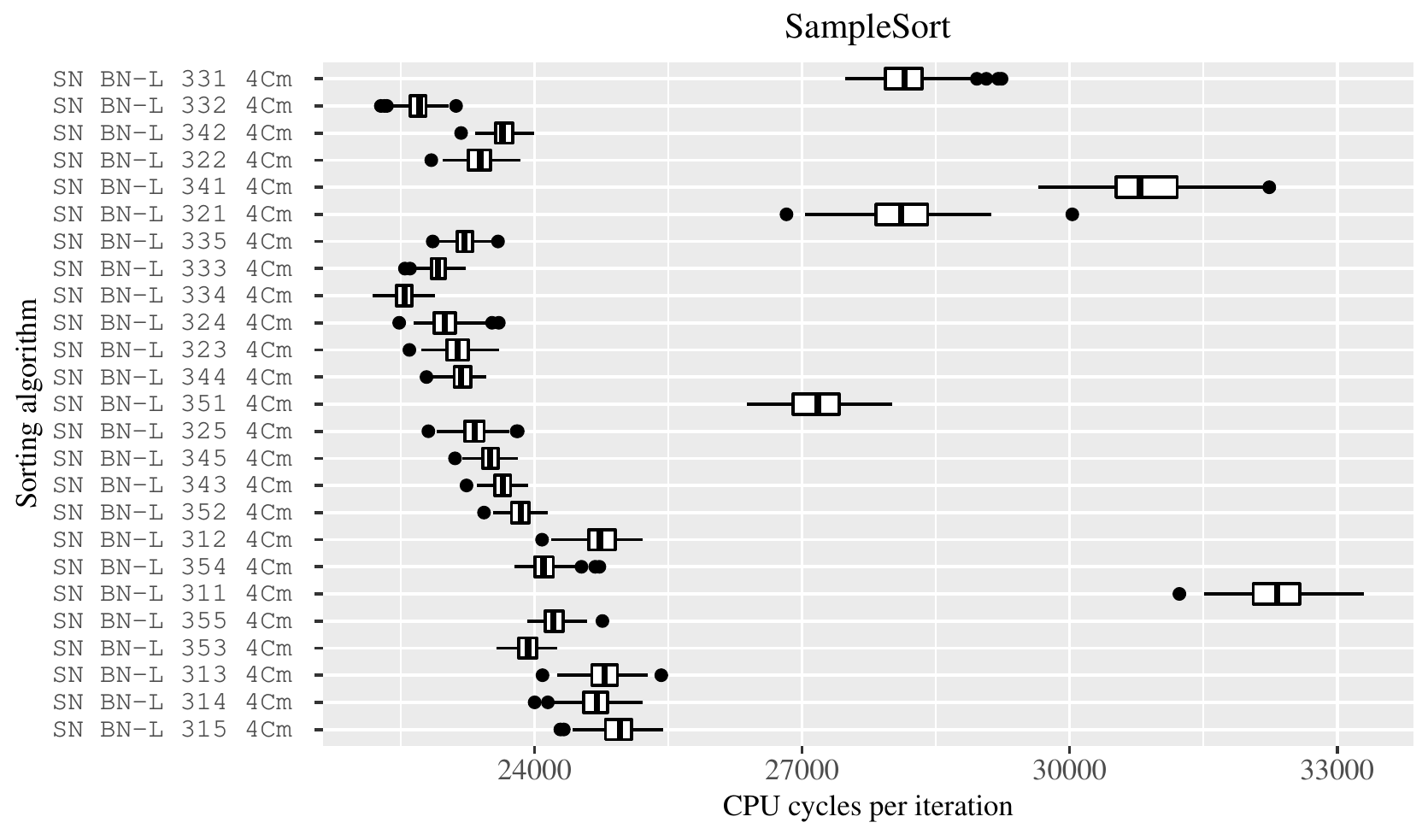}
  \caption{Register Sample Sort on machine Intel-2670 with 256 items and different configurations}\label{plot:samplesort:bonel:B}
\end{figure}

\begin{figure}
  \includegraphics{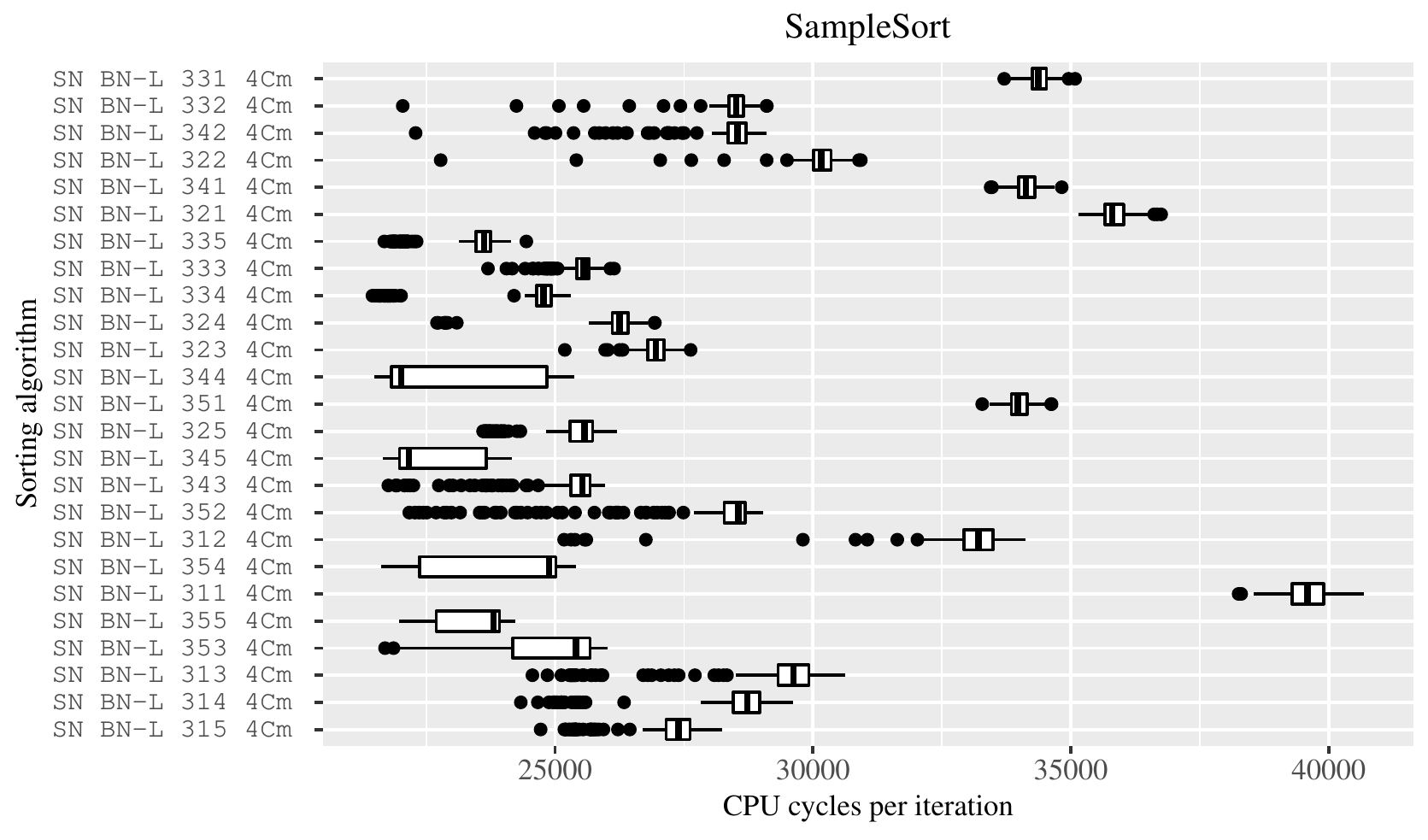}
  \caption{Register Sample Sort on machine Ryzen-1800X with 256 items and different configurations}\label{plot:samplesort:bonel:C}
  \bigskip
  \includegraphics{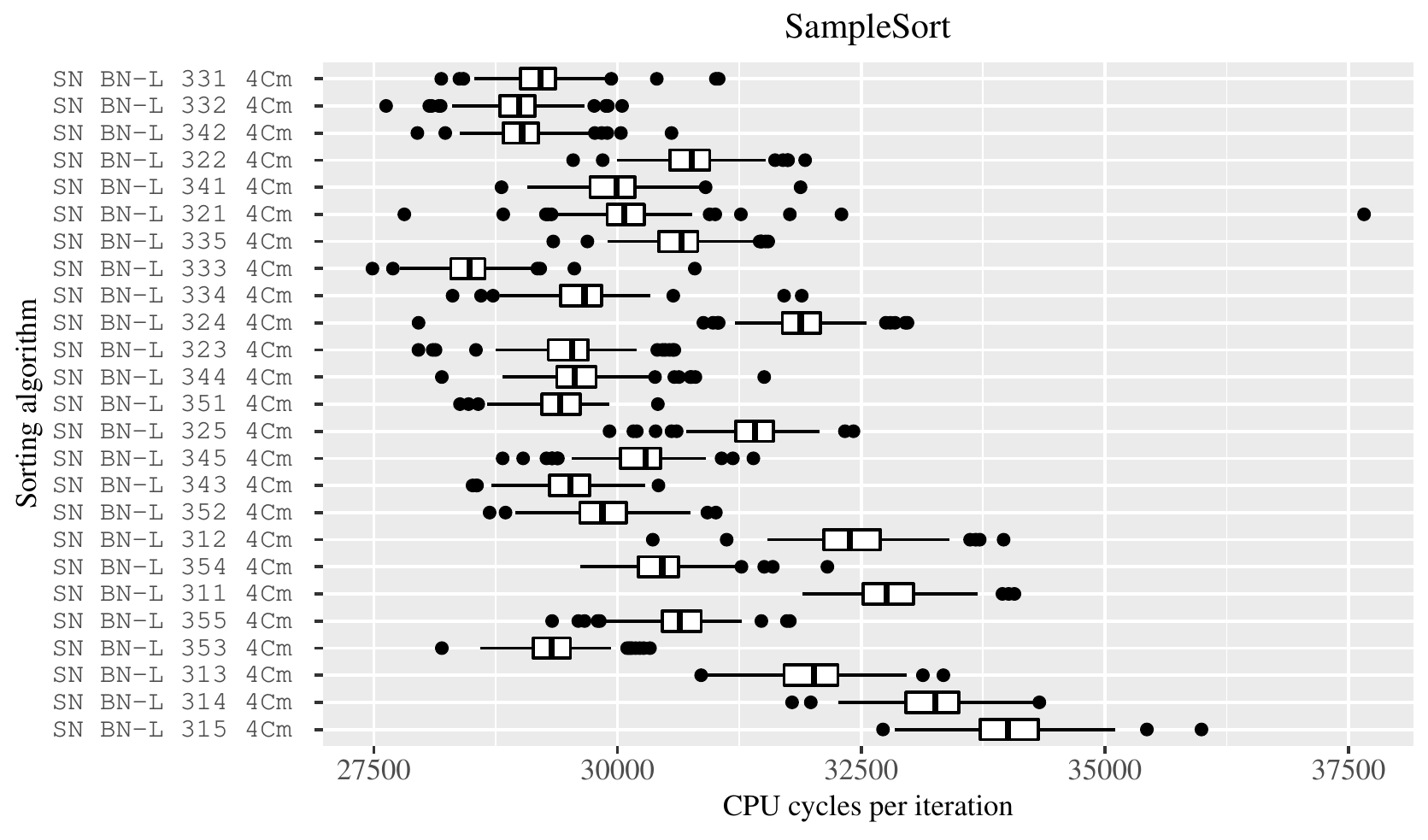}
  \caption{Register Sample Sort on machine RK3399 with 256 items and different configurations}\label{plot:samplesort:bonel:D}
\end{figure}

Figures~\ref{plot:samplesort:bonel:A}, \ref{plot:samplesort:bonel:B}, \ref{plot:samplesort:bonel:C}, and \ref{plot:samplesort:bonel:D} show box plots of running times of Register Sample Sort with $\texttt{numberOfSplitters} = 3$ and locality-optimized Bose-Nelson networks as base case on 256 items.
To be able to compare the results on the different machines, the configurations in all plots were ordered based on their speed on machine Intel-2650.
We measured larger variances and got a lot more outliers, so choosing a \enquote{best} configuration was not so easy.

An oversampling factor of 3 performed best with respect to the median on machine Intel-2650, Intel-2670, and RK3399, while 4 was best on machine Ryzen-1800X.
The best results with respect to $\texttt{blockSize}$ are also interesting, because these depend on the number of general purpose registers available.
On machine Intel-2650 $\texttt{blockSize} = 1$ was best with oversampling factor 3, while machine Intel-2670 allowed a $\texttt{blockSize} = 4$, with 3 and 2 close behind.
On machine Ryzen-1800X block sizes larger than 2 performed better (on average) along with an oversampling factors of 3 or greater.
When looking at the other networks and insertion sort as base case, consistently well performing parameters are an oversampling factor of 3 and a block size of 4, but with very little lead over other configurations.

That is interesting to see because all three machines run x86\_64 assembly instructions and have the same number of publicly visible general purpose registers.
What comes into play here are hidden virtual registers and the size of the instruction cache: Machine Ryzen-1800X has double the amount of L1 instruction cache of what machines Intel-2650 and Intel-2670 have.
We can only assume that the instructions for classifying three elements need more space than the smaller 32 KiB instruction caches can provide, while the 64 KiB instruction cache in machine Ryzen-1800X can fit the instructions for classifying four and/or almost five elements at once, considering that block size 5 also performs well.

Machine RK3399 however is an ARM chip, which shows a much larger variance in the results of Register Sample Sort, but is also more robust against the parameters.
On machine RK3399 oversampling factor 3 with $\texttt{blockSize} = 3$ was best, which indicates a similar number of general purpose registers as in the x86\_64 machines.
However, the larger robustness can actually be interpreted as that the ARM chip incorporates \emph{fewer} unpredictable running time optimizations such as speculative and out-of-order execution of instructions.

\begin{figure}
  \includegraphics{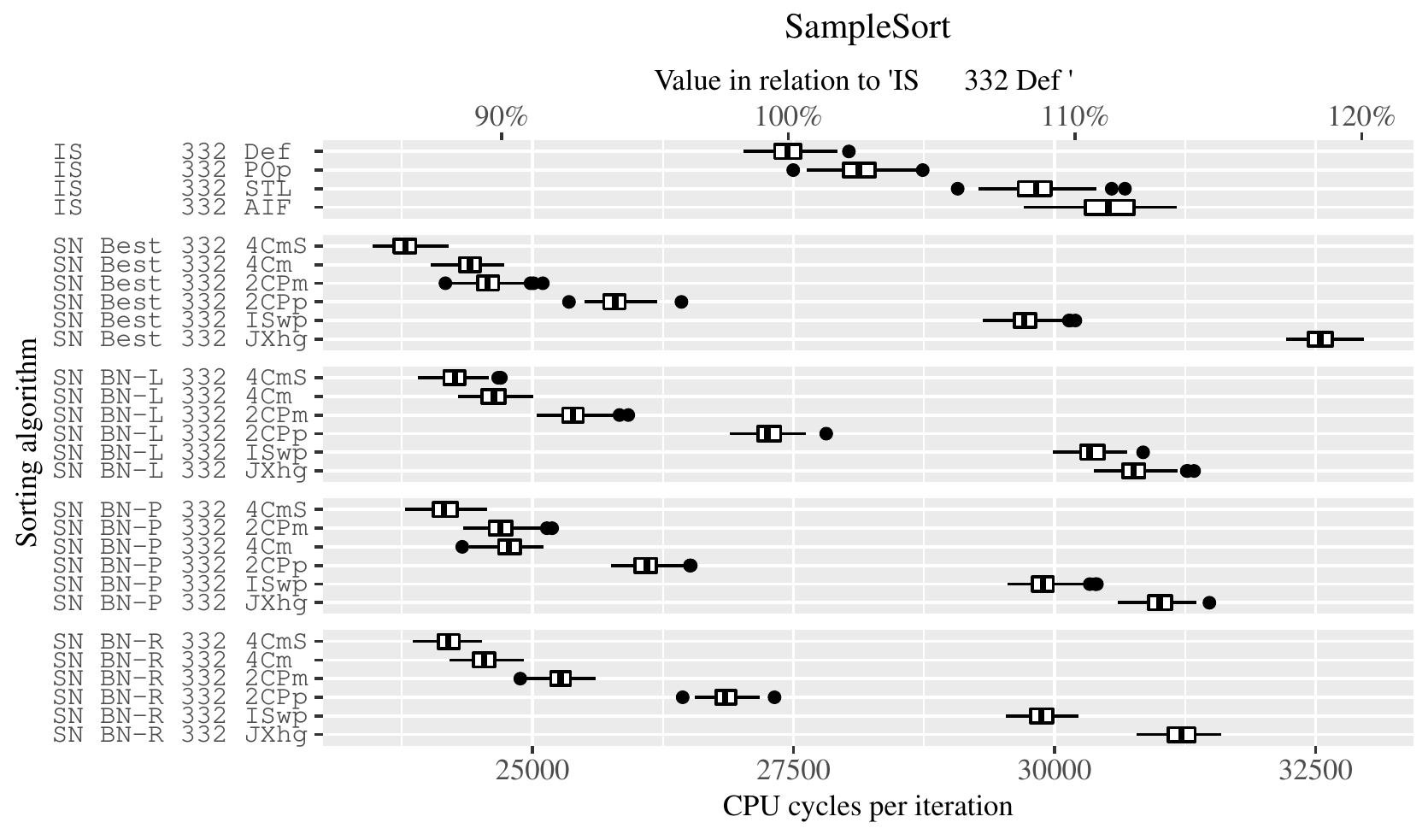}
  \caption{Register Sample Sort \texttt{332} with different base cases on machine Intel-2650} \label{plot:samplesort:s332:A}
  \bigskip
  \includegraphics{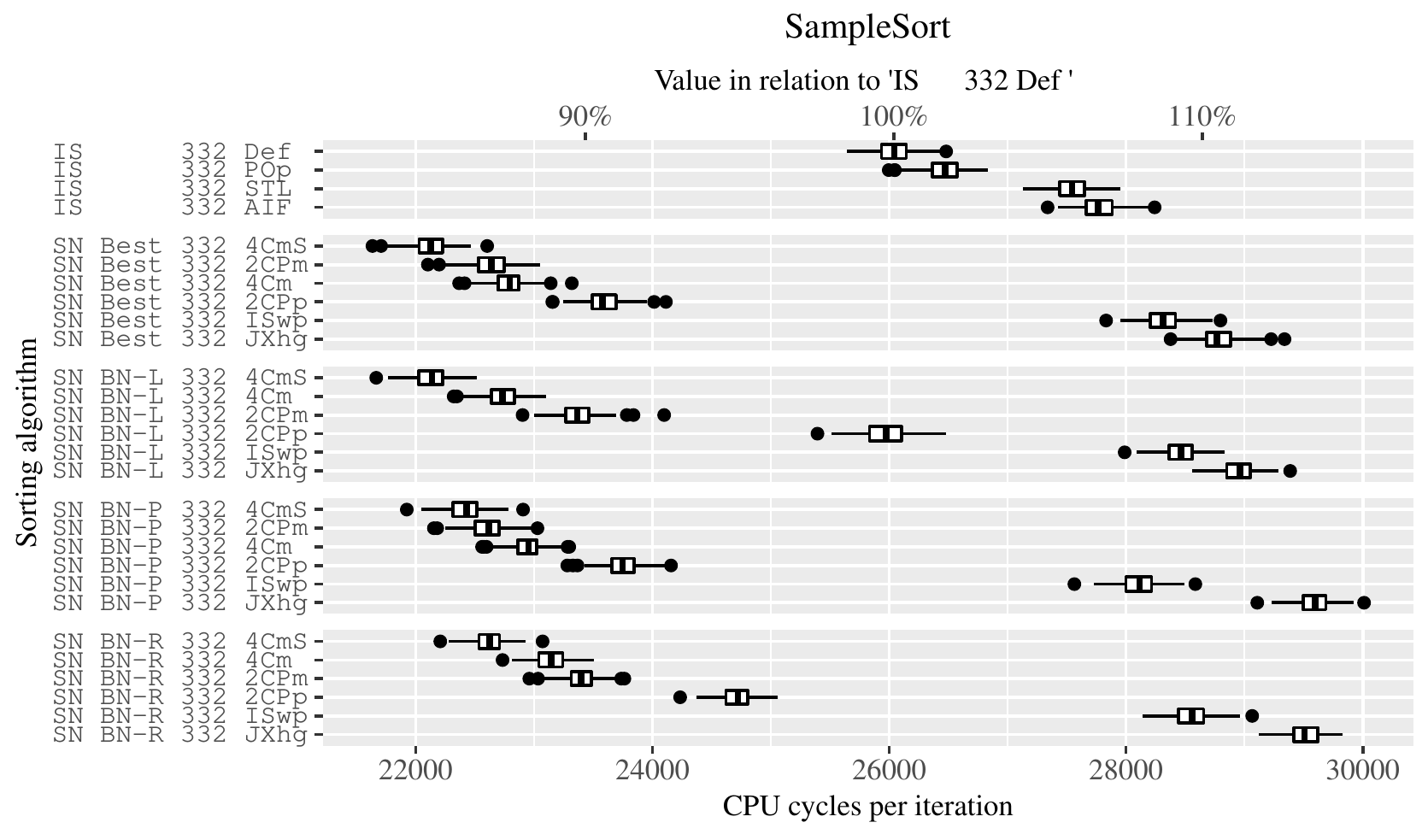}
  \caption{Register Sample Sort \texttt{332} with different base cases on machine Intel-2670} \label{plot:samplesort:s332:B}
\end{figure}

\begin{figure}
  \includegraphics{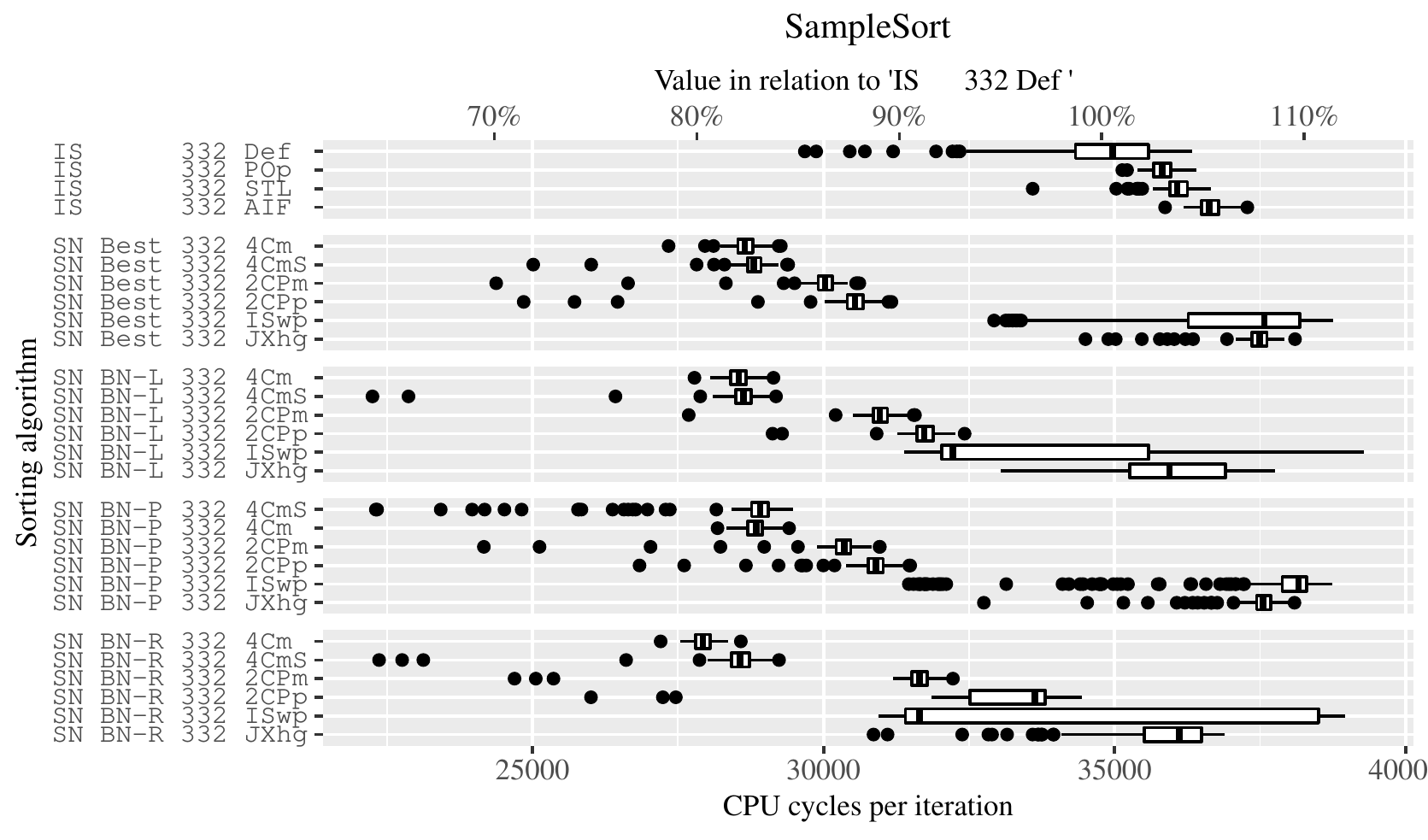}
  \caption{Register Sample Sort \texttt{332} with different base cases on machine Ryzen-1800X} \label{plot:samplesort:s332:C}
  \bigskip
  \includegraphics{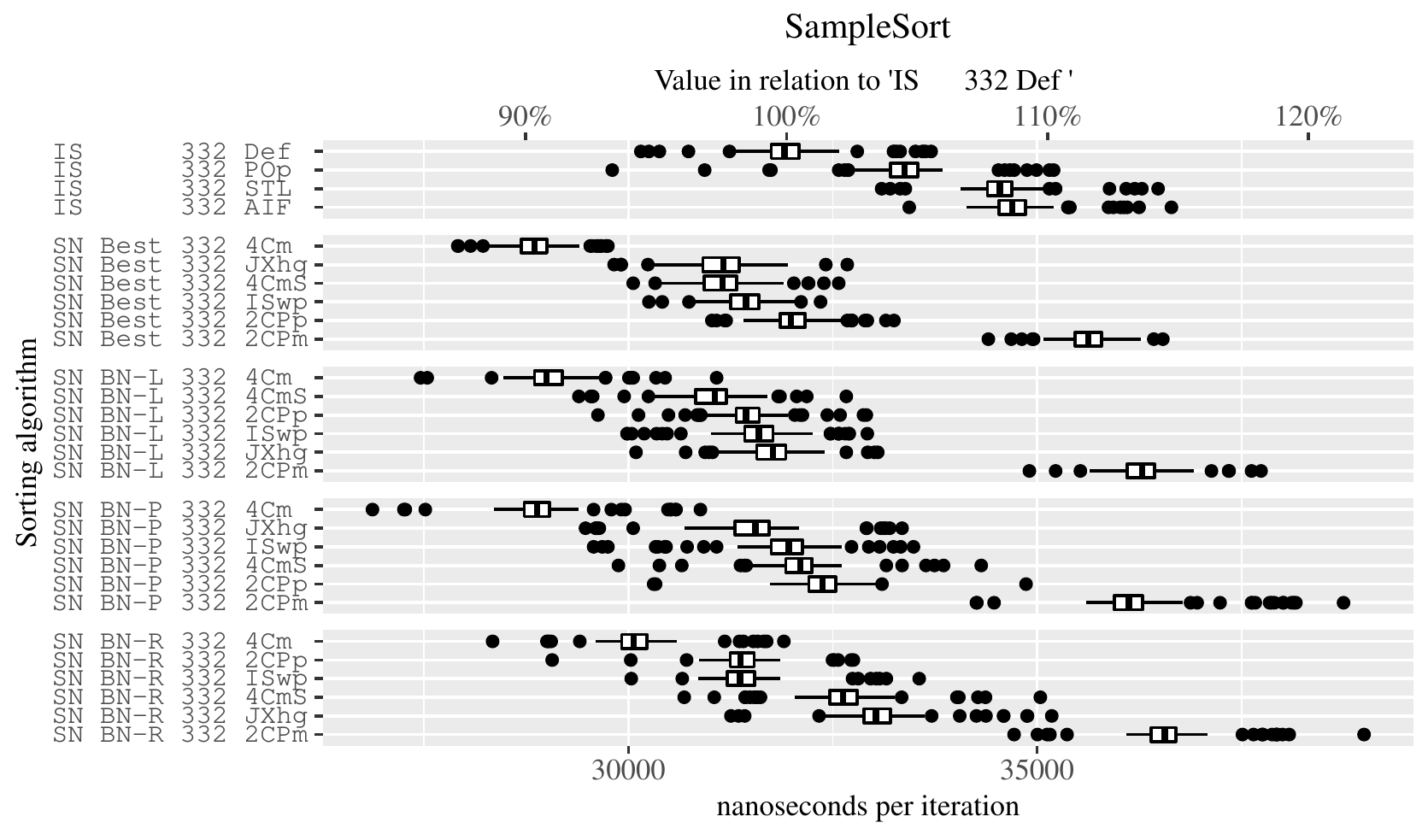}
  \caption{Register Sample Sort \texttt{332} with different base cases on machine RK3399} \label{plot:samplesort:s332:D}
\end{figure}

The second goal was to see if the results from Sections~\ref{section:experiments:normal}, \ref{section:experiments:inrow}, and \ref{section:experiments:quicksort} would relate to using sample sort with the sorting networks as base cases.
These results can be seen in Figures~\ref{plot:samplesort:s332:A}, \ref{plot:samplesort:s332:B}, \ref{plot:samplesort:s332:C}, and \ref{plot:samplesort:s332:D} for the \texttt{332} configuration.
All measurements were made with a base case limit of 16.

The achieved speedups of using the sorting networks are summarized in Table~\ref{table:samplesort:speedups}.
On the left we see Register Sample Sort with insertion sort as base case \texttt{I Def} and unmodified \texttt{std::sort} that was also measured sorting 256 elements.
On the top we see the best performing network \enquote{\texttt{SN Best 332 4CmS}} as a base case for Register Sample Sort on all three machines.
The number indicates the speedup of Register Sample Sort \emph{with} the sorting network over Register Sample Sort with insertion sort and over std::sort.

Again we see that due to machine Ryzen-1800X having a larger L1 instruction cache the performance gain is greater than for the other machines.
And the results from ARM machine RK3399 again have a larger variance than the others with conditional swap \texttt{4Cm} performing better than \texttt{4CmS}.
Unlike in the previous section though we achieved much greater speedups as a result of using the sorting networks as a base case.
That comes from the fact that Register Sample Sort has no unpredictable branches classifying the elements, as opposed to Quicksort having to deal with conditional branches during the partitioning, while both need to invest the same time to sort all the base cases.
So with Register Sample Sort, the base case sorting takes up a larger time portion of the whole execution than it does with Quicksort.
From further results we also see that we can get up to a factor of 1.4 faster than \texttt{std::sort} for sets of 256 items with very few conditional branches.



\begin{table}
  \caption{Average speedups of the fastest sorting network over the fastest insertion sort as base case in Register Sample Sort and unmodified \texttt{std::sort}} \label{table:samplesort:speedups}
  \centering
  \begin{tabular}{ c | c | c | c | c }
                       & Intel-2650:           & Intel-2670:           & Ryzen-1800X:           & RK3399:              \\
    		       & \texttt{SN Best 4CmS} & \texttt{SN Best 4CmS} & \texttt{SN BN-R 4Cm } & \texttt{SN Best 4Cm} \\ \hline
    \texttt{I Def}     & 13.3\% 	       & 15\% 		       & 19.7\% 	       & 9.6\%                \\
    \texttt{std::sort} & 28.7\% 	       & 32.5\% 	       & 21.5\% 	       & 2.8\%                \\
  \end{tabular}
\end{table}

\subsection{Sorting a Large Set of Items with \ipso} \label{section:experiments:ipso}

With our efficient implementation of Register Sample Sort for medium-sized sets we can now incorporate our new base case sorters into a complex sorting algorithm.
We evaluated them in Axtmann et. al's \textit{In-Place Parallel Super Scalar Samplesort} (\ipso) \cite{axtmann2017inplace} without additionally introducing parallelism into our experiments.
The \ipso algorithm has many parameters that can be adjusted.
For our evaluation the most important parameter was \texttt{IPS4oBaseCaseSize}, which specifies which base case size to \emph{aim} for.
Even though this number is the goal, \ipso may output larger base cases because it is a large-size sorter.
This was the reason we developed Register Sample Sort to break these medium-sized sets down into sizes that can be sorted using the sorting networks.

We initially started the measurement using the best combination of Register Sample Sort from Section~\ref{section:experiments:samplesort} as a base case for \ipso, together with the default \texttt{IPS4oBaseCaseSize} = 16.
However, this combination turned out to perform worse than just insertion sort.

In preliminary measurements we found that 51\% of base cases are at most 16 items, and 78\% are at most 32 items with \texttt{IPS4oBaseCaseSize} = 16.
For \texttt{IPS4oBaseCaseSize} = 32, 23\% of base cases are at most 16 items, and 47\% are at most 32 items.
From that it was evident that in most of the instances with parameter \texttt{IPS4oBaseCaseSize} = 16  the base case sorter was being invoked on sets smaller than even 32 elements.
That also meant that Register Sample Sort had to deal with inputs just slightly larger than 16, which incurs a larger overhead than plain insertion sort and is not justified by the larger amount of items.

In addition to that the size of the instruction cache that had already had a great influence on the measurements of Quicksort seemed to be another factor for the bad performance of Register Sample Sort as a base case.

That is why we decided to measure the following setups:
\begin{enumerate}[nosep]
	\item[(a)]{pure insertion sort as base case (\texttt{IS}) with $\texttt{IPS4oBaseCaseSize} \in \{ 16, 32 \}$,}
	\item[(b)] {Register Sample Sort as base case (\texttt{SS+SN}) with $\texttt{IPS4oBaseCaseSize} \in \{ 16, 32, 64 \}$, tree configurations \enquote{\texttt{331}} and \enquote{\texttt{332}}, and \texttt{Best} networks and Bose-Nelson networks optimizing locality (\texttt{BN-L}) and recursion (\texttt{BN-R}) for base case size 16 of Register Sample Sort with conditional swap \texttt{4CmS}, and}
	\item[(c)] {a combination of the sorting networks and insertion sort (\texttt{IS+SN}) without Register Sample Sort.}
\end{enumerate}
Variant (c) was introduced because the base case sizes were often smaller than 16, and we wanted to make use of that by using the sorting networks, while not having to rely on Register Sample Sort with its larger overhead for the slightly larger base cases, hence \texttt{IS+SN} uses the Bose-Nelson networks optimizing locality if the set had 16 elements or less, and insertion sort otherwise.

\begin{figure}
  \includegraphics{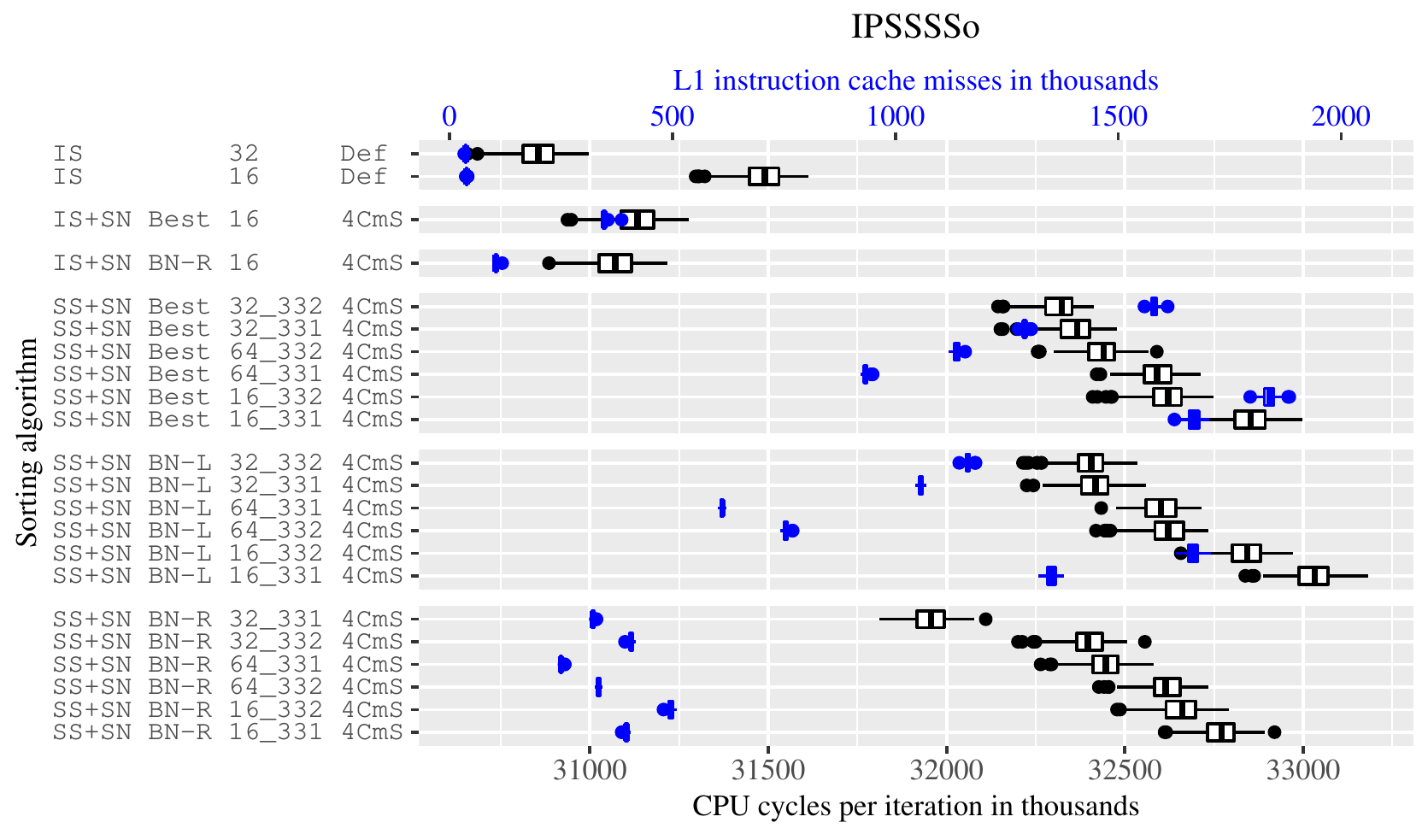}
  \caption{Sorting times for \ipso on machine Intel-2650 with different base cases and base case sizes} \label{plot:ipso:A}
  \bigskip
  \includegraphics{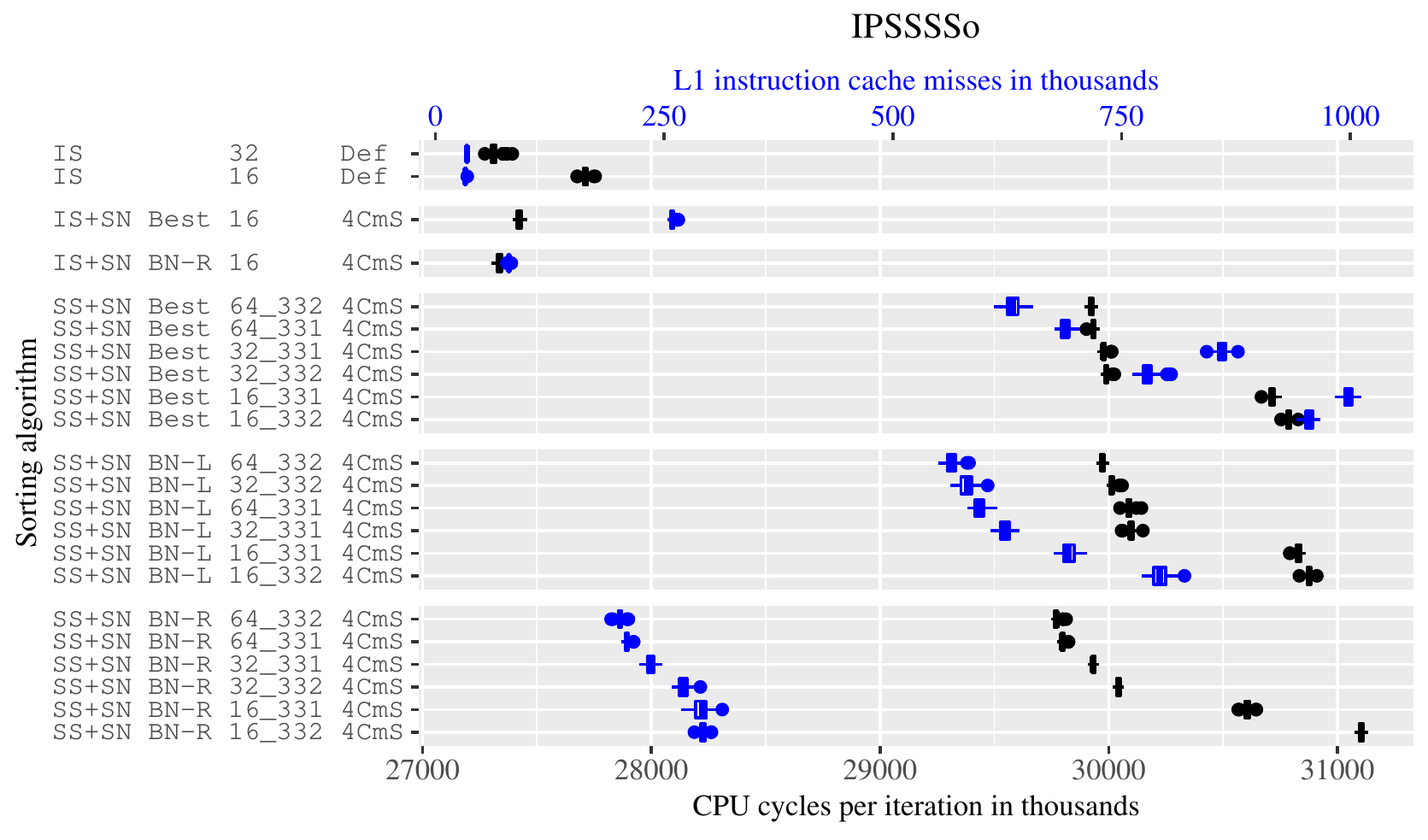}
  \caption{Sorting times for \ipso on machine Intel-2670 with different base cases and base case sizes} \label{plot:ipso:B}
\end{figure}

\begin{figure}
  \includegraphics{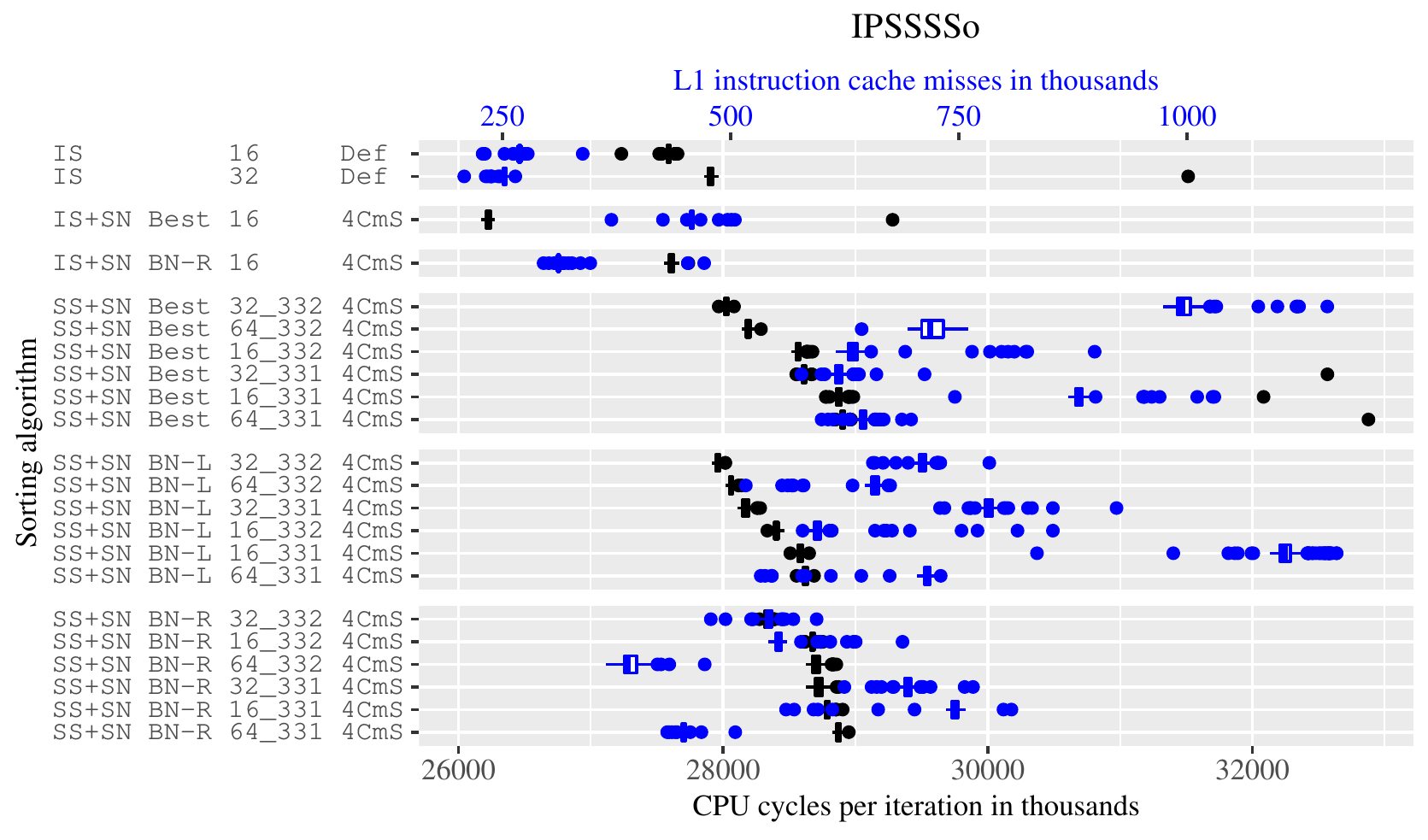}
  \caption{Sorting times for \ipso on machine Ryzen-1800X with different base cases and base case sizes} \label{plot:ipso:C}
  \bigskip
  \includegraphics{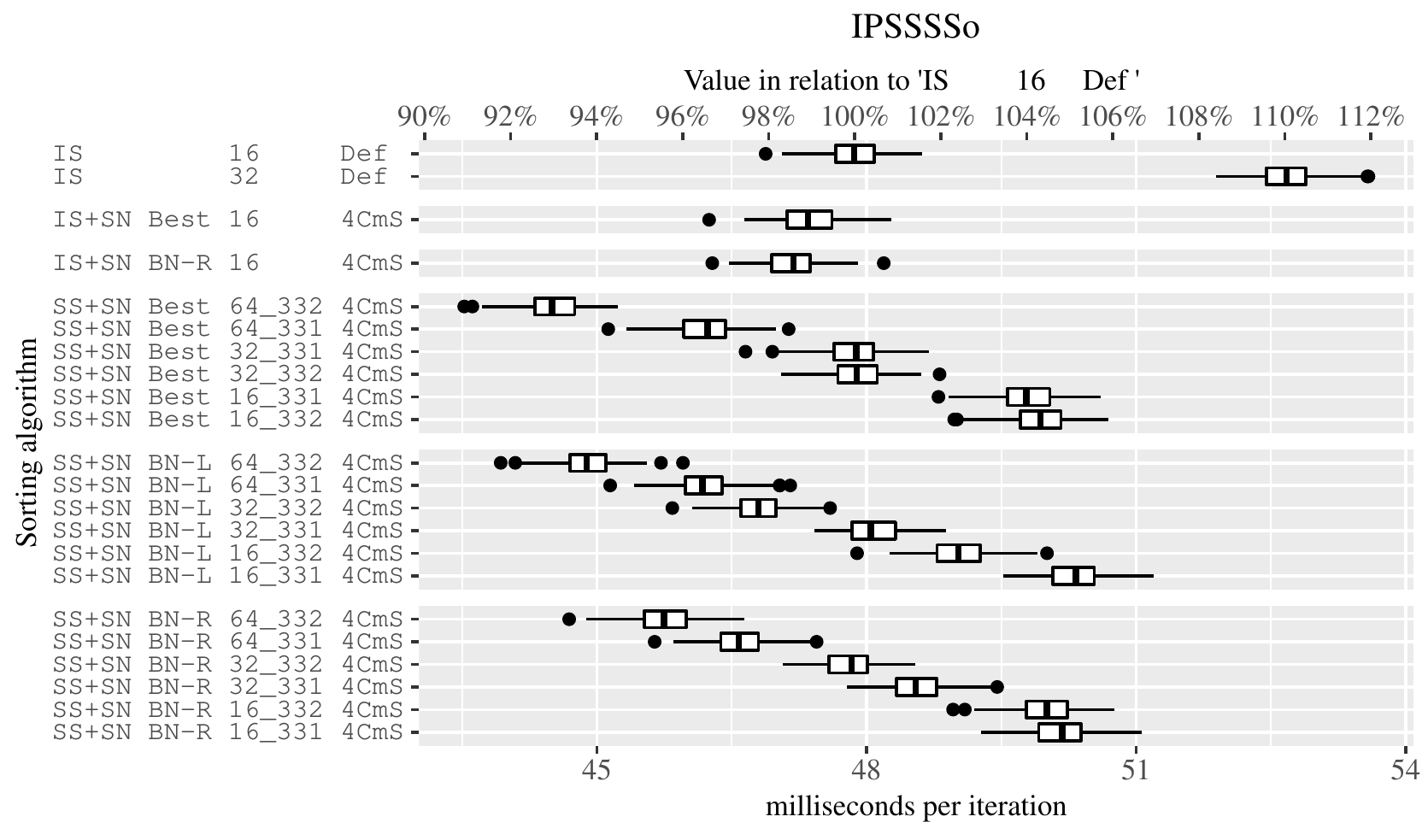}
  \caption{Sorting times for \ipso on machine RK3399 with different base cases and base case sizes} \label{plot:ipso:D}
\end{figure}

Figures~\ref{plot:ipso:A}, \ref{plot:ipso:B}, \ref{plot:ipso:C}, and \ref{plot:ipso:D} display the results from the measurements with the four variants above.
The $\texttt{IPS4oBaseCaseSize} \in \{ 16, 32 \}$  was added to the variant's label along with an underscore followed by the Register Sample Sort configuration.
The OneArrayRepeat experiment loop was used with parameters $\mathtt{numberOfIterations} = 50$, $\mathtt{numberOfMeasures} = 200$, and $\mathtt{arraySize} = 2^{18} = 262144$.
On machine Intel-2650 one outlier for \enquote{\texttt{SS+SN BN-R 32\_331 4CmS}} with the value 38454084 cycles was removed from the plot for better readability.

As already seen in Axtmann et. al's publication\cite{axtmann2017inplace}, we measured an average speedup factor of 2.3 over \texttt{std::sort} with unchanged \ipso across all machines.
However, on machine Intel-2650, none of our variants (\texttt{IS}, \texttt{SS+SN}, \texttt{IS+SN}) led to an improvement in sorting speed over the default use of insertion sort with \texttt{IPS4oBaseCaseSize} = 32.
For machine Intel-2670, interestingly, using Register Sample Sort did not lead to an improvement, but the combination of insertion sort and Bose-Nelson networks did manage to reach par with the default implementation.
For machine Ryzen-1800X, Register Sample Sort also did not lead to an improvement, but \texttt{IS+SN} outperformed the default insertion sort by about 5\%.
Only on machine RK3399 we see a reasonable speedup of Register Sample Sort with sorting networks of up to 7\% in the best configuration.

Figures~\ref{plot:ipso:A}, \ref{plot:ipso:B}, and \ref{plot:ipso:C} also show the number of L1 instruction cache misses in blue plotted on the top x-axis.
These were measured using the performance counters in the CPU using the Linux \texttt{perf} interface, which were only available on Intel and AMD machines, and not on the RK3399 system.
In Figure~\ref{plot:ipso:A} we can recognize some expected behavior: \ipso with insertion sort code (\texttt{IS}) is small enough to fit into the L1 cache and thus has the lowest cache miss count.
The versions with sorting networks are larger, with the ``best'' networks (\texttt{Best}) incurring the most instruction cache misses.
The recursively defined Bose-Nelson networks (\texttt{BN-R}) require less than half as many cache misses as the locality-aware (\texttt{BN-L}) variants.
The same expected behavior is also visible on machine Intel-2670 in Figure~\ref{plot:ipso:B}.
There appears to be a relationship between time (CPU cycles) and L1 instruction cache misses, this relationship is neither perfect nor linear, but the smaller sorter code \texttt{IS} and \texttt{BN-R} is also faster on these systems.
On the AMD Ryzen-1800X in Figure~\ref{plot:ipso:C} the first striking difference is the wide range of results for the the L1 cache misses.
There are many more outliers on the AMD system, possibly due to a different hardware performance counters or sampling thereof.
The actual body of the boxplots however are usually very small, which leads us to believe that these outliers are more measuring errors than actually a concern.
The absolute number of L1 instruction cache misses on the AMD machine starts at around 250k while they are much smaller on the Intel machines. This is probably due to the different hardware measurement methods.
The L1 instruction cache misses again appear to have a relationship to CPU cycles, however it is less clear than on the Intel machines.
For example, the first of the  two insertion sorts (\texttt{IS}) is faster, but has more cache misses than the second one.
The same is the case for the insertion sorts with sorting networks (\texttt{IS}+\texttt{SN}) in the third and fourth line.
The order within the sorting networks is also strange: faster networks often have higher instruction cache miss counts.
The correlation between cycles and cache misses is much clearer in Figure~\ref{plot:ipso:B}.
As stated before, there are no cache miss numbers in Figure~\ref{plot:ipso:D}, because the hardware or software did not support measuring them.
We can conclude that measuring L1 cache misses yields further evidence that the sorting networks are large in code size, there appears to be relationship with running time, but it is less clear and definitely not linear.

\begin{table}
  \caption{Average speedups of the fastest sorting network over the fastest insertion sort as base case in \ipso and unmodified \texttt{std::sort}} \label{table:ipso:speedups}
  \centering\scriptsize\def\arraystretch{1.2}
  \begin{tabular}{ c | c | c | c | c }
                       & Intel-2650:                 & Intel-2670:                 & Ryzen-1800X:                & RK3399:                     \\
                       & \texttt{IS+SN BN-R 16 4CmS} & \texttt{IS+SN BN-R 16 4CmS} & \texttt{IS+SN BN-R 16 4CmS} & \texttt{SS+SN Best 64 4CmS} \\ \hline
    \texttt{IS 16 Def} & 1.3\%                       & 1.36\%			   & 4.9\%                       & 7\%                         \\
    \texttt{IS 32 Def} & -0.7\%                      & -0.08\%                     & 6\%                         & 15.4\%                      \\
    \texttt{std::sort} & 58.2\%                      & 62.2\%                      & 66.6\%                      & 38.4\%                      \\
  \end{tabular}
\end{table}


\section{Conclusion} \label{section:conclusion}

\subsection{Results and Assessment}
We have seen that for sorting sets of up to 16 elements it can be viable to use sorting algorithms other than insertion sort.
We looked at sorting networks in particular, paying special attention to the implementation of the conditional swap and giving multiple alternative ways of realizing it.

After seeing that the sorting networks outperform insertion sort each on their own for a specific array size in Section~\ref{section:experiments:normal} and \ref{section:experiments:inrow}, we saw in Section~\ref{section:experiments:quicksort} that this improvement does not necessarily transfer to sorting networks being used as base case sorter in Quicksort.
Because the networks have a larger code size, the code for Quicksort is removed from the instruction cache and the advantage of not having conditional branches is impaired by that larger code size.
But we also saw that for machines with larger instruction caches using sorting networks with Quicksort can lead to visible improvements (about 13\%), but we were still able to reach up to 9\% improvement over \texttt{std::sort} on machine Intel-2650.

Then we integrated the sorting networks into a more advanced sorter, \ipso, which was possible by adding an intermediate sorter into the procedure.
For that we created Register Sample Sort, which is an implementation of Super Scalar Sample Sort that holds the splitters in general-purpose registers instead of an array. When measuring \ipso with Register Sample Sort as a base case, we found that the instruction cache makes even more of a difference, because we now add the code size for Register Sample Sort on top of the code size for the sorting networks.

We proposed an additional alternative to Register Sample Sort, using a combination of insertion sort and sorting networks: For base cases of 16 elements or less, we used the sorting network, for any size above that insertion sort.

On machine Intel-2650 with a smaller instruction cache of 32 KiB we could not achieve a speedup with any of the variants, on machine Intel-2670 the combination of insertion sort and sorting networks led to a running time on par with insertion sort as base case.
The only substantial improvement we achieved with \ipso was on the ARM machine, where using Register Sample Sort led to an improvement of 7\% over the best insertion sort variant.

In closing, we want to mention that this particular implementation only compiles when using the gcc C++ compiler due to compiler-dependent inline-assembly statements.
This also means that the code is probably not as fast as it could be due to the inline-assembly not being optimized by the compiler.
However, as of today, there is no consistent method to generate conditional move operations from C++ without assembly.
The complete project is available on github at \url{https://github.com/JMarianczuk/SmallSorters}.

\subsection{Experiences and Hurdles}
The greatest hurdle we encountered during this project was, as mentioned in Section~\ref{section:parameters}, the fact that the compiler reduces its optimizations with increasing compilation effort, when compiling only a single source file.
That can lead to performance variations that happen for no \enquote{apparent} reason, and is especially tricky when dealing with templated methods that can not be moved from header files into source files.
The solution was to use code generation and to include all logically coherent method invocations in one wrapper method that is then placed in its own source file, to not have different parts of the program influencing each other over the decision which one gets to be optimized and which one not.

\subsection{Future Work}
Empirically, an important improvement would be
to mitigate the instruction cache faults that limited performance in
our experiments. Besides looking at sorting networks with smaller
memory footprint, one could separate and delay solving the base cases from the
remaining logics of a divide-and-conquer sorter such as Quicksort or
sample sort. By also bucket-sorting the base cases by their size, one
could further improve instruction-cache locality -- one could then solve
batches of subproblems of identical size.  The obvious downside
of worsened data locality could be mitigated by performing this
separation not globally but on subproblems that fit into an
appropriate level of the cache hierarchy (e.g., L3 cache).

Identical subproblem sizes are also possible when implementing the
base case of merge sort. For large data sets, this would likely imply
using merge sort for intermediate input sizes that fit into the cache
and one would want to have a highly tuned implementation that avoids
branch mispredictions, e.g., based on \cite{elmasry2012branch}. One
could even consider merging circuits as yet another intermediate
stage.

One can also look further into implementation techniques for sorting
networks. One would like to explore further possibilities to implement
the conditional swap for the sorting networks, as well as seeing which
of the C++ compilers generate conditional moves when using portable
C++ code instead of compiler- and architecture-dependent inline-assembly. That also
includes looking at conditional swaps for elements that differ from
the 64-bit key and reference value pair that we looked at in this
paper.  We also have not investigated yet how the instruction
scheduling policies of the compilers interact with our
implementations.  For some data types, it would certainly be
interesting to consider SIMD instructions.

Going beyond sorting networks, one could look for small case sorters
not based on compare-and-swap primitives. These might involve fewer
instructions or data-dependencies.

\bibliography{references}


\begin{table}[p]
  \caption{Average number of CPU cycles per iteration of OneArrayRepeat experiment on machine Intel-2650.}\label{table:normalsort:avg:A}
  \scriptsize\centering
  \def\arraystretch{1.1}
  \setlength\tabcolsep{12pt}
  \input{normalSortAvgTable129.tex}
\end{table}

\begin{table}[p]
  \caption{Average number of CPU cycles per iteration of OneArrayRepeat experiment on machine Intel-2670.}\label{table:normalsort:avg:B}
  \scriptsize\centering
  \def\arraystretch{1.1}
  \setlength\tabcolsep{12pt}
  \input{normalSortAvgTable130.tex}
\end{table}

\begin{table}[p]
  \caption{Average number of CPU cycles per iteration of OneArrayRepeat experiment on machine Ryzen-1800X.}\label{table:normalsort:avg:C}
  \scriptsize\centering
  \def\arraystretch{1.1}
  \setlength\tabcolsep{12pt}
  \input{normalSortAvgTable133.tex}
\end{table}

\begin{table}[p]
  \caption{Average number of nanoseconds per iteration of OneArrayRepeat experiment on machine RK3399.}\label{table:normalsort:avg:D}
  \scriptsize\centering
  \def\arraystretch{1.1}
  \setlength\tabcolsep{12pt}
  \input{normalSortAvgTable144.tex}
\end{table}

\end{document}

%% file: speedupTableNormalAllGeoms.tex
\begin{tabular}{ l | r|l|r|l|r|l|r|l|r|l |}
Sorter & \multicolumn{2}{c|}{Overall}&\multicolumn{2}{c|}{Intel-2650}&\multicolumn{2}{c|}{Intel-2670}&\multicolumn{2}{c|}{Ryzen-1800X}&\multicolumn{2}{c|}{RK3399} \\
 & Rank&GeoM&Rank&GeoM&Rank&GeoM&Rank&GeoM&Rank&GeoM\\ \hline
\verb+SN BN-L 4Cm +&1&1.09&4&1.12&1&1.08&1&1.08&3&1.07\\
\verb+SN Best 4Cm +&2&1.12&6&1.14&5&1.10&5&1.19&2&1.04\\
\verb+SN BN-P 4Cm +&3&1.13&8&1.17&8&1.14&4&1.18&1&1.03\\
\verb+SN BN-R 4Cm +&4&1.15&3&1.11&4&1.09&3&1.15&4&1.25\\
\verb+SN BN-L 4CmS+&5&1.18&2&1.09&3&1.09&2&1.14&5&1.45\\
\verb+SN Best 4CmS+&6&1.23&1&1.09&2&1.09&6&1.23&15&1.57\\
\verb+SN BN-P 4CmS+&7&1.27&5&1.13&7&1.14&7&1.24&18&1.64\\
\verb+SN BN-R 4CmS+&8&1.33&7&1.15&6&1.12&8&1.29&21&1.89\\
\verb+SN Best 2CPp+&9&1.53&12&1.35&13&1.43&12&1.86&6&1.50\\
\verb+SN BN-P 2CPp+&10&1.55&13&1.38&14&1.45&13&1.87&7&1.54\\
\verb+SN Best 2CPm+&11&1.55&9&1.19&9&1.18&10&1.62&23&2.54\\
\verb+SN BN-P 2CPm+&12&1.58&10&1.21&10&1.22&9&1.61&27&2.62\\
\verb+SN BN-L 2CPm+&13&1.68&11&1.32&11&1.31&11&1.82&24&2.56\\
\verb+SN BN-L 2CPp+&14&1.68&15&1.54&15&1.61&15&2.10&11&1.55\\
\verb+SN BN-R 2CPp+&15&1.81&16&1.67&16&1.73&16&2.38&16&1.57\\
\verb+SN BN-R 2CPm+&16&1.82&14&1.43&12&1.41&14&2.06&28&2.66\\[\smallskipamount]
\verb+SN BN-P ISwp+&17&2.32&17&2.03&18&2.13&26&4.32&8&1.54\\
\verb+SN Best ISwp+&18&2.33&19&2.05&17&2.10&30&4.41&12&1.55\\
\verb+SN BN-L ISwp+&19&2.37&23&2.15&21&2.19&27&4.33&10&1.54\\
\verb+SN BN-R ISwp+&20&2.40&24&2.17&25&2.25&29&4.37&13&1.56\\
\verb+SN BN-L JXhg+&21&2.41&28&2.29&27&2.38&17&3.92&14&1.57\\
\verb+SN Best JXhg+&22&2.45&27&2.29&28&2.41&22&4.24&9&1.54\\
\verb+IS      Def +&23&2.45&22&2.13&22&2.21&19&4.14&20&1.84\\
\verb+SN BN-P JXhg+&24&2.49&29&2.32&29&2.45&20&4.23&17&1.60\\
\verb+SN BN-R JXhg+&25&2.53&30&2.38&30&2.50&18&4.10&19&1.68\\
\verb+IS      POp +&26&2.62&26&2.24&26&2.32&21&4.23&22&2.13\\
\verb+SN Best Tie +&27&2.83&18&2.05&19&2.16&28&4.34&30&3.32\\
\verb+SN BN-R Tie +&28&2.86&25&2.20&23&2.22&25&4.31&29&3.17\\
\verb+SN BN-L Tie +&29&2.86&21&2.09&24&2.22&23&4.24&31&3.40\\
\verb+SN BN-P Tie +&30&2.93&20&2.08&20&2.17&24&4.25&32&3.85\\
\verb+IS      STL +&31&3.02&31&2.46&31&2.59&31&5.09&26&2.59\\
\verb+IS      AIF +&32&3.08&32&2.48&32&2.69&32&5.27&25&2.57\\
\end{tabular}

%% file: speedupTableNormalAll.tex
\begin{tabular}{l | r|r|r|r|r|r|r|r|r|r|r|r|r|r|r|@{ }|r|}
Machine & 2&3&4&5&6&7&8&9&10&11&12&13&14&15&16 & Avg\\ \hline
Intel-2650 & 2.37&1.91&2.16&2.79&2.54&2.62&2.44&2.16&2.20&1.95&2.08&1.84&1.76&1.77&1.78& 2.16\\
Intel-2760 & --&3.15&3.16&3.02&3.23&2.84&2.77&2.47&2.27&2.04&2.06&1.97&1.86&1.79&1.80& 2.46\\
Ryzen-1800X & 4.51&5.28&5.29&5.17&5.61&7.11&4.86&4.26&4.03&3.72&3.60&3.35&3.70&3.25&3.25& 4.47\\
RK3399 & 1.08&1.16&1.39&1.50&1.85&1.92&1.92&2.00&2.05&1.91&1.94&2.05&2.00&2.03&2.05& 1.79\\
\hline
Average & 2.65&2.88&3.00&3.12&3.31&3.62&3.00&2.72&2.64&2.40&2.42&2.30&2.33&2.21&2.22&2.72
\end{tabular}

%% file: inrowSortAvgTableAll.tex
\hfill
\begin{tabular}[t]{ l | l }
Sorter&GeoM\\ \hline
\verb+SN BN-L 4CmS+&1.030\\
\verb+SN Best 4CmS+&1.041\\
\verb+SN BN-L 4Cm +&1.041\\
\verb+SN Best 4Cm +&1.053\\
\verb+SN BN-R 4Cm +&1.076\\
\verb+SN BN-P 4CmS+&1.080\\
\verb+SN BN-P 4Cm +&1.082\\
\verb+SN BN-R 4CmS+&1.145\\
\end{tabular}
\hfill
\begin{tabular}[t]{ l | l }
Sorter&GeoM\\ \hline
\verb+SN Best 2CPm+&1.277\\
\verb+SN BN-P 2CPm+&1.308\\
\verb+SN Best 2CPp+&1.443\\
\verb+SN BN-L 2CPm+&1.451\\
\verb+SN BN-P 2CPp+&1.483\\
\verb+SN BN-R 2CPm+&1.537\\
\verb+SN BN-L 2CPp+&1.685\\
\verb+SN BN-R 2CPp+&1.773\\
\end{tabular}
\hfill
\begin{tabular}[t]{ l | l }
Sorter&GeoM\\ \hline
\verb+SN Best Tie +&2.239\\
\verb+SN Best ISwp+&2.244\\
\verb+SN BN-P ISwp+&2.244\\
\verb+SN BN-L Tie +&2.244\\
\verb+SN BN-P Tie +&2.248\\
\verb+IS      Def+&2.300\\
\verb+SN BN-R Tie +&2.302\\
\verb+SN BN-L ISwp+&2.329\\
\end{tabular}
\hfill
\begin{tabular}[t]{ l | l }
Sorter&GeoM\\ \hline
\verb+IS      POp+&2.330\\
\verb+SN BN-R ISwp+&2.338\\
\verb+SN BN-L JXhg+&2.415\\
\verb+SN Best JXhg+&2.449\\
\verb+SN BN-P JXhg+&2.457\\
\verb+SN BN-R JXhg+&2.460\\
\verb+IS      STL+&2.487\\
\verb+IS      AIF+&2.514\\
\end{tabular}
\hfill

%% file: speedupTableInrowAll.tex
\begin{tabular}{l | r|r|r|r|r|r|r|r|r|r|r|r|r|r|r|@{ }|r|}
Machine & 2&3&4&5&6&7&8&9&10&11&12&13&14&15&16 & Avg\\ \hline
Intel-2650 & 1.50&1.88&1.82&2.01&2.29&2.47&2.52&2.36&2.35&2.25&2.23&2.08&2.06&1.96&1.90& 2.11\\
Intel-2760 & 1.69&2.14&2.06&2.14&2.46&2.54&2.53&2.41&2.36&2.23&2.23&2.07&2.09&1.91&2.14& 2.20\\
Ryzen-1800X & 1.44&2.11&2.52&2.84&3.09&3.37&3.53&3.48&3.50&3.36&3.44&3.19&3.20&3.18&3.21& 3.03\\
RK3399 & 1.03&1.07&1.23&1.46&1.61&1.70&1.77&1.78&1.95&1.94&2.04&1.99&1.97&2.03&2.12& 1.71\\
\hline
Average & 1.42&1.80&1.91&2.11&2.36&2.52&2.59&2.51&2.54&2.45&2.49&2.34&2.33&2.27&2.34&2.26
\end{tabular}

%% file: normalSortAvgTable129.tex
\begin{tabular}{l | r @{~~} r | r@{~~}r@{~~}r@{~~}r@{~~}r@{~~}r@{~~}r@{~~}r@{~~}r@{~~}r@{~~}r@{~~}r@{~~}r@{~~}r@{~~}r@{~~}r|}
 & \multicolumn{2}{c|}{Overall} & \multicolumn{15}{c}{Array Size} \\
 & Rank & GeoM & 2&3&4&5&6&7&8&9&10&11&12&13&14&15&16\\ \hline
\verb+IS      Def+ & 25 & 2.17 & 15.4&41.9&78.8&119&170&207&258&293&354&394&452&487&561&582&650\\
\verb+IS      POp+ & 28 & 2.29 & 17.1&45.4&83.1&128&180&219&268&302&367&409&476&503&582&607&681\\
\verb+IS      STL+ & 31 & 2.41 & 24.9&53.0&90.1&133&184&222&275&310&379&411&489&511&587&608&678\\
\verb+IS      AIF+ & 32 & 2.63 & 23.8&51.0&90.9&150&215&264&319&357&416&461&534&564&629&650&723\smallskip \\
\verb+SN BN-L 4CmS+ & 1 & 1.09 & 12.7&21.5&37.6&53.7&70.9&83.5&\textbf{105}&135&164&199&238&267&315&322&\textbf{365}\\
\verb+SN BN-L 4Cm + & 3 & 1.11 & 9.46&24.3&\textbf{35.3}&55.3&72.2&73.4&107&136&174&213&249&289&346&351&405\\
\verb+SN BN-L 2CPm+ & 11 & 1.31 & 8.95&24.6&38.5&62.4&87.6&127&146&187&227&252&275&335&386&396&433\\
\verb+SN BN-L 2CPp+ & 15 & 1.53 & 6.78&35.3&40.5&78.4&113&141&166&213&262&322&350&396&449&483&504\\
\verb+SN BN-L Tie + & 21 & 2.09 & 19.7&39.0&66.4&97.0&137&171&213&267&328&372&446&525&617&657&764\\
\verb+SN BN-L ISwp+ & 22 & 2.10 & 19.8&42.1&74.8&96.3&136&164&213&262&348&366&488&499&635&623&708\\
\verb+SN BN-L JXhg+ & 26 & 2.25 & 18.6&40.3&69.7&108&144&186&236&307&380&408&492&583&678&714&777\smallskip \\
\verb+SN BN-P 4CmS+ & 6 & 1.13 & 12.8&\textbf{21.5}&37.5&49.0&67.0&83.1&110&137&185&215&246&288&333&358&395\\
\verb+SN BN-P 4Cm + & 7 & 1.15 & 9.00&24.4&35.4&\textbf{44.0}&70.7&84.0&112&133&195&226&270&319&360&396&452\\
\verb+SN BN-P 2CPm+ & 10 & 1.20 & 6.83&25.2&39.1&65.3&85.2&110&136&175&207&222&249&285&338&353&397\\
\verb+SN BN-P 2CPp+ & 13 & 1.37 & \textbf{6.70}&34.9&40.3&84.3&101&128&147&210&218&250&282&326&376&395&447\\
\verb+SN BN-P Tie + & 18 & 2.03 & 18.2&38.9&66.2&96.1&134&162&210&251&320&369&440&478&587&662&763\\
\verb+SN BN-P ISwp+ & 19 & 2.04 & 21.0&42.4&74.1&98.6&136&168&210&250&321&346&425&480&589&573&676\\
\verb+SN BN-P JXhg+ & 29 & 2.31 & 18.4&39.0&72.1&119&153&192&244&299&379&451&506&586&682&726&825\smallskip \\
\verb+SN BN-R 4Cm + & 4 & 1.11 & 9.00&24.0&36.1&54.1&71.8&\textbf{72.9}&110&153&199&209&\textbf{228}&289&324&351&408\\
\verb+SN BN-R 4CmS+ & 8 & 1.17 & 13.0&23.0&36.4&54.1&\textbf{66.5}&79.8&124&164&219&227&267&280&315&344&408\\
\verb+SN BN-R 2CPm+ & 14 & 1.45 & 11.4&28.1&46.6&74.0&113&126&150&182&237&268&303&349&406&423&491\\
\verb+SN BN-R 2CPp+ & 16 & 1.69 & 13.8&31.2&52.0&91.0&126&135&161&260&290&304&355&384&477&501&575\\
\verb+SN BN-R Tie + & 23 & 2.12 & 19.2&38.9&68.6&101&134&169&215&291&346&397&463&531&595&657&751\\
\verb+SN BN-R ISwp+ & 24 & 2.15 & 19.9&39.4&75.6&100&140&168&231&299&349&409&453&537&596&641&707\\
\verb+SN BN-R JXhg+ & 30 & 2.34 & 18.9&40.5&72.6&110&152&190&262&311&411&446&530&570&682&708&811\smallskip \\
\verb+SN Best 4CmS+ & 2 & 1.10 & 13.4&21.7&37.4&52.8&70.5&83.5&105&136&174&\textbf{196}&234&258&324&331&371\\
\verb+SN Best 4Cm + & 5 & 1.11 & 9.27&24.2&35.4&53.8&72.5&79.7&107&\textbf{123}&\textbf{163}&211&256&291&363&377&418\\
\verb+SN Best 2CPm+ & 9 & 1.18 & 9.33&24.3&38.7&62.5&88.0&128&146&149&178&210&234&\textbf{250}&\textbf{311}&\textbf{319}&372\\
\verb+SN Best 2CPp+ & 12 & 1.35 & 7.10&35.2&40.9&78.9&113&142&166&195&191&251&274&291&350&363&412\\
\verb+SN Best ISwp+ & 17 & 2.02 & 19.4&43.6&74.0&102&149&171&221&262&299&346&416&442&554&558&667\\
\verb+SN Best Tie + & 20 & 2.04 & 19.7&39.7&66.4&99.8&139&183&217&262&311&348&426&487&568&613&724\\
\verb+SN Best JXhg+ & 27 & 2.28 & 19.0&39.7&72.2&115&158&191&248&290&351&405&512&573&663&703&806\\
\end{tabular}

%% file: normalSortAvgTable130.tex
\begin{tabular}{l | r @{~~} r | r@{~~}r@{~~}r@{~~}r@{~~}r@{~~}r@{~~}r@{~~}r@{~~}r@{~~}r@{~~}r@{~~}r@{~~}r@{~~}r@{~~}r@{~~}r|}
 & \multicolumn{2}{c|}{Overall} & \multicolumn{15}{c}{Array Size} \\
 & Rank & GeoM & 2&3&4&5&6&7&8&9&10&11&12&13&14&15&16\\ \hline
\verb+IS      Def+ & 22 & 2.21 & 8.67&33.9&66.0&103&135&169&205&246&281&321&377&421&466&500&526\\
\verb+IS      POp+ & 26 & 2.32 & 10.4&37.3&69.4&110&143&181&218&255&294&337&383&423&478&523&566\\
\verb+IS      STL+ & 31 & 2.59 & 18.0&52.7&78.2&117&159&198&239&276&322&387&446&486&522&572&611\\
\verb+IS      AIF+ & 32 & 2.69 & 17.1&51.0&78.5&131&179&222&260&298&334&396&446&493&524&562&612\smallskip \\
\verb+SN BN-L 4Cm + & 1 & 1.08 & 2.70&13.1&22.6&39.0&45.6&\textbf{61.5}&\textbf{76.9}&105&126&178&204&262&300&312&348\\
\verb+SN BN-L 4CmS+ & 3 & 1.09 & 3.39&21.2&\textbf{22.1}&41.7&49.1&64.5&77.8&111&124&172&190&240&256&282&299\\
\verb+SN BN-L 2CPm+ & 11 & 1.31 & \textbf{-0.93}&21.3&26.5&49.8&60.0&90.0&103&153&168&201&223&278&303&330&345\\
\verb+SN BN-L 2CPp+ & 15 & 1.61 & -0.05&30.6&31.9&67.0&79.0&110&125&189&214&246&273&347&360&392&415\\
\verb+SN BN-L ISwp+ & 21 & 2.19 & 11.0&37.5&62.8&82.4&105&144&171&226&285&311&410&441&551&593&677\\
\verb+SN BN-L Tie + & 24 & 2.22 & 12.0&34.3&55.7&79.5&119&142&181&243&281&318&431&479&551&626&691\\
\verb+SN BN-L JXhg+ & 27 & 2.38 & 11.0&34.0&61.7&99.0&128&172&199&264&313&337&421&504&579&657&693\smallskip \\
\verb+SN BN-P 4CmS+ & 7 & 1.14 & 3.62&20.9&22.2&38.8&45.7&64.6&79.9&112&148&182&200&256&285&330&339\\
\verb+SN BN-P 4Cm + & 8 & 1.14 & 2.59&12.8&22.4&\textbf{37.6}&\textbf{43.3}&61.8&79.4&110&161&195&221&282&318&356&394\\
\verb+SN BN-P 2CPm+ & 10 & 1.22 & -0.12&26.5&25.3&47.0&51.0&78.0&89.4&137&147&182&195&246&285&309&335\\
\verb+SN BN-P 2CPp+ & 14 & 1.45 & 0.07&30.6&32.4&61.6&67.1&96.0&107&164&178&213&232&293&311&358&390\\
\verb+SN BN-P ISwp+ & 18 & 2.13 & 10.4&38.6&61.3&85.2&104&140&173&217&258&304&382&428&510&558&631\\
\verb+SN BN-P Tie + & 20 & 2.17 & 12.0&33.7&55.0&81.5&105&139&192&229&266&317&407&472&541&604&664\\
\verb+SN BN-P JXhg+ & 29 & 2.45 & 10.5&39.1&62.4&96.5&119&159&203&259&327&396&423&556&590&687&755\smallskip \\
\verb+SN BN-R 4Cm + & 4 & 1.09 & 2.60&\textbf{12.0}&22.7&38.8&45.4&63.0&77.8&139&158&172&\textbf{176}&255&305&306&314\\
\verb+SN BN-R 4CmS+ & 6 & 1.12 & 4.00&13.2&22.3&41.4&49.7&63.8&94.0&130&170&188&228&242&\textbf{234}&285&312\\
\verb+SN BN-R 2CPm+ & 12 & 1.41 & 4.31&28.4&30.0&53.5&64.9&93.7&109&198&174&204&226&288&311&344&367\\
\verb+SN BN-R 2CPp+ & 16 & 1.73 & 4.60&33.7&37.4&72.4&83.3&114&132&237&222&260&279&346&374&425&463\\
\verb+SN BN-R Tie + & 23 & 2.22 & 12.1&33.0&53.6&88.0&109&143&179&248&287&343&404&466&542&625&707\\
\verb+SN BN-R ISwp+ & 25 & 2.25 & 10.9&36.3&61.1&85.7&117&142&194&244&282&338&403&490&541&595&676\\
\verb+SN BN-R JXhg+ & 30 & 2.50 & 10.3&36.1&60.8&103&135&167&212&288&341&381&475&527&582&670&731\smallskip \\
\verb+SN Best 4CmS+ & 2 & 1.09 & 3.66&20.7&22.6&42.3&49.6&64.8&78.2&109&\textbf{121}&\textbf{165}&181&228&274&286&308\\
\verb+SN Best 4Cm + & 5 & 1.10 & 2.95&13.7&22.1&38.9&46.1&61.6&77.8&\textbf{105}&125&181&212&267&309&339&359\\
\verb+SN Best 2CPm+ & 9 & 1.18 & -0.72&21.7&26.4&49.4&60.2&90.0&103&130&132&169&189&\textbf{215}&253&\textbf{278}&\textbf{294}\\
\verb+SN Best 2CPp+ & 13 & 1.43 & -0.12&30.9&32.6&67.4&78.3&110&125&157&157&193&218&255&298&324&358\\
\verb+SN Best ISwp+ & 17 & 2.10 & 10.8&35.7&63.3&90.3&114&152&181&223&240&300&345&389&478&533&609\\
\verb+SN Best Tie + & 19 & 2.16 & 12.0&35.1&56.4&82.7&110&150&193&234&253&306&388&449&504&599&632\\
\verb+SN Best JXhg+ & 28 & 2.41 & 10.9&34.9&63.3&104&133&182&204&250&285&358&421&495&588&661&726\\
\end{tabular}

%% file: normalSortAvgTable133.tex
\begin{tabular}{l | r @{~~} r | r@{~~}r@{~~}r@{~~}r@{~~}r@{~~}r@{~~}r@{~~}r@{~~}r@{~~}r@{~~}r@{~~}r@{~~}r@{~~}r@{~~}r@{~~}r|}
 & \multicolumn{2}{c|}{Overall} & \multicolumn{15}{c}{Array Size} \\
 & Rank & GeoM & 2&3&4&5&6&7&8&9&10&11&12&13&14&15&16\\ \hline
\verb+IS      Def+ & 19 & 4.14 & 15.0&46.9&80.3&125&167&220&258&290&327&369&397&446&464&510&559\\
\verb+IS      POp+ & 21 & 4.23 & 10.9&45.8&84.9&134&183&228&266&298&339&382&414&462&495&555&591\\
\verb+IS      STL+ & 31 & 5.09 & 21.1&63.3&103&150&200&255&298&340&392&433&485&536&577&629&695\\
\verb+IS      AIF+ & 32 & 5.27 & 22.3&59.5&103&149&201&264&314&364&415&459&516&557&599&662&747\smallskip \\
\verb+SN BN-L 4Cm + & 1 & 1.08 & 6.13&10.3&14.8&27.4&35.7&\textbf{44.8}&56.8&\textbf{70.0}&\textbf{82.4}&\textbf{99.5}&113&\textbf{145}&156&163&\textbf{181}\\
\verb+SN BN-L 4CmS+ & 2 & 1.14 & 6.72&13.6&15.1&28.4&35.3&53.9&60.0&73.8&84.0&109&\textbf{113}&145&154&\textbf{154}&186\\
\verb+SN BN-L 2CPm+ & 11 & 1.82 & 2.86&17.6&26.3&52.7&54.6&88.3&86.3&133&167&200&222&272&285&317&335\\
\verb+SN BN-L 2CPp+ & 15 & 2.10 & 3.15&27.2&25.6&70.1&72.0&90.8&94.5&162&201&236&251&295&313&341&351\\
\verb+SN BN-L JXhg+ & 17 & 3.92 & 15.1&37.0&65.9&91.8&125&161&205&275&325&372&425&497&561&619&691\\
\verb+SN BN-L Tie + & 23 & 4.24 & 15.1&40.0&76.3&103&133&189&223&292&337&405&451&527&590&671&791\\
\verb+SN BN-L ISwp+ & 27 & 4.33 & 14.2&41.0&76.8&104&130&176&219&304&370&404&494&541&661&724&782\smallskip \\
\verb+SN BN-P 4Cm + & 4 & 1.18 & 6.69&10.9&\textbf{14.4}&25.7&\textbf{33.4}&46.6&\textbf{55.2}&75.0&99.4&106&127&173&178&214&231\\
\verb+SN BN-P 4CmS+ & 7 & 1.24 & 6.65&13.1&15.2&\textbf{24.3}&39.0&51.8&59.1&80.2&93.7&130&125&154&183&240&231\\
\verb+SN BN-P 2CPm+ & 9 & 1.61 & \textbf{2.55}&17.0&20.3&49.9&48.8&79.0&92.4&127&140&171&186&227&237&270&284\\
\verb+SN BN-P 2CPp+ & 13 & 1.87 & 3.12&26.8&25.2&68.2&61.6&84.1&90.4&147&156&185&206&242&266&295&311\\
\verb+SN BN-P JXhg+ & 20 & 4.23 & 14.4&42.6&68.9&99.1&128&168&227&277&350&420&466&549&588&733&802\\
\verb+SN BN-P Tie + & 24 & 4.25 & 16.2&34.1&63.2&97.7&140&199&226&295&347&402&461&535&631&725&790\\
\verb+SN BN-P ISwp+ & 26 & 4.32 & 13.9&40.3&73.9&106&143&191&237&309&338&418&467&538&622&715&752\smallskip \\
\verb+SN BN-R 4Cm + & 3 & 1.15 & 5.87&11.2&14.8&25.6&35.0&48.5&56.2&93.4&93.7&113&124&148&\textbf{153}&187&202\\
\verb+SN BN-R 4CmS+ & 8 & 1.29 & 6.06&10.9&15.2&28.8&35.8&59.8&71.8&89.8&132&157&161&147&156&197&216\\
\verb+SN BN-R 2CPm+ & 14 & 2.06 & 8.12&19.1&26.8&55.5&71.8&87.0&93.9&140&179&209&225&287&298&325&347\\
\verb+SN BN-R 2CPp+ & 16 & 2.38 & 8.72&31.5&28.3&75.2&75.2&92.6&107&180&209&244&256&299&321&354&390\\
\verb+SN BN-R JXhg+ & 18 & 4.10 & 19.4&42.4&57.6&96.5&132&177&221&289&334&382&442&504&557&635&667\\
\verb+SN BN-R Tie + & 25 & 4.31 & 15.6&37.1&61.9&111&146&193&228&322&372&426&464&553&607&684&709\\
\verb+SN BN-R ISwp+ & 29 & 4.37 & 14.3&41.3&80.4&99.8&141&189&252&328&359&415&468&529&576&702&790\smallskip \\
\verb+SN Best 4Cm + & 5 & 1.19 & 8.05&\textbf{10.2}&14.8&28.8&36.1&45.2&59.3&74.3&85.9&104&126&164&188&206&233\\
\verb+SN Best 4CmS+ & 6 & 1.23 & 7.30&13.6&16.1&29.7&36.4&55.2&63.9&74.7&87.6&103&117&162&185&211&224\\
\verb+SN Best 2CPm+ & 10 & 1.62 & 3.27&17.5&20.4&53.3&55.2&80.8&86.5&130&129&162&202&200&226&264&276\\
\verb+SN Best 2CPp+ & 12 & 1.86 & 3.84&27.0&25.9&70.5&72.1&91.1&94.7&142&139&170&196&214&249&274&292\\
\verb+SN Best JXhg+ & 22 & 4.24 & 16.0&39.8&70.8&102&137&180&223&278&333&397&437&543&634&714&774\\
\verb+SN Best Tie + & 28 & 4.34 & 16.6&39.2&72.5&106&138&205&242&294&337&386&455&529&598&715&814\\
\verb+SN Best ISwp+ & 30 & 4.41 & 17.2&37.5&76.1&120&149&194&247&309&339&410&457&517&608&681&777\\
\end{tabular}

%% file: normalSortAvgTable144.tex
\begin{tabular}{l | r @{~~} r | r@{~~}r@{~~}r@{~~}r@{~~}r@{~~}r@{~~}r@{~~}r@{~~}r@{~~}r@{~~}r@{~~}r@{~~}r@{~~}r@{~~}r@{~~}r|}
 & \multicolumn{2}{c|}{Overall} & \multicolumn{15}{c}{Array Size} \\
 & Rank & GeoM & 2&3&4&5&6&7&8&9&10&11&12&13&14&15&16\\ \hline
\verb+IS      Def+ & 20 & 1.84 & 17.6&38.4&56.5&86.9&114&138&172&205&249&284&326&369&409&465&513\\
\verb+IS      POp+ & 22 & 2.13 & 26.7&39.5&61.9&94.8&127&162&195&235&284&331&373&433&485&539&595\\
\verb+IS      AIF+ & 25 & 2.57 & 42.6&69.3&88.3&122&154&184&221&271&315&363&408&467&522&583&647\\
\verb+IS      STL+ & 26 & 2.59 & 37.4&62.2&94.1&123&155&194&232&268&324&369&422&472&531&592&656\smallskip \\
\verb+SN BN-L 4Cm + & 3 & 1.07 & 22.2&31.5&\textbf{38.8}&51.7&63.0&\textbf{68.2}&87.4&106&129&150&167&189&202&239&256\\
\verb+SN BN-L 4CmS+ & 5 & 1.45 & 26.1&36.2&45.5&66.3&83.0&103&110&146&180&205&233&270&321&377&384\\
\verb+SN BN-L ISwp+ & 10 & 1.54 & 23.4&34.4&48.0&66.7&87.0&103&121&176&205&246&266&308&341&372&417\\
\verb+SN BN-L 2CPp+ & 11 & 1.55 & 21.1&35.9&50.8&68.3&88.1&103&125&172&205&241&266&304&339&374&419\\
\verb+SN BN-L JXhg+ & 14 & 1.57 & 27.4&33.7&53.6&75.2&84.2&105&130&169&204&241&260&303&329&372&405\\
\verb+SN BN-L 2CPm+ & 24 & 2.56 & 24.7&52.0&66.9&113&141&188&222&307&359&428&461&561&613&675&728\\
\verb+SN BN-L Tie + & 31 & 3.40 & 31.1&56.5&82.4&129&174&219&256&427&511&651&717&829&904&1028&1040\smallskip \\
\verb+SN BN-P 4Cm + & 1 & 1.03 & \textbf{17.1}&\textbf{28.6}&40.4&51.7&\textbf{59.6}&72.8&\textbf{79.6}&109&126&151&\textbf{163}&190&204&233&249\\
\verb+SN BN-P 2CPp+ & 7 & 1.54 & 22.9&33.3&46.1&70.1&86.1&110&123&171&212&233&266&300&346&372&399\\
\verb+SN BN-P ISwp+ & 8 & 1.54 & 22.0&32.0&47.3&70.8&86.0&111&130&169&206&240&266&299&350&370&401\\
\verb+SN BN-P JXhg+ & 17 & 1.60 & 30.4&36.9&51.2&73.8&93.8&114&130&174&195&232&283&292&329&369&406\\
\verb+SN BN-P 4CmS+ & 18 & 1.64 & 30.2&37.5&49.2&63.9&79.5&105&118&156&198&248&289&347&409&463&497\\
\verb+SN BN-P 2CPm+ & 27 & 2.62 & 26.2&50.5&72.5&108&143&188&216&309&369&448&495&576&637&716&758\\
\verb+SN BN-P Tie + & 32 & 3.85 & 46.4&67.5&95.4&171&221&283&323&465&552&654&717&846&935&1056&939\smallskip \\
\verb+SN BN-R 4Cm + & 4 & 1.25 & 27.6&29.7&41.8&52.6&61.7&70.6&86.7&129&185&228&271&244&258&256&274\\
\verb+SN BN-R ISwp+ & 13 & 1.56 & 21.5&31.6&48.6&68.7&87.2&107&133&180&218&248&275&316&344&386&416\\
\verb+SN BN-R 2CPp+ & 16 & 1.57 & 25.3&34.3&46.1&66.8&86.1&103&132&184&218&254&272&304&342&384&413\\
\verb+SN BN-R JXhg+ & 19 & 1.68 & 27.5&42.9&52.6&69.4&86.6&109&152&189&229&269&281&315&347&392&440\\
\verb+SN BN-R 4CmS+ & 21 & 1.89 & 31.5&38.6&48.7&61.1&84.0&125&169&198&243&290&346&426&484&574&634\\
\verb+SN BN-R 2CPm+ & 28 & 2.66 & 29.5&49.9&72.7&108&142&191&237&304&364&427&479&576&647&730&800\\
\verb+SN BN-R Tie + & 29 & 3.17 & 36.5&69.0&85.3&134&174&221&261&372&444&533&572&706&747&798&847\smallskip \\
\verb+SN Best 4Cm + & 2 & 1.04 & 25.6&29.5&41.3&\textbf{50.8}&61.7&69.4&84.7&\textbf{106}&\textbf{120}&\textbf{149}&164&\textbf{165}&\textbf{198}&\textbf{217}&\textbf{236}\\
\verb+SN Best 2CPp+ & 6 & 1.50 & 22.8&39.6&46.7&72.5&87.0&107&125&165&187&242&253&276&303&341&379\\
\verb+SN Best JXhg+ & 9 & 1.54 & 29.1&37.2&50.7&73.7&86.3&105&130&162&201&221&245&283&315&355&377\\
\verb+SN Best ISwp+ & 12 & 1.55 & 31.7&37.4&49.0&74.8&91.2&107&137&166&178&239&253&280&305&346&381\\
\verb+SN Best 4CmS+ & 15 & 1.57 & 26.4&41.8&49.8&61.9&86.3&107&110&154&183&229&270&304&379&423&476\\
\verb+SN Best 2CPm+ & 23 & 2.54 & 30.4&52.9&70.3&110&142&193&219&283&341&408&453&527&597&659&701\\
\verb+SN Best Tie + & 30 & 3.32 & 34.0&60.3&89.6&130&175&224&260&423&462&591&670&705&830&956&1002\\
\end{tabular}

%% file: paper.bbl
\begin{thebibliography}{10}
\providecommand \doibase [0]{http://dx.doi.org/}%

\bibitem{karypis1998fast}
Karypis G, Kumar V. A Fast and High Quality Multilevel Scheme for Partitioning
  Irregular Graphs. {\it {SIAM} Journal on Scientific Computing} 1998\string;
  20(1)\string: 359--392.

\bibitem{bose1962sorting}
Bose RC, Nelson RJ. A Sorting Problem. {\it Journal of the ACM (JACM)}
  1962\string; 9(2)\string: 282--296.

\bibitem{codish2017optimizing}
Codish M, Cruz-Filipe L, Nebel M, Schneider-Kamp P. Optimizing sorting
  algorithms by using sorting networks. {\it Formal Aspects of Computing}
  2017\string; 29(3)\string: 559--579.

\bibitem{sanders2004super}
Sanders P, Winkel S. Super Scalar Sample Sort. In:  {\it 12th European
  Symposium on Algorithms (ESA)}. . 3221 of {\it LNCS}. Springer; 2004\string:
  784--796.

\bibitem{axtmann2017inplace}
Axtmann M, Witt S, Ferizovic D, Sanders P. In-Place Parallel Super Scalar
  Samplesort (IPS$^4$o). In:  {\it 25th European Symposium on Algorithms
  (ESA)}. . 87 of {\it LIPIcs}. Schloss Dagstuhl. ; 2017\string: 9:1--9:14.
\newblock preprint arXiv:1705.02257.

\bibitem{marianczuk2019engineering}
Marianczuk J. Engineering Faster Sorters for Small Sets of Items. Bachelor
  Thesis. Karlsruhe Institute of Technology, Germany. arXiv:1908.08111;  2019.

\bibitem{knuth1998sorting}
Knuth DE. {\it The Art of Computer Programming, Volume 3: Sorting And
  Searching}.
\newblock Addison Wesley Longman Publishing Co., Inc.
\newblock 2~ed. 1998.

\bibitem{mahmoud2011sorting}
Mahmoud HM. {\it Sorting: A Distribution Theory}.
\newblock John Wiley \& Sons .
\newblock 2000.

\bibitem{batcher1968sorting}
Batcher KE. Sorting Networks and Their Applications. In:  {\it American
  Federation of Information Processing Societies (AFIPS)}. . 32 of {\it {AFIPS}
  Conference Proceedings}. Thomson Book Company, Washington {D.C.};
  1968\string: 307--314.

\bibitem{ajtai1983nlogn}
Ajtai M, Koml{\'{o}}s J, Szemer{\'{e}}di E. An $O(n \log n)$ Sorting Network.
  In:  {\it 15th {ACM} Symposium on Theory of Computing (STOC)}. ACM;
  1983\string: 1--9.

\bibitem{inoue2012high}
Inoue H, Moriyama T, Komatsu H, Nakatani T. A high-performance sorting
  algorithm for multicore single-instruction multiple-data processors. {\it
  Software \& Practice and Experience} 2012\string; 42(6)\string: 753--777.

\bibitem{furtak2007using}
Furtak T, Amaral JN, Niewiadomski R. Using {SIMD} registers and instructions to
  enable instruction-level parallelism in sorting algorithms. In:  {\it 10th
  Annual ACM Symposium on Parallel Algorithms and Architectures (SPAA)}. {ACM};
  2007\string: 348--357.

\bibitem{xiaochen2013register}
Xiaochen T, Rocki K, Suda R. Register level sort algorithm on multi-core {SIMD}
  processors. In:  {\it 3rd Workshop on Irregular Applications: Architectures
  and Algorithms}. ACM; 2013\string: 1--8.

\bibitem{bramas2017novel}
Bramas B. A Novel Hybrid Quicksort Algorithm Vectorized using {AVX-512} on
  Intel Skylake. {\it International Journal of Advanced Computer Science and
  Applications} 2017\string; 8(10).
\newblock arXiv preprint arXiv:1704.08579.

\bibitem{hou2018framework}
Hou K, Wang H, Feng WC. A framework for the automatic vectorization of parallel
  sort on x86-based processors. {\it {IEEE} Transactions on Parallel and
  Distributed Systems (TPDS)} 2018\string; 29(5)\string: 958--972.

\bibitem{hibbard1963empirical}
Hibbard TN. An empirical study of minimal storage sorting. {\it Communications
  of the ACM} 1963\string; 6(5)\string: 206--213.

\bibitem{yin2019efficient}
Yin Z, Zhang T, M{\"{u}}ller A, et al. Efficient Parallel Sort on AVX-512-Based
  Multi-Core and Many-Core Architectures. In:  {\it 21st {IEEE} International
  Conference on High Performance Computing and Communications; 17th {IEEE}
  International Conference on Smart City; 5th {IEEE} International Conference
  on Data Science and Systems (HPCC/SmartCity/DSS)}. IEEE. ; 2019\string:
  168--176.

\bibitem{mueller2012sorting}
M{\"{u}}ller R, Teubner J, Alonso G. Sorting networks on {FPGAs}. {\it The VLDB
  Journal} 2012\string; 21(1)\string: 1--23.

\bibitem{sklyarov2014high}
Sklyarov V, Skliarova I. High-performance implementation of regular and easily
  scalable sorting networks on an {FPGA}. {\it Microprocessors and
  Microsystems} 2014\string; 38(5)\string: 470--484.

\bibitem{hoare1962quicksort}
Hoare CAR. Quicksort. {\it The Computer Journal} 1962\string; 5(1)\string:
  10--16.

\bibitem{sedgewick1983algorithms}
Sedgewick R. {\it Algorithms}.
\newblock Addison-Wesley .
\newblock 1983.

\bibitem{parberry1991computer}
Parberry I. A Computer-Assisted Optimal Depth Lower Bound for Nine-Input
  Sorting Networks. {\it Mathematical Systems Theory} 1991\string;
  24(1)\string: 101--116.

\bibitem{bundala2014optimal}
Bundala D, Zavodny J. Optimal Sorting Networks. In:  {\it 8th International
  Conference on Language and Automata Theory and Applications (LATA)}. . 8370
  of {\it LNCS}. Springer; 2014\string: 236--247.

\bibitem{codish2014twentyfive}
Codish M, Cruz-Filipe L, Frank M, Schneider-Kamp P. Twenty-Five Comparators Is
  Optimal When Sorting Nine Inputs (and Twenty-Nine for Ten). In:  {\it 26th
  {IEEE} International Conference on Tools with Artificial Intelligence
  (ICTAI)}. IEEE. ; 2014\string: 186--193.

\bibitem{ehlers2015new}
Ehlers T, M{\"{u}}ller M. New Bounds on Optimal Sorting Networks. In:  {\it
  11th Conference on Computability in Europe (CiE)}. . 9136 of {\it LNCS}.
  Springer. ; 2015\string: 167--176.

\bibitem{dobbelaere2018sorterhunter}
Dobbelaere B. SorterHunter. 2017.
\newblock \url{https://github.com/bertdobbelaere/SorterHunter}, Website
  accessed 2019.

\bibitem{GccInlineAssembly}
{Free Software Foundation} . Using the GNU Compiler Collection (GCC): How to
  Use Inline Assembly Language in C Code.
  \url{https://gcc.gnu.org/onlinedocs/gcc/Using-Assembly-Language-with-C.html},
  Website accessed 2019.

\bibitem{gamble2019networks}
Gamble JM. Sorting network generator.
  \url{http://pages.ripco.net/~jgamble/nw.html}, Website accessed 2019.

\bibitem{frazer1970samplesort}
Frazer WD, McKellar AC. Samplesort: A Sampling Approach to Minimal Storage Tree
  Sorting. {\it Journal of the ACM (JACM)} 1970\string; 17(3)\string: 496--507.

\bibitem{blelloch1991comparison}
Blelloch GE, Leiserson CE, Maggs BM, Plaxton CG, Smith SJ, Zagha M. A
  Comparison of Sorting Algorithms for the Connection Machine {CM-2}. In:  {\it
  3rd Symposium on Parallel Algorithms and Architectures (SPAA)}. ACM;
  1991\string: 3--16.

\bibitem{gerbessiotis1994direct}
Gerbessiotis AV, Valiant LG. Direct Bulk-Synchronous Parallel Algorithms. {\it
  Journal of Parallel and Distributed Computing} 1994\string; 22(2)\string:
  251--267.

\bibitem{axtmann2017robust}
Axtmann M, Sanders P. Robust Massively Parallel Sorting. In:  {\it 19th
  Workshop on Algorithm Engineering \& Experiments (ALENEX)}. SIAM. ;
  2017\string: 83--97.
\newblock preprint arXiv:1606.08766.

\bibitem{leischner2010gpu}
Leischner N, Osipov V, Sanders P. {GPU} sample sort. In:  {\it 24th {IEEE}
  International Symposium on Parallel \& Distributed Processing (IPDPS)}. IEEE.
  ; 2010\string: 1--10.

\bibitem{bingmann2013parallel}
Bingmann T, Sanders P. Parallel String Sample Sort. In:  {\it 21th European
  Symposium on Algorithms (ESA)}. . 8125 of {\it LNCS}. Springer; 2013\string:
  169--180.
\newblock preprint arXiv:1305.1157.

\bibitem{bingmann2018scalable}
Bingmann T. {\it Scalable String and Suffix Sorting: Algorithms, Techniques,
  and Tools}. PhD thesis. Karlsruhe Institute of Technology, Germany,  2018.

\bibitem{musser1997introspective}
Musser DR. Introspective Sorting and Selection Algorithms. {\it Software \&
  Practice and Experience} 1997\string; 27(8)\string: 983--993.

\bibitem{elmasry2012branch}
Elmasry A, Katajainen J, Stenmark M. Branch mispredictions don't affect
  mergesort. In:  {\it Symposium on Experimental Algorithms (SEA)}. . 7276 of
  {\it LNCS}. Springer. ; 2012\string: 160--171.

\end{thebibliography}
